\documentclass[fleqn,10pt,a4paper]{book}
\usepackage[latin1]{inputenc}
\usepackage{graphicx}
\usepackage{amssymb}
\usepackage{amsmath}
\usepackage{psfrag}
\usepackage{epsfig}
\usepackage{caption2}
\usepackage{fancyhdr}

 \fancyheadoffset{0cm}
\newcommand{\MSbar}{\overline{\mbox{MS}}}

\newcommand{\p}{\partial}
\newcommand{\oc}{\overline{c}}

\newcommand{\omu}{\overline{\mu}}
\newcommand{\lms}{\Lambda_{\overline{\mbox{\tiny{MS}}}}}
\newcommand{\Evac}{E_{\textrm{\tiny{vac}}}}
\newcommand{\mf}{\mathcal{F}}

\newcommand{\om}{\overline{m}}

\newcommand{\og}{\overline{g}}

\newcommand{\ol}{\overline{\lambda}}
\newcommand{\wrho}{\widehat{\varrho}}
\newcommand{\wm}{\widehat{m}}
\newcommand{\wl}{\widehat{\lambda}}

\renewcommand{\normalfont}{\sffamily}

\renewcommand{\leq}{\leqslant}

\begin{document}
\normalfont

\thispagestyle{empty}

\vspace*{-3cm}

\begin{center}

{\LARGE Rodrigo Ferreira Sobreiro}

\vspace*{6.5cm}

{\huge\bf Non-perturbative aspects of  } \\
\vspace{2mm} {\huge\bf Yang-Mills theories}

\vspace{5cm}

{\large Programa de Pós-Graduação em Física \\
Instituto de Física \\
Universidade do Estado do Rio de Janeiro}

\vspace{6cm}

Physics PhD Thesis
\\\vspace*{0.2cm}
February 2007 \\\vspace*{0.2cm}Advisor: Prof.\ Dr.\ Silvio Paolo
Sorella
\end{center}

\newpage
\thispagestyle{empty}
\noindent \textit{``Wise is the one who knows the limits of own ignorance.'' \\
    \bfseries Socrates}

\vspace{17.5cm}
\begin{flushright}

\noindent \textit{``Keep walking.'' \\
    \bfseries John Walker}

\end{flushright}

\thispagestyle{empty}

\

\vfill

\begin{flushright}
\hfill \textit{Dedicated to my daughter\\
and to the memories of my\\
father and grandmother.}
\end{flushright}

\vspace*{1cm}

\vphantom{lul}

\thispagestyle{empty}

\newpage

\thispagestyle{empty}

\enlargethispage*{28cm}

\begin{center}
 \emph{{\Large Acknowledgements}}
 \end{center}

Gostaria de expresar meus mais sinceros agradecimentos para todos
aqueles que, de uma forma ou de outra, contribuiram para a
realização deste trabalho. Não necessariamente na ordem de
importância, são os que cito.

Agradeço, portanto, a minha mãe, Maria de Fátima Ferreira, e a meu
avô, Enéas Ferreira, pelo apoio incondicional, moral e fianceiro, ao
longo de toda minha jornada pela Física e pela vida, não existem
palavras ou atos que possam retribuir-lhes o que fizeram por mim. À
minha filha, Tainara Sobreiro, que foi o motivo de eu amadurecer e
seguir com seriedade uma carreira, que eu lhe sirva de bom exemplo
para seu futuro. À Thais Rodrigues, mulher da minha vida, que eu
tanto admiro e amo, que tem me ajudado e apoiado a conseguir meus
objetivos. Sem ela, muito eu não teria conseguido neste fim de
projeto.

Não obstante agradeço ao meu orientador, Silvio Sorella, pela
orientação e paciência, ele que me mostrou os verdadeiros valores da
pesquisa científica e me ensinou a ver a Física de um jeito belo,
simples e eficiente. Sou muito grato ao Vitor Lemes pela
colaboração, basicamente uma coorientação e pelo bom humor
inabalável. Também ao Marcelo Sarandy pela colaboração na primeira
parte de meu doutorado. Agradeço também ao Ronaldo Thibes e Marcio
Capri pela colaboração na segunda parte de meu doutorado. Sentirei
falta dos cafés de discussão.

Ainda com relação aos colaboradores venho a agradecer ao {\it
Departamento de Física Matemática e Astronomia} da {\it Universidade
de Ghent}, Bélgica, pela hospitalidade e apoio financeiro em 2004 e
em 2005, pelos períodos que lá fiquei trabalhando em conjunto com
David Dudal e Henri Verschelde, aos quais serei sempre grato pela
colaboração. Em especial ao David também pela amizade. Igualmente a
John Gracey, pela colaboração na parte numérica e computacional do
trabalho, confirmando muitos dos resultados aqui presentes.

À CAPES, SBF, FAPERJ e CNPq pelo apoio financeiro direto ou
indireto.

Não devo deixar de citar os professores que foram importantes para
minha formação, José Sá Borges, Jim Skea, Jaime Rocha, Cesar
Linhares, Sergio Makler, Mirian Bracco, Silvio Sorella, Mauro
Tonasse e Vitor Lemes.

Agradeço ainda ao {\it Programa de Pós-Graduação em Física} da {\it
Universidade do Estado do Rio de Janeiro - UERJ}. Em particular aos
coordenadores do PPGF, Henrique Oliveira e Marcelo Chiapparini, bem
como às secretárias, Fernanda, Flávia, Katia, Keyla e Laurimar e ao
secretário Rogério.

Aos meus grandes amigos (não vou citar nenhum pois sei que
esquecerei alguém) e colegas. Ao Beto, agradeço também por me
auxiliar bastante com as figuras desta tese.

Ao Johnnie Walker, um dos responsáveis por eu continuar caminhando,
mesmo nos momentos mais difíceis.

À Ambev, pela amizade e companheirismo nos momentos de alegria e
tristeza, a única que nunca me abandonou ou magoou, a única que
sempre estava lá quando eu precisava, era só atravesar a rua
(qualquer uma). Sou eternamente grato e fiel a ela.

Também à Coca-Cola pela companhia nas manhãs, tardes e madrugadas de
trabalho em casa juntamente com Beethoven, Mozart, Shumann, Sonic
Youth, Stravinsky, Wagner, Verdi, Villa-Lobos, Vivaldi, entre
outros.

E, finalmente, a todos aqueles que por um motivo ou outro não estão
nesta injusta carta de agradecimento.

\newpage
\vphantom{lul}\thispagestyle{empty}\thispagestyle{empty}
\enlargethispage*{28cm}
\begin{center}
 \emph{{\Large Preface}}
 \end{center}

O tema escolhido para esta tese, os \emph{aspectos não perturbativos
das teorias de Yang-Mills}, não é um tema qualquer, constitui um dos
maiores desafios da Física atual. De fato, este problema foi
classificado pelo Instituto de Matemática CLAY como um dos sete
maiores problemas do milênio. Quem resolver o setor infravermelho
das teorias de Yang-Mills será agraciado com o prêmio de um milhão
de Dólares americanos.

Esta tese consiste numa compilação de toda a pesquisa feita nestes
cinco anos de doutorado no Instituto de Física da UERJ. Os
resultados desta pesquisa podem ser encontrados nos artigos e demais
trabalhos publicados
\cite{Ribeiro:2003yn,Dudal:2003pe,Dudal:2003tc,Dudal:2003np,Dudal:2003by,Sobreiro:2004us,Dudal:2004rx,Sobreiro:2004yj,Dudal:2004ch,Dudal:2005bk,
Dudal:2005na,Sobreiro:2005ec,Dudal:2005zr,Sobreiro:2005vn,Capri:2005tj,Capri:2005vw,Capri:2005dy,Capri:2005zj,Browne:2006uy,Capri:2006vv,Capri:2006ne,
Capri:2006bj,Dudal:2006ib,Dudal:2006tp,Capri:2006cz,Sorella:2006pj,Dudal:2006xd,Lemes:2006aw,Dudal:2006ip}.

Durante este período, foram feitas pesquisas em diversos calibres
importantes para o estudo do problema infravermelho, \emph{i.e.}, o
calibre de Landau, os calibres lineares covariantes, o calibre de
Curci-Ferrari e o calibre máximo Abeliano. Contudo, nesta tese, nos
fixamos no caso específico do calibre de Landau. Esta escolha foi
motivada pelo fato de este ser o calibre covariante mais simples
para quantizar as teorias de Yang-Mills. Ademais, fomos capazes de
identificar efeitos notáveis neste calibre. Não obstante,
eventualmente, ao longo do texto, pequenos comentários são feitos
sobre os demais calibres.

A estrutura de texto escolhida foi a seguinte: Em primeiro lugar,
não nos preocupamos em fornecer explicações sobre os tópicos da
Teoria Quântica de Campos. Para tal nos referimos a literatura
existente \cite{Itzykson,Peskin,Ryder,Weinberg}. Desta forma, esta
tese foi escrita para o leitor já experiente em TQC e, em
particular, em teorias de calibre. Por outro lado, o corpo do texto
foi desenvolvido de forma a deixar a leitura fluir, com o mínimo de
detalhes técnicos possíveis. Uma vez que todo aparato técnico pode
ser encontrado nos artigos acima citados, nos prendemos apenas à
Física envolvida. Neste sentido, muitas discussões físicas estarão
disponíveis.

As técnicas utilizadas para extrair nossos resultados foram alocadas
em apêndices. Desta forma, todo aparato sofisticado, alguns de
difícil acesso na literatura, foram compilados nestes apêndices.
Isso facilitará a vida do leitor mais interessado.

Desnecessário dizer, esta tese está longe de ser o fim desta
pesquisa e muito ainda deve ser investigado para uma compreensão
completa sobre a região infravermelha das teorias de Yang-Mills.

Esperamos que o leitor se divirta com esta tese tanto quanto nós nos
divertimos para desenvolvê-la.

\vfill
\begin{flushright}

\noindent Rio de Janeiro, 03 de Janeiro de 2007.\\
Rodrigo Ferreira Sobreiro

\end{flushright}

\newpage
\vphantom{lul}\thispagestyle{empty}\thispagestyle{empty}
\enlargethispage*{28cm}
\begin{center}
 \emph{{\Large Abstract}}
 \end{center}

Some nonperturbative aspects of Euclidean Yang-Mills theories in
four dimensions, quantized in the Landau gauge, are analytically
studied. In particular, we investigate the dynamical mass generation
for the gluons due to the presence of dimension two condensates.
This study is performed in the framework of the local composite
operator technique in the case of the Yang-Mills action as well as
in the case of the Gribov-Zwanziger action. Further, an
investigation of the Gribov ambiguities in the linear covariant
gauges is presented.

In the case of the Yang-Mills action, we perform a detailed analysis
of the local composite operator formalism when the operators
$A_\mu^aA_\mu^a$ and $f^{abc}\bar{c}^bc^c$ are simultaneously
introduced. Using the algebraic renormalization theory, we prove the
renormalizability of the model trough all orders in perturbation
theory. With the help of the renormalization group equations, a
quantum effective action for the condensates
$\langle{A}_\mu^aA_\mu^a\rangle$ and
$\langle{f}^{abc}\bar{c}^bc^c\rangle$ is constructed. This
construction is performed by means of the dimensional regularization
in the $\MSbar$ renormalization scheme. We show that nonvanishing
condensates values are dynamically favored, independently of the
chosen scale. Explicit one loop computations are then presented,
providing estimates for the condensates as well as for the vacuum
energy. With the help of suitable Ward identities, we are able to
formally show that the presence of the condensate
$\langle{f}^{abc}\bar{c}^bc^c\rangle$, is responsible for the
breaking of the transversality of the vacuum polarization. However,
the gluon propagator remains transverse, trough all orders in
perturbation theory. Finally, we end this analysis with the explicit
computation of the one loop correction to the effective gluon mass.
The result is that, due to the condensate
$\langle{f}^{abc}\bar{c}^bc^c\rangle$, the Abelian and the non
Abelian sectors acquire different masses. Since the non Abelian mass
is larger than the Abelian one, this feature can be interpreted as
an evidence for the Abelian dominance principle in the Landau gauge.

With respect to the Gribov-Zwanziger action, we start our analysis
without taking into account the condensate. We show that, at one
loop order, the vacuum energy is always positive, independently of
the renormalization scheme and scale. We also present attempts to
solve the one and two loops horizon condition in the $\MSbar$
scheme, and the respective failure of it.

Later, using the algebraic renormalization theory, we show, to all
orders in perturbation theory, the renormalizability of the
Gribov-Zwanziger action when the composite operator $A_\mu^aA_\mu^a$
is included in the framework of the local composite operator
technique. Attempts to solve the one loop gap equations in the
$\MSbar$ scheme are then presented. We show that there are no
solutions for $\langle{A}_\mu^aA_\mu^a\rangle<0$ while for
$\langle{A}_\mu^aA_\mu^a\rangle>0$ there is only one possible
solution for the horizon condition. However, in the $\MSbar$ scheme,
we were not able to find explicitly that solution. As an
improvement, an optimization of the renormalization scheme is
performed in the sense of minimizing the renormalization scheme
dependence. In that approach the $\MSbar$ limit solution was found
as well as the solution with minimal dependence on the
renormalization scheme. In both cases, the vacuum energy shows
itself to be positive while the condensate
$\langle{A}_\mu^aA_\mu^a\rangle>0$. A discussion of the consequences
of a nonvanishing Gribov parameter is then provided.

Finally, we present a study of the Gribov ambiguities in the linear
covariant gauges for the case $\alpha\ll1$, where $\alpha$ is the
gauge parameter. After the identification of a region in field space
where there are no close Gribov copies, we perform the respective
restriction in the path integral. As a tree level effect, the
transverse gluon propagator turns out to be infrared suppressed, due
to the presence of the Gribov horizon. The longitudinal component
shows itself to be infrared suppressed due to the dynamical mass,
associated with the condensate $\langle{A}_\mu^aA_\mu^a\rangle$.
Further, differently of the Landau gauge, the ghost propagator is
not related to the appearance of long range forces. Instead, an
infrared singular Green function related to the Gribov horizon is
identified. This Green function can be associated with long range
forces. At the end, all Landau gauge results can be recovered taking
the limit $\alpha\rightarrow0$.

\tableofcontents

\part{INTRODUCTION}

\chapter{Introduction}\label{cap1}

O tema principal desta tese consiste no estudo de fenômenos ainda
incompreendidos das chamadas teorias de Yang-Mills,
\cite{Yang:1954ek}. Desta forma, neste capítulo introdutório,
comecemos por discutir suas principais propriedades. Faremos uma
breve introdução sobre sua origem seguida da discussão sobre seus
aspectos físicos conhecidos mais interessantes. Nesta discussão
iremos começar a definir algumas quantidades e idéias importantes
para esta tese. Todos os detalhes técnicos podem ser encontrados nos
apêndices \ref{ap_nota} e \ref{ap_ferram}. Após esta discussão
apresentaremos um sumário sobre os efeitos não perturbativos das
teorias de Yang-Mills, em particular abordaremos a questão da massa
do glúon e o confinamento da cor, principais temas desta tese.

Uma recente e completa compilação sobre o entendimento das teorias
de Yang-Mills nos últimos cinquenta anos pode ser encontrado em
\cite{'tHooft:2005cq}.

Para uso futuro apresentaremos algumas fórmulas ao longo deste
capítulo. Abriremos mão de incluir matéria (férmions) nas discussões
que se seguem, uma vez que os campos de calibre já introduzem
complicações suficientes, como veremos ao longo desta tese. Desta
forma trabalharemos exclusivamente com as chamadas teorias puras de
Yang-Mills\footnote{Por simplicidade nos referiremos às
\emph{teorias de Yang-Mills puras} simplesmente por \emph{teorias de
Yang-Mills}. Da mesma forma, independentemente de estarmos no caso
$SU(N)$, muitas vezes nos referiremos aos \emph{campos de calibre}
como \emph{gluons}.}. Não obstante, nos referiremos aos quarks
sempre que for necessário e/ou um fenômeno se aplicar ao mesmo.

Ademais, não vamos adentrar nas questões de validade não
perturbativa da rotação de Wick \cite{Itzykson,Peskin}. Ao mesmo
tempo, queremos fazer cálculos explícitos na QCD. Desta forma
estaremos sempre imersos em um espaço-tempo quadrimensional
Euclideano.

\section{Yang-Mills theories}

\subsection{Origin}

Para entender a origem das teorias de Yang-Mills, podemos começar
pela bem co-nhecida Eletrodinâmica Quântica (QED\footnote{Do inglês:
\emph{Quantum Electrodynamics}.}),
\cite{Feynman:1949zx,Feynman:1950ir}. A QED é tida hoje em dia como
a teoria mais bem sucedida da Física, um título forte o suficiente
para tê-la como inspiração para outras teorias. Na metade do século
XX, suas predições já excediam espectativas e seu acordo com o
experimento era, e ainda é, surpreendente. Fisicamente falando, a
QED descreve a teoria quântica do eletromagnetismo, onde os elétrons
interagem entre si através dos fótons que, por sua vez, não
interagem entre si.

Além da beleza física, a QED possui uma bela estrutura matemática.
De fato, a QED pode ser entendida geometricamente como uma teoria de
calibre, \emph{i.e.}, a ação que descreve a teoria é invariante por
transformações de calibre associadas ao grupo de Lie $U(1)$. De
maneira formal, a QED é a teoria que descreve a dinâmica de uma
conexão imersa num fibrado principal com grupo de simetria Abeliano
$U(1)$, veja \cite{Bertlmann}. É o fato de a QED ser uma teoria
Abeliana que implica na propriedade de o fóton não ser capaz de
interagir consigo mesmo.

Analizando o sucesso da QED com relação aos resultados experimentais
bem como sua beleza e simplicidade teóricas, sua generalização para
o caso não Abeliano $SU(N)$ surge naturalmente no intuito de
explicar outras interações fundamentais da Natureza. É a
generalização da QED para o caso não Abeliano que define, nada mais
nada menos que as teorias de Yang-Mills. Nas teorias de Yang-Mills,
o análogo ao fóton, \emph{i.e.}, os campos de calibre interagem
entre si.

As teorias de Yang-Mills foram então utilizadas, com bastante
sucesso, na tentativa de explicar as interações nuclear fraca e
nuclear forte. A base para explicar as interações fracas foi baseada
num esquema de quebra espontânea de simetria $SU(2)$,
\cite{Weinberg:1967tq,Glashow:1970gm}. Desta forma os campos de
calibre adquirem massa através da interação com o chamado campo de
Higgs. Apesar do sucesso teórico para a geração de massa para os
bósons vetoriais, o Higgs ainda não foi encontrado
experimentalmente.

O segundo caso, a interação nuclear forte é descrita pela
Cromodinâmica Quântica (QCD\footnote{Do inglês: \emph{Quantum
Chromodynamics}.}), veja por exemplo
\cite{Gross:1973br,Politzer:1973br}. Neste modelo, o grupo de
simetria é o grupo $SU(3)$. Neste caso temos oito campos de calibre
que mediam a interação entre os quarks (cons-tituíntes dos
nucleons). Tais mediadores, denominados gluons\footnote{Do inglês:
\emph{glue}.}, devido a imensa força que eles geram entre os quarks,
são responsáveis pelo fenômeno do confinamento de quarks e gluons.
Sendo este o foco principal desta tese, vamos nos abster agora de
mais detalhes uma vez que estes serão expostos exaustivamente ao
longo da tese.

Resumidamente, as teorias de Yang-Mills surgiram no intuito de
descrever algumas das interações fundamentais da Natureza,
\emph{i.e.}, as forças nucleares e o eletromagetismo (caso Abeliano
limite). Não podemos deixar de comentar o fato da Gravitação não
estar incluída nesta lista. De fato a união da Gravitação com a
Mecânica Quântica consiste num dos maiores problemas da Física
atual, se equiparando ao problema do confinamento de quarks e
gluons. Contudo, nada mais falaremos sobre o assunto.

\subsection{Quantization}

Comecemos por definir um pouco mais refinadamente as teorias de
Yang-Mills. A ação que descreve a dinâmica dos campos de calibre é
dada por
\begin{equation}
S_{YM}=\frac{1}{4}\int{d^4x}F^a_{\mu\nu}F^a_{\mu\nu}\;.\label{1ym1}
\end{equation}
Classicamente esta ação já possui inúmeras propriedades
interessantes. Por exemplo, a invariância de calibre já elimina a
possibilidade de se encontrar soluções das equações de campos
consistentemente. Desta forma, mesmo classicamente, a fixação de
calibre se faz necessária. Ademais, devido aos termos de
autointeração entre gluons, as equações de campo associadas são não
lineares, permitindo soluções topológicas como, por exemplo,
monopólos e vórtices \cite{'tHooft:2005cq}. Contudo, estamos mais
interessados nos aspectos quânticos das teorias de Yang-Mills.

O método mais elegante e eficiente de se quantizar covariantemente
uma teoria é através da integral de caminho de Feynman,
\cite{Itzykson,Peskin}. Fisicamente, ao se calcular um valor
esperado com a integral de caminho, estamos sempre somando sobre
todas as configurações possíveis. Contudo, a simetria de calibre
destrói a interpretação probabilística de uma teoria quântica. A
simetria de calibre diz que um campo possui infinitas configurações
equivalentes. Desta forma, a simetria de calibre faz com que o mesmo
campo seja contado inúmeras vezes através de seus equivalentes. A
solução é fixar o calibre introduzindo um vínculo para o campo de
calibre. A fixação de calibre supostamente elimina a simetria de
calibre da ação (\ref{1ym1}).

Ao contrário da QED, a introdução de um vínculo de calibre na
integral de caminho não é simples. Um eficiente método de se
introduzir um vínculo na integral de caminho é o chamado
\emph{ansätz} de Faddeev-Popov, \cite{Faddeev:1967fc}. Com a
quantização de Faddeev-Popov, um vínculo pode ser introduzido na
integral de caminho com o custo de se introduzir campos extras, os
chamados campos fantasmas de Faddeev-Popov. Tais campos são
escalares que obedecem a estatística de Fermi. De acordo com o
teorema spin-estatística \cite{Wightman}, estes campos violariam a
causalidade. Contudo, os campos fantasmas aparecem apenas em laços
fechados e não são, portanto, excitações observáveis da teoria.
Notavelmente, estes campos são responsáveis pela unitariedade da
teoria, seus diagramas cancelam graus de liberdade não físicos da
série perturbativa \cite{Kugo:1979aa}.

De acordo com a quantização de Faddeev-Popov, \cite{Faddeev:1967fc},
o termo de fixação de calibre covariante mais simples, a ser
adicionado a ação de Yang-Mills (\ref{1ym1}), pode ser escrito como
\begin{equation}
S_{gf}=\int{d^4x}\left(b^a\partial_\mu{A}_\mu^a+\bar{c}^a\partial_\mu{D}_\mu^{ab}c^b\right)\;,\label{1landau1}
\end{equation}
onde $b^a$ é o chamado campo de Lautrup-Nakanishi, que funciona como
um multiplicador de Lagrange para o vínculo. Os campos $c^a$ e
$\bar{c}^a$ são os campos de Faddeev-Popov. Este vínculo define o
chamado calibre de Landau\footnote{A menos que o contrário seja
dito, calibre de Landau será o calibre utilizado nesta tese para
quantizar as teorias de Yang-Mills. Obviamente esta não é a única
possibilidade, veja Apêndice \ref{ap_ferram}.}, cujo vínculo filtra
o espaço de configurações contando apenas os campos transversos
\begin{equation}
\partial_\mu{A}_\mu^a=0\;.\label{1landau2}
\end{equation}
A ação com calibre fixado é, portanto,
\begin{equation}
S=S_{YM}+S_{gf}\;.\label{1action1}
\end{equation}

Ao fixar o calibre como descrito anteriormente a simetria de calibre
é quebrada. Contudo, surge uma nova simetria na ação
(\ref{1action1}), a chamada simetria BRST,
\cite{Becchi:1975nq,Tyutin:1975qk}. Definindo o operador BRST como
$s$, a ação $S$ se mostra invariante sob as seguintes transformações
\begin{eqnarray}
sA_\mu^a&=&-D_\mu^{ab}c^b\;,\nonumber\\
sc^a&=&\frac{g}{2}f^{abc}c^bc^c\;,\nonumber\\
s\bar{c}^a&=&b^a\;,\nonumber\\
sb^a&=&0\;.\label{1brs1}
\end{eqnarray}
ou seja,
\begin{equation}
sS=0\;.\label{1inv1}
\end{equation}
A simetria BRST é o tipo mais simples de supersimetria. Ainda, esta
simetria é extremamente útil para se mostrar a renormalizabilidade
de uma teoria. Outra importante propriedade de $s$ é o fato de ser
um operador nilpotente, $s^2=0$. Esta propriedade permite provar a
unitariedade da matriz de espalhamento de uma teoria, dando o
sentido físico final a teoria, \cite{Kugo:1979aa}. Não é de se
admirar que a simetria BRST terá um papel fundamental nesta tese.

Uma extraordinária propriedade das teorias de Yang-Mills consiste em
sua interpretação geométrica \cite{Bertlmann}. De fato, a ação de
Yang-Mills descreve a dinâmica da conexão $A_\mu^a$ definida num
fibrado principal, não trivial, de grupo de simetria $SU(N)$, e
espaço de imersão $\mathbb{R}^4$. Neste contexto, os campos
fantasmas são as uma-formas de Maurer-Cartan enquanto que a variação
BRST é isomorfa a derivada exterior. A fixação de calibre consiste
na definição de uma seção no fibrado. Contudo, a Topologia diz que
definir uma seção global em um fibrado não trivial não é algo
possível. Veremos adiante que este é um dos problemas fundamentais
das teorias de calibre.

\subsection{Renormalization}

Como foi observado anteriormente, a simetria BRST é muito útil para
a renormali-zação de uma teoria. De acordo com a teoria de
renormalização algébrica\footnote{Veja apêndice \ref{ap_ferram}.},
\cite{book}, para trabalhar a renormalizabilidade das teorias de
Yang-Mills introduzimos mais um termo à ação (\ref{1action1}), com
fontes externas acopladas às transformações BRST não lineares, de
forma a podermos descrever a simetria BRST compativelmente ao
princípio de ação quântica,
\cite{Schwinger:1951xk,Lowenstein:1971vf,Lowenstein:1971jk,Lam:1972mb,Clark:1976ym}.
Este termo é dado por
\begin{eqnarray}
S_{ext}&=&s\int{d^4x}\left(-\Omega^a_\mu{A}_\mu^a+L^ac^a\right)\nonumber\\
 &=&\int{d^4x}\left(-\Omega^a_\mu{D}_\mu^{ab}c^b+\frac{g}{2}f^{abc}L^ac^bc^c\right)\;,\label{1ext1}
\end{eqnarray}
onde as fontes externas são $s$-invariantes,
\begin{equation}
s\Omega_\mu^a=sL^a=0\;.\label{1brs2}
\end{equation}
Desta forma, a ação apropriada para se estudar a renormalizabilidade
das teorias de Yang-Mills é simplesmente
\begin{equation}
\Sigma=S+S_{ext}\;.\label{1action2}
\end{equation}
Esta ação, e suas generalizações, serão amplamente utilizadas nos
capítulos que se seguem. De fato, a ação (\ref{1action2}) é
renormalizável, com apenas duas divergências a serem renormalizadas,
uma associada a constante de acoplamento $g$ e outra asscociada ao
campo de calibre $A_\mu^a$. A renormalizabilidade é feita de maneira
recursiva e vale a todas as ordens em teoria de perturbações.

\subsection{Asymptotic freedom and the infrared problem}

Como resultado da renormalizabilidade, as teorias de Yang-Mills
obedecem também ao grupo de renormalização \cite{Wilson:1974sk}.
Fisicamente, existe uma invariância de escala na teoria. Desta
forma, após a renormalização de uma certa quantidade, teremos a
dependência na escala de renormalização $\mu$. Em particular, a
constante de acoplamento renormalizada perturbativamente, a um laço,
apresenta a forma
\begin{equation}
g^2(\mu)=\frac{1}{\frac{11N}{16\pi^2}\ln\frac{\mu^2}{\Lambda_{QCD}^2}}\;,\label{1g1}
\end{equation}
onde $N$ é a dimensão do grupo e
$\Lambda_{QCD}\approx237\mathrm{MeV}$ um corte de escala. Desta
expressão se extrai o conceito de \emph{liberdade assintótica}
\cite{Gross:1973br,Politzer:1973br}. Para tal, analizando a
expressão (\ref{1g1}) vemos que para escalas de energia grandes a
constante de acoplamento tende a zero. Neste regime teríamos quarks
e gluons quase livres.

Enquanto $g$ for pequeno a teoria de perturbações utilizada na
expansão em laços, que possui $g$ como parâmetro de expansão, tem
sentido. Vemos que isso ocorre para altas energias. Neste caso
dizemos que estamos no \emph{regime perturbativo} ou na \emph{região
ultravioleta}. Neste regime, a QCD se mostra muito bem sucedida uma
vez que é capaz de fazer predições com precisão experimental, dando
credibilidade nas teorias de Yang-Mills como teoria fundamental para
as interações fortes.

Por outro lado, conforme a energia diminui a constante de
acoplamento começa a crescer. Até que a escala atinge o valor
$\mu^2=\Lambda^2_{QCD}\exp\{16\pi^2/11N\}$ fazendo com que $g=1$. A
partir daí a expansão perturbativa deixa de fazer sentido e a teoria
entra no chamado \emph{regime não perturbativo} ou na \emph{região
infravermelha}. Contudo, podemos ser levianamente teimosos, de forma
que, diminuindo mais ainda a energia, chegamos ao chamado \emph{polo
de Landau}, $\mu=\Lambda_{QCD}$. Neste ponto a constante de
acoplamento explode, indicando uma escala em que não podemos definir
a teoria, nem mesmo levianamente. Porém, se a energia diminuir mais
ainda, a constante de acopalmento passa a ser imaginária e a teoria
perde completamente o sentido.

Podemos interpretar este problema como um limite de validade da
quantização das teorias de Yang-Mills. De fato, este limite está
vinculado à validade da teoria de perturbações. A quantização como a
conhecemos pode ser caracterizada como uma quantização perturbativa.
Infelizmente um método não perturbativo capaz de suplantar a teoria
de perturbações não está disponível. Desta forma, a região
infravermelha ainda permanece insolúvel. A este problema se dá o
nome de \emph{problema infravermelho}.

Devemos ressaltar que se acredita que as teorias de Yang-Mills
descrevam realmente a QCD a baixas energias. Essa crença vem do
sucesso da teoria no setor de altas energias confirmando a
existência de quarks, da liberdade assintótica e toda a predição em
Física hadrônica que a QCD é capaz de fazer. Mesmo que não na forma
(\ref{1ym1}), é de lá que devemos partir para chegar a uma descrição
consistente da QCD a baixas energias em termos de hadrons e bolas de
gluons\footnote{Do inglês: \emph{glueballs}.}.

\subsection{Non-perturbative methods}

Como discutido anteriormente, não conhecemos um algorítimo capaz de
resolver uma teoria quântica no regime de acoplamento forte, o que
se faz neste caso são aproximações, tentativas de aprimoração da
teoria de perturbações etc... Neste estudo são encontradas muitas
evidências de fenômenos não perturbativos que possuem influência na
região infravermelha.

Em particular, o melhor método não perturbativo é a chamada
\emph{QCD na rede}, \cite{Wilson:1974sk}, onde o espaço-tempo
Euclideano é discretizado formando uma rede de pontos espaciais.
Desta forma o espaçamento entre os pontos da rede é capaz de
regularizar tanto as divergências ultravioletas como as
infravermelhas. Contudo, este método exige um poderoso aparato
computacional para efetuar suas simulações numéricas de forma a
resolver diretamente a integral de caminho. Tratamentos analíticos
na rede também existem, mas não possuem o poder das simulações
numéricas. Obviamente existem problemas na rede a serem enfrentados,
por exemplo, o limite ao contínuo.

Um outro método de análise não perturbativa muito utilizado é o
estudo através das equações de Schwinger-Dyson,
\cite{vonSmekal:1997is,vonSmekal:1997is2,Atkinson:1997tu,Atkinson:1998zc,Alkofer:2000wg,Watson:2001yv,Zwanziger:2001kw,
Lerche:2002ep,Fischer:2006ub}. Neste método são feitos \emph{ansätz}
para os propagadores e suas propriedades, tais \emph{ansätz} são
substituídos nas equações de Schwinger-Dyson. Para resolver estas
equações são utilizados métodos numéricos em união com considerações
fenomenológicas.

Métodos puramente analíticos também são utilizados para o
endendimento dos aspectos infravermelhos das teorias de Yang-Mills.
Por exemplo, são feitos os chamados estudos semi-perturbativos,
através das equações do grupo de renormalização. Neste tratamento a
teorias de Yang-Mills são estudadas no âmbito dos fenômenos
críticos, pontos fixos e transições de fase. Muitas vezes a
temperatura se torna útil para estudar estas duas fases da QCD,
ultravioleta e infravermelha.

Outra aplicação analítica no setor infravermelho é o estudo de
soluções clássicas estáveis, os bem conhecidos instantons
\cite{'tHooft:2005cq}. Tais configurações de vácuo são utilizadas
para tentar definir o vácuo da QCD e utilizá-lo na expansão
perturbativa. Ainda se tratando do vácuo da QCD, existe o interesse
recente em operadores com dimensão de massa, cuja condensação pode
influenciar o vácuo da QCD \cite{Dudal:2003tc}. Tais condensados são
estudados através de diferentes técnicas, desde a expansão em
produtos de operadores (OPE\footnote{Do inglês: \emph{Operator
Product Expansion}.}) até a técnica de operadores compostos locais
(Técnica LCO\footnote{Do inglês: \emph{Local Composite Operator
Technique}.}) \cite{Verschelde:2001ia}. Em particular, a técnica LCO
é um dos métodos empregados nesta tese para o estudo de operadores
locais de dimensão dois.

Devemos lembrar que operadores de dimensão dois não são os únicos de
interesse na QCD, em particular condensados de vácuo com outras
dimensões possuem um importante papel no comportameno infravermelho
das teorias de Yang-Mills. Exemplos famosos são o condensado de
gluons $\left\langle F_{\mu\nu}^2\right\rangle$ e o condensado de
quarks $\left\langle \overline{q}q\right\rangle$ associado a geração
de massa dos quarks, veja \cite{Shifman:1978bx} para mais detalhes.

Ainda, existem estudos baseados em uma dualidade entre as teorias de
Yang-Mills e outras teorias que descreveriam a QCD a baixas energias
\cite{'tHooft:2005cq}. Em particular, uma dualidade bastante
promissora consiste no caso AdS/CFT, veja por exemplo
\cite{Boschi-Filho:2006fi}, onde a QCD a baixas energias seria
descrita por uma teoria efetiva em termos de monopólos e vórtices.
Mesmo no supercondutor Abeliano existe uma dualidade semelhante onde
a teoria efetiva pode ser descrita em termos de vórtices e pares de
Cooper, veja \cite{Ribeiro:2003yn} para o caso de três dimensões.

Finalmente, um dos mais importante tratamentos consiste no
entendimento da própria quantização das teorias de Yang-Mills. De
fato, como discutimos acima, a quantização parece falhar no limite
de baixas energias, desta forma se faz necessária a aprimoração do
método de quantização das teorias de Yang-Mills. Este problema está
associado às chamadas \emph{ambigüidades de Gribov}
\cite{Gribov:1977wm,Sobreiro:2005ec}. Esta questão será discutida na
próxima seção e em detalhes na parte III desta tese.

De qualquer forma existe um consenso entre todos os métodos
empregados: O pro-blema infravermelho, ou regime não perturbativo,
está associado a dois fenômenos físicos tão complexos quanto
interessantes, o \emph{confinamento de quarks e gluons} e a
\emph{geração dinâmica de massa}. É fácil entender que estes dois
efeitos não sejam totalmente independentes entre si. Estes dois
efeitos compoem o tema desta tese, vamos entrar em mais detalhes na
próxima seção. Ainda, veremos um terceiro fenômeno, que dará suporte
aos dois previamente citados, o chamado \emph{princípio de
dominância Abeliana}.

\section{Some non-perturbative effects of QCD}

\subsection{Quark and gluon confinement}

Através da análise feita da constante de acoplamento renormalizada
(\ref{1g1}), vimos que atravessar o polo de Landau sugere um tipo de
transição de fase QCD. Uma fase onde quarks e gluons estariam
confinados devido ao grande valor da constante de acoplamento. Na
transição de fase ocorreria a formação de hadrons e bolas de gluons.
Esta hipótese é confirmada sob o ponto de vista experimental uma vez
que quarks e gluons não podem ser observados isoladamente. Não
obstante, de acordo com a rede, o laço de Wilson possui a
intepretação de um parâmetro de ordem das fases da QCD. Sob o ponto
de vista da rede é indiscutível a confirmação da QCD confinante a
baixas energias.

Uma forma ilustrativa simples de se entender o confinamento é
imaginando um estado ligado de dois quarks a baixas energias. Se
tentamos separá-los, a energia entre eles tende a aumentar devido a
seu acoplamento forte. Quando mais energia damos para separá-los
mais energia precisamos. Finalmente, chegamos a uma quantidade de
energia capaz de criar um novo par quark-antiquark. Neste momento
conseguimos separar os dois quarks iniciais, contudo, cada um deles
se liga a um dos novos quarks criados com a energia fornecida.
Terminamos com dois estados ligados novamente confinados.

Técnicamente, o confinamento caracteriza a impossibilidade de
existirem observáveis carregando índices de cor, ou seja,
quantidades que carreguem carga de cor. Todos as quantidades
observáveis devem ser sem cor, ou seja, sem índices livres na
álgebra do grupo $SU(N)$. Fisicamente, todos os observáveis a baixas
energias seriam hadrons e bolas de gluons, lembrando que gluons
carregam carga de cor, e portanto não são observáveis.

Existem duas propostas principais para se explicar o confinamento, o
mecanismo de supercondutividade dual e o cenário de
Gribov-Zwanziger. Nenhum destes tem a palavra final sobre o
confinamento e é muito possível que sejam equivalentes ou ainda,
necessários um ao outro para uma compreensão completa do fenômeno.

\subsubsection{Dual superconductivity mecanism}

Como o próprio nome sugere, o confinamento através da
supercondutividade dual, \cite{Nambu,'tHooft,Mandelstam}, é uma
idéia inspirada no supercondutor Abeliano do tipo II. Suponhamos que
além do supercondutor, temos a presença de monopólos magnéticos.
Podemos pensar num tipo de fase supercondutora da QED quando existem
monopólos de Dirac. Neste sistema, a condensação dos pares de Cooper
produzem tubos de fluxo magnéticos cuja energia é proporcional a seu
comprimento. Tais tubos possuem como fontes as cargas magnéticas de
Dirac, que, por sua vez, ficam aprisionadas através dos tubos
magnéticos. Temos assim, o confinamento de cargas magnéticas através
de tubos de fluxo magnéticos.

Ainda, o sistema descrito acima possui uma segunda propriedade
extraordinária, e-xiste uma simetria dual entre os setores elétrico
e magnético. Desta forma, na teoria dual, teríamos cargas elétricas
confinadas através de tubos de fluxo elétricos.

A generalização deste mecanismo para o caso não Abeliano pode prover
uma explicação para o confinamento de quarks e gluons. Assim, cargas
cromoelétricas se confina-riam através de tubos cromoelétricos. A
vantagem é que no caso não Abeliano, monopólos constituem
configurações de vácuo estáveis, ou seja, não necessitamos
incluí-los na mão, como no caso da QED.

Ainda, este mecanismo sugere o desacoplamento entre os setores
\emph{elétrico} (Abeliano) e \emph{magnético} (não Abeliano) da
teoria. Tal decomposição deu origem ao estudo dos chamados calibres
Abelianos, onde a fixação de calibre é feita de forma diferente nos
setores Abeliano e não Abeliano \cite{'tHooft:1981ht}.

\subsubsection{Gribov-Zwanziger scenario}

Outro mecanismo que poderia explicar o confinamento está relacionado
à própria quantização das teorias de Yang-Mills. Em
\cite{Gribov:1977wm}, Gribov chamou a atenção para o fato de que
fixar o calibre não é suficiente para eliminar a simetria de
calibre. De fato, uma simetria de calibre residual sobrevive ao
processo de quantização perturbativa e seus efeitos se tornam
evidentes na região infravermelha. Este problema, chamado
ambigüidades de Gribov, é na verdade uma patologia das teorias de
Yang-Mills, e não de um calibre específico; As ambigüidades de
Gribov existirão em qualquer calibre, \cite{Singer:1978dk}.

Uma quantização mais eficiente no caso do calibre de Landau, dando
conta das ambigüidades de Gribov, colocaria a teoria no verdadeiro
vácuo da QCD. Neste vácuo gluons deixam de fazer parte do espectro
físico da teoria e pelo menos uma grande parte das divergências
infravermelhas são eliminadas. Mais recentemente foram encontradas
evidências de que configurações topológicas de vácuo habitam o vácuo
de Gribov-Zwanziger \cite{Maas:2005qt}. Ainda, neste vácuo, pode-se
identificar o surgimento de forças de longo alcance, indicando
confinamento.

Um ponto importante é o fato de que algumas divergências
infravermelhas sobreviventes podem vir a possuir interpretações
físicas fundamentais para o próprio mecanismo de confinamento,
contudo, mudando o calibre, tais interpretações se mostram um tanto
obscuras
\cite{Sobreiro:2005vn,Capri:2005tj,Capri:2006vv,Dudal:2006ib,Capri:2006cz}.
Este problema indica que o mecanismo de Gribov-Zwanziger de
confinamento pode variar de calibre para calibre.

Este mecanismo vincula o confinamento e as divergências
infravermelhas ao problema mais fundamental de uma teoria, o
problema de quantizá-la consistentemente. Sendo um dos principais
temas desta tese, este problema será discutido em detalhes na
terceira parte desta tese. Note ainda que, mesmo que este cenário
não seja capaz de dar uma completa explicação ao confinamento e/ou
dar conta das divergências infravermelhas, o tratamento das
ambigüidades de Gribov se faz necessário para uma teoria quântica
consistente e completa.

\subsection{Dynamical mass generation}

Como vimos na seção anterior, as teorias de Yang-Mills estão
impregnadas de divergências infravermelhas. Este não é um problema
específico do caso quadridimensional, por exemplo, em três dimensões
a teoria é superenormalizável, contudo as divergências
infravermelhas são ainda mais patológicas. Surge então a pergunta:
Se as teorias de Yang-Mills realmente descrevem o setor
infravermelho, como a própria teoria se livra das divergências?

Vimos anteriormente que o tratamento das ambigüidades de Gribov
resolvem grande parte do problema infravermelho, contudo, as
divergências residuais possuem um caráter estranho devido a
dependência no calibre escolhido. Uma hipótese bastante aceita que
resolveria completamente o problema das divergências infravermelhas
é que exista o chamado \emph{gap de massa}, ou seja, de alguma
forma, existe uma massa dentro da teoria que regularizaria as
divergências infravermelhas de maneira invariante de calibre.

Note ainda, que, devido ao confinamento e ao fato de os gluons
interagirem entre si, podem existir estados ligados compostos
puramente por gluons, as bolas de gluons. É muito difícil aceitar
que tais estados sejam formados por excitações não massivas, ou
seja, deveríamos ter excitações viajando na velocidade da luz
formando um estado ligado com uma massa considerável, da ordem de
pelo menos $\approx1.63\mathrm{GeV}$ \cite{Boschi-Filho:2006fi}.

Assim, esta massa apareceria após o processo de renormalização da
teoria, devido a efeitos quânticos internos. Portanto, a teoria
seria renormalizável, unitária e, ainda, livre de divergências
infravermelhas.

Muitos dos métodos não perturbativos discutidos na seção anterior
mostram evidências da massa dinâmica. No caso da QCD na rede,
parâmetros de massa são utilizados para ajustar os dados obtidos
\cite{Cucchieri:1999sz,Bonnet:2001uh,Langfeld:2001cz,Cucchieri:2003di,Bloch:2003sk,Furui:2003jr,Furui:2004cx}.
Contudo, na rede, é muito difícil determinar a origem destes
parâmetros. No âmbito fenomenológico, temos as equações de
Schwinger-Dyson, onde parâmetros de massa são utilizados nos
propagadores como \emph{ansätz} para resolver as equações. Em
particular, o estudo de condensados de dimensão dois pode dar muitas
evidências da existência dessa massa, \cite{Dudal:2003tc}. Sendo
este último um principal tema desta tese, vamos deixar a discussão
para a próxima parte.

É importante ter em mente que simplesmente uma teoria de Yang-Mills
massiva não é a resposta para esta questão. Tal modelo possui
problemas de renormalizabilidade, de fato, este modelo só é
renormalizável no calibre de Landau, \cite{Dudal:2002pq}, e mesmo
assim, devido ao fato de as transformações BRST não serem
nilpotentes, o modelo não é unitário.

\subsection{Abelian dominance}

Para terminar este capítulo introdutório, vamos entender o princípio
de dominância Abeliana \cite{Ezawa:bf}. Tal princípio diz que, a
baixas energias, a QCD seria descrita apenas por graus de liberdade
Abelianos. Note que existem confirmações deste fenômeno também na
rede \cite{Suzuki:1989gp,Suzuki:1992gz,Hioki:1991ai}.

Esta idéia não só é compatível com a idéia de confinamento via
supercondutividade dual e da geração dinâmica de massa, como reforça
tais idéias. Entender esta idéia é muito simples. No caso do
confinamento via supercondutividade dual, vimos que este mecanismo
requer um desacoplamento entre os setores Abeliano e não Abeliano.
Da mesma forma, a dominância Abeliana requer tal desacoplamento.

Com relação à geração dinâmica de massa a questão seria um pouco
mais elaborada. Suponhamos que exista uma massa dinâmica para o
glúon. Neste caso, se houvesse a quebra de degenerescência desta
massa, teríamos o desacoplamento entre gluons Abelianos e não
Abelianos. Ainda, a massa não Abeliana deveria  ser maior que a
massa Abeliana. Desta forma, para energias menores que a massa não
Abeliana, este setor desacoplaria do espectro uma vez que não
haveria energia para criar pares com tal massa. Finalmente, a teoria
estaria composta apenas por graus de liberdade Abelianos.

\chapter{This thesis}

Vimos na introdução como os efeitos não perturbativos são
importantes para se determinar o comportamento infravermelho das
teorias de Yang-Mills. Obviamente, a discussão apresentada é
suficiente para motivar toda e qualquer pesquisa sobre efeitos não
perturbativos das teorias de Yang-Mills. No entanto, vamos ser um
pouco mais específicos neste sentido e enumerar objetivamente as
motivações deste estudo. Em seguida apresentaremos nossas propostas
e os respectivos resultados.

\section{Motivations}

As pricipais motivações para se estudar os \emph{Aspectos não
perturbativos das teorias de Yang-Mills} podem ser resumidos em um
único: O entendimento da região infravermelha da QCD. Contudo,
apesar de um grande estudo ter sido feito, muitos aspectos foram
deixados de lado em benefício de uma profunda pesquisa pesquisa em
dois aspectos particulares: A geração dinâmica de massa devido a
condensação de operadores de dimensão dois e as ambigüidades de
Gribov.

A principal motivação para a massa dinâmica vem da hipótese sobre a
existência de um regularizador infravermelho. Ainda, temos o apoio e
incentivo proveniente de simulações numéricas na rede. Na QCD na
rede, é muito difícil determinar a origem física de parâmetros de
massa utilizados no ajuste de dados
\cite{Cucchieri:1999sz,Bonnet:2001uh,Langfeld:2001cz,Cucchieri:2003di,Bloch:2003sk,Furui:2003jr,Furui:2004cx}.
Desta forma, um estudo puramente analítico se faz necessário para se
compreender tais parâmetros. Da mesma forma, massas são utilizadas
para parametrizar propagadores no estudo através das equações de
Schwinger-Dyson. Introduzidos através de \emph{ansätze}, tais
parâmetros são interpretados como uma geração dinâmica de massa,
contudo, sua origem física permanece obscura
\cite{vonSmekal:1997is,vonSmekal:1997is2,Atkinson:1997tu,Atkinson:1998zc,Alkofer:2000wg,Watson:2001yv,Zwanziger:2001kw,
Lerche:2002ep,Fischer:2006ub}.

Desta forma o estudo puramente analítico é muito bem vindo no
intuito de dar evidências sobre a origem física destes parâmetros.
Em particular, para encontrar evidências analíticas da massa
dinâmica optamos por um estudo detalhado de operadores de dimensão
dois para possível condensação. Tal escolha é motivada pelo fato de
haver um método analítico bastante completo para estudar estes
condensados, o chamado método LCO \cite{Verschelde:2001ia}.

O outro aspecto não perturbativo amplamente discutido nesta tese
dispensa muitos motivos. As ambigüidades de Gribov estão
relacionadas com a própria quantização das teorias de Yang-Mills e
seu comportamento infravermelho. Ademais, existe ainda uma motivação
extra, o fato de existir uma ação local e renormalizável que
descreve as teorias de Yang-Mills no calibre de Landau com a
eliminação de um grande número das ambi-guidades de Gribov,
\cite{Zwanziger:1989mf,Zwanziger:1992qr,Maggiore:1993wq}, a chamada
\emph{ação de Gribov-Zwanziger}. Com esta ação a integral de caminho
possui domínio de integração restrito a uma região finita do espaço
funcional das configurações de calibre, a chamada região de Gribov.
Tal ação fornece o cenário ideal para se fazer cálculos explícitos.

Ainda, devido ao fato de o estudo sobre as ambigüidades de Gribov
ter sido desenvolvido exclusivamente nos calibres de Landau e
Coulomb, temos bons motivos para iniciar discussões sobre as
ambigüidades de Gribov em outros calibres.

\section{Proposals and results}

Nossa proposta consiste no estudo de condensados de dimensão dois
através do método LCO de forma a estudar a geração dinâmica de massa
para os gluons nas teorias de Yang-Mills. Com este método, somos
capazes de analizar a renormalizabilidade de um operador de massa e
calcular o potencial efetivo associado a este operador. Desta forma
podemos fazer uma análise deste potencial através do grupo de
renormalização e procurar soluções que favoreçam dinamicamente um
condensado não trivial. Tal condensado pode ser relacionado a uma
massa gerada dinamicamente para os gluons.

De fato, partimos de um estudo de diversos operadores de dimensão
dois invariantes de calibre, não locais. Mostramos que existem duas
classes distintas destes operadores. Tais operadores são
incompatíveis com o método LCO\footnote{Veja apêndice
\ref{ap_ferram}.}. Contudo, mostramos que as duas classes estão
relacionadas, no calibre de Landau, com o operador local
$A_\mu^aA_\mu^a$ que não é invariante de calibre. Desta forma, o
estudo da condensação deste operador pode dar evidências importantes
sobre a geração de massa e pistas sobre um regularizador invariante
de calibre proveniente da condensação de operadores invariantes de
calibre.

Uma outra classe de operadores compostos locais de dimensão dois são
os operadores fantasmas cujo efeito é gerar massa taquiônica para o
glúon \cite{Dudal:2002xe,Dudal:2003dp}. Nesta tese apresentamos um
estudo conjunto do operadores $A_\mu^aA_\mu^a$ e
$gf^{abc}\bar{c}^bc^c$ através do método LCO, apresentando
resultados formais e resultados explícitos bem como interessantes
efeitos físicos. No caso de cálculos explícitos, nos limitamos ao
caso $SU(2)$, por motivos de simplicidade. Esta análise, disposta no
Capítulo 4, tem como resultados, publicados em \cite{Capri:2005vw},
os que se seguem:

\begin{itemize}

\item O formalismo LCO para os operadores $A_\mu^aA_\mu^a$ e
$f^{abc}\bar{c}^bc^c$ é renormalizável a todas as ordens em teoria
de perturbações. A prova foi feita em conjunto com a teoria de
renormalização algébrica.

\item Os operadores compostos $A_\mu^aA_\mu^a$ e
$f^{abc}\bar{c}^bc^c$ não possuem dimensão anômala independente. De
fato, devido às relações entre os fatores de renormalização
multiplicativa,
\begin{eqnarray}
Z_{A^2}&=&Z_gZ_A^{-1/2}\;,\nonumber\\
Z_{f\bar{c}c}&=&Z_A^{1/2}\;,\label{2r1}
\end{eqnarray}
temos que as dimensões anômalas correspondentes são dadas por
\begin{eqnarray}
\gamma_{A^2}(g^2)&=&-\left[\frac{\beta(g^2)}{2g^2}+\gamma_A(g^2)\right]\;,\nonumber\\
\gamma_{f\bar{c}c}(g^2)&=&\gamma_A(g^2)\;,\label{2r2}
\end{eqnarray}
preservando os resultados anteriores, onde os operadores foram
estudados separadamente. Em (\ref{2r2}) $\beta(g^2)$ e
$\gamma_A(g^2)$ são, respectivamente, as dimensões anômalas da
constante de acoplamento e do campo de calibre.

\item Demonstramos explicitamente\footnote{A menos que o contrário seja especificado,
todas os resultados apresentados daqui para frente, nesta lista,
foram obtidos para o caso $SU(2)$ a um laço no esquema de
renormalização $\MSbar$ utilizando regularização dimensional.} que a
ação quântica efetiva obedece a uma equação homogênea do grupo de
renormalização.

\item A existência de condensados não trivias $\langle A_\mu^aA_\mu^a\rangle$ e $\langle f^{abc}\bar{c}^bc^c\rangle$ é favorecida dinamicamicamente,
uma vez que a energia do vácuo correspondente é sempre negativa,
independentemente da escala escolhida.

\item Calculamos de forma consistente com o grupo de renormalização,
a energia do vácuo, bem como os valores dos condensados $\langle
A_\mu^aA_\mu^a\rangle\propto\overline{m}^2$ e
$\langle\varepsilon^{abc}\bar{c}^bc^c\rangle\propto\widetilde{\omega}$.
Estes são,
\begin{eqnarray}
\overline{m}^2&\approx&\nonumber3.07\lms^2\;,\nonumber\\
\widetilde{\omega}&\approx&18.48\lms^2\;,\nonumber\\
\Evac&\approx&-1.15\lms^4\;.\label{2r3}
\end{eqnarray}

\item Foi mostrado para o caso geral $SU(N)$ que o propagador do
glúon permanece transverso a todas as ordens em teoria de
perturbações.

\item Da mesma forma, foi demonstrado que a polarização do vácuo não
é trasnversa devido, exclusivamente, a presença do condensado
$\langle\varepsilon^{abc}\bar{c}^bc^c\rangle$.

\item Calculamos explicitamente os valores da massa do glúon gerada
dinamicamente por estes condensados levando em consideração as
correções a um laço, encontrando os valores
\begin{eqnarray}
m_{\mathrm{Ab}}^2&\approx&1.66\lms^2\;,\nonumber\\
m_{\mathrm{nAb}}^2&\approx&2.34\lms^2\;,\label{2r4}
\end{eqnarray}
de onde identificamos a quebra da degenerescência da massa do glúon,
associada a quebra da transversalidade da polarização do vácuo. Este
resultado pode ser interpretado como uma evidência da dominância
Abeliana no calibre de Landau.

\end{itemize}

No Capítulo 6, fizemos um estudo detalhado do operador
$A_\mu^aA_\mu^a$ quando levamos em conta o horizonte de Gribov.
Incorporamos o operador $A_\mu^aA_\mu^a$ na ação de Gribov-Zwanziger
através do método LCO. Os resultados desta análise foram publicados
em \cite{Sobreiro:2004us,Sobreiro:2004yj,Dudal:2005na}. Inicialmente
fizemos uma análise a um laço da ação de Gribov-Zwanziger sem
considerar o operador $A_\mu^aA_\mu^a$. Os resultados provenientes
desteste estudo são os que se seguem:

\begin{itemize}

\item Devido a renormalizbilidade da ação de Gribov-Zwanziger o parâmetro de Gribov $\gamma$ não possuem dimensão anômala
independente. De fato, devido à relação,
\begin{equation}
Z_{\gamma^2}=Z_g^{-1/2}Z_A^{-1/4}\;,\label{2r5}
\end{equation}
temos que a dimensão anômala correspondente é dada por
\begin{eqnarray}
\gamma_{\gamma^2}(g^2)=-\frac{1}{2}\left[\frac{\beta(g^2)}{2g^2}-\gamma_A(g^2)\right]\;,\label{2r6}
\end{eqnarray}

\item Demonstramos explicitamente\footnote{A menos que o contrário seja especificado,
todos os resultados apresentados daqui para frente foram obtidos na
aproximação de um laço no esquema de renormalização $\MSbar$
utilizando regularização dimensional.} que a ação quântica efetiva
obedece a uma equação homogênea do grupo de renormalização.

\item Foi demonstrado formalmente que, independentemente do esquema de
renormali-zação e da escala, a energia do vácuo na presença do
horizonte é sempre positiva.

\item Não foi possível encontrar uma solução para $\gamma$
consistente com as equações do grupo de renormalização pois não foi
possível encontrar um parâmetro de expansão suficientemente pequeno
para dar sentido a série perturbativa. Esse \emph{resultado} foi
obtido a um e dois laços.

\end{itemize}

Após esta análise inicial, foi feita a inclusão do operador
$A_\mu^aA_\mu^a$, provendo os seguintes resultados

\begin{itemize}

\item O formalismo LCO para o operador $A_\mu^aA_\mu^a$ na ação de Gribov-Zwanziger é renormalizável a todas as ordens em teoria
de perturbações. A prova foi feita em conjunto com a teoria de
renormalização algébrica.

\item O operador composto $A_\mu^aA_\mu^a$ e
o parâmetro de Gribov $\gamma$ não possuem dimensão anômala
independente. De fato, devido às relações,
\begin{eqnarray}
Z_{A^2}&=&Z_gZ_A^{-1/2}\;,\nonumber\\
Z_{\gamma^2}&=&Z_g^{-1/2}Z_A^{-1/4}\;,\label{2r7}
\end{eqnarray}
temos que as dimensões anômalas correspondentes são dadas por
\begin{eqnarray}
\gamma_{A^2}(g^2)&=&-\left[\frac{\beta(g^2)}{2g^2}+\gamma_A(g^2)\right]\;,\nonumber\\
\gamma_{\gamma^2}(g^2)&=&-\frac{1}{2}\left[\frac{\beta(g^2)}{2g^2}-\gamma_A(g^2)\right]\;,\label{2r8}
\end{eqnarray}
preservando os resultados anteriores, onde o horizonte e o operador
$A_\mu^aA_\mu^a$ foram estudados separadamente.

\item Demonstramos explicitamente\footnote{A menos que o contrário seja especificado,
todos os resultados apresentados daqui para frente foram obtidos na
aproximação de um laço no esquema de renormalização $\MSbar$
utilizando regularização dimensional.} que a ação quântica efetiva
obedece a uma equação homogênea do grupo de renormalização.

\item Foi demonstrado, independentemente da
escala escolhida, que não é possível existir uma solução com $m^2>0$
para equação do gap definindo o horizonte. Para o caso $m^2>0$
apenas uma solução é permitida. Ainda assim, uma solução consistente
não foi encontrada no esquema $\MSbar$.

\item Providenciamos para que as equações renormalizadas fossem
escritas sob um cenário otimizado, no sentido de que reduzimos a
dependência no esquema de renormali-zação escolhido. A dependência é
descrita por apenas um parâmetro, $b_0$, associado a renormalização
da cosntante de acoplamento $g$.

\item No cenário otimizado, o esquema $\MSbar$ corresponde ao caso $b_0=0$. A solução neste caso, com parâmetro de expansão
relativamente pequeno, $\frac{N}{16\pi^2x}\approx0.340$, é dada
por\footnote{O parâmetro $\lambda$ nada mais é que o próprio
parâmetro de Gribov, com diferente normalização,
$\lambda\propto\gamma$.}
\begin{eqnarray}
\wl^4x^{-2b}&\approx&15.66\lms^4\;,\nonumber\\
\wm^2x^{-a}&\approx&-1.40\lms^2\;,\nonumber\\
E_\mathrm{vac}&\approx&0.11\lms^4\;.\label{2r9}
\end{eqnarray}
Para o caso de dependência mínima no esquema de renormalização, caso
$b_0\approx0.425$, a solução é dada por
\begin{eqnarray}
\wl^4x^{-2b}&\approx&2.07\lms^4\;,\nonumber\\
\wm^2x^{-a}&\approx&-0.23\lms^2\;,\nonumber\\
E_\mathrm{vac}&\approx&0.019\lms^4\;,\label{2r10}
\end{eqnarray}
onde o parâmetro de expansão é satisfatoriamente pequeno,
$\frac{N}{16\pi^2 x}\approx0.047$.

\end{itemize}

Por fim, ainda no Capítulo 6, foi fornecida uma discussão sobre as
consequências da presença do parâmetro de massa dinâmica e do
parâmetro de Gribov:

\begin{itemize}

\item O propagador do glúon, no nível árvore, se apresenta com o comportamento tipo
Stingl \cite{Stingl:1985hx}
\begin{equation}
D_{\mu\nu}^{ab}(q)=\delta^{ab}\frac{q^2}{q^4+m^2q^2+\frac{\lambda^4}{4}}\left(\delta_{\mu\nu}-\frac{q_\mu{q}_\nu}{q^2}\right)\;.\label{2r11}
\end{equation}

\item O propagador do glúon é suprimido no limite infravermelho.
Sendo que $D(0)=0$.

\item O propagador do glúon viola o \emph{princípio de
positividade}, exceto na ausência do horizonte, $\lambda=0$.

\item Devido a condição de horizonte, o propagador dos campos de Faddeev-Popov, apre-sentam
comportamento singular infravermelho tipo Gribov
\begin{equation}
\frac{\delta^{ab}}{N^2-1} \left\langle
c^a\oc^b\right\rangle_{q\approx0}\approx\frac{1}{q^4}\;.\label{2r12}
\end{equation}

\end{itemize}

Finalmente, no Capítulo 7, fornecemos o primeiro estudo analítico
das ambigüidades de Gribov nos calibres lineares covariantes. Os
resultados são os que seguem:

\begin{itemize}

\item Foi identificada uma região no espaço funcional das
configurações de calibre que pode ser utilizada para eliminar as
cópias de Gribov infinitesimais\footnote{Todos resultados que se
seguem foram obtidos exclusivamente para o caso em que o parâmetro
de calibre é pequeno, $\alpha\ll1$.},
\begin{equation}
\Omega\equiv\left\{A_\mu^a\;\big|\;A_\mu^a=\;A_\mu^{aT}+A_\mu^{aL},\;\mathcal{M}^{abT}>0\right\}\;.\label{2r13}
\end{equation}
onde,
\begin{equation}
\mathcal{M}^{abT}=-\partial_\mu\left(\partial_\mu-gf^{abc}A_\mu^{cT}\right)\;.\label{2r14}
\end{equation}
e $A^T$ é a componente transversa do campo de calibre e $A^L$ a
componente longitudinal do mesmo.

\item O efeito da restrição aparece nos propagadores da teoria.

\item O propagador do glúon é dado por
\begin{equation}
D_{\mu\nu}^{ab}(k)=\delta^{ab}\left[\frac{k^2}{k^4+2Ng^2\gamma^4}\left(\delta_{\mu\nu}-
\frac{k_{\mu}k_\nu}{k^2}\right)+\frac{\alpha}{k^2}\frac{k_\mu{k}_\nu}{k^2}\right]\;.\label{2r15}
\end{equation}
onde $\gamma$ é o parâmetro de Gribov, determinado através de uma
equação de gap idêntica ao caso do calibre de Landau, a um laço.

\item Quando a geração dinâmica de massa é levada em consideração o
propagador encontrado é na forma
\begin{equation}
D_{\mu\nu}^{ab}(k)=\delta^{ab}\left[\frac{k^2}{k^4+m^2k^2+2Ng^2\gamma^4}\left(\delta_{\mu\nu}-
\frac{k_{\mu}k_\nu}{k^2}\right)+
\frac{\alpha}{k^2+\alpha{m}^2}\frac{k_{\mu}k_\nu}{k^2}\right]\;.\label{2r16}
\end{equation}
com equação de gap dada por
\begin{equation}
\frac{3}{4}Ng^2\int\frac{d^{4}q}{\left(2\pi\right)^4}\frac{1}{q^4+m^2q^2+2Ng^2\gamma^4}=1\;.
\label{2r172}
\end{equation}

\item Os setores transverso e logitudinal são suprimidos no limite
infravermelho. O parâmetro de Gribov é responsável pela supressão do
setor transverso enquanto a massa força a supressão no setor
longitudinal.

\item O propagador dos campos de Faddeev-Popov não apresentam o comportamento
$1/k^4$.

\item Devido a equação do gap, a função de Green
$(\mathcal{M}^T)^{-1}=\left[-\partial_\mu\left(\delta^{ab}\partial_\mu-gf^{abc}A_\mu^{cT}\right)\right]^{-1}$
apresenta comportamento singular forte no limite infravermelho,
\begin{equation}
\frac{1}{N^2-1}\left[\left(\mathcal{M}^T(k)\right)^{-1}\right]^{aa}\bigg|_{k\rightarrow0}\propto\frac{1}{k^{4}}\;.
\label{2r18}
\end{equation}

\item O calibre de Landau e suas propriedades são recuperadas no
limite $\alpha\rightarrow0$.

\end{itemize}

\part{DYNAMICAL MASS}

\chapter{Condensation of dimension two operators}

Neste capítulo vamos iniciar o estudo de operadores de dimensão
dois\footnote{Como frizamos na primeira parte desta tese, tais
operadores também são chamados de operadores de massa.} como
motivação para se entender a possível condensação destes operadores
de forma a gerar uma massa dinâmica para os gluons.

Discutiremos os operadores de massa invariantes de calibre.
Começando com o caso mais simples, ou seja, o caso Abeliano, veremos
que existem quatro tipos de operadores de massa invariantes de
calibre. Contudo, estes operadores se mostram equivalentes entre si,
ao nível clássico. Ao se generalizar tais operadores para o caso não
Abeliano, veremos que existem duas classes distintas. Ainda, estes
operadores não Abelianos se mostram não locais e/ou não
renormalizáveis quando introduzidos na ação de Yang-Mills acoplados
a parâmetros de massa. Os detalhes desta análize podem ser
encontrados em \cite{Capri:2005dy}.

Em seguida, mostraremos que, no caso específico do calibre de
Landau, podemos relacionar as duas classes de operadores invariantes
de calibre com o operador local $A_\mu^aA_\mu^a$, motivando assim o
estudo deste operador. Lembramos que este operador tem sido
extensamente estudado nos últimos anos, veja
\cite{Gubarev:2000nz,Gubarev:2000eu,Boucaud:2001st,Boucaud:2005rm,Kondo:2001nq,Verschelde:2001ia,Knecht:2001cc,Dudal:2002pq,Browne:2003uv,
Dudal:2003vv,Dudal:2003pe,Dudal:2003by,Dudal:2004rx,Browne:2004mk,Gracey:2004bk,Dudal:2005na,Furui:2005bu,Gubarev:2005it,
Slavnov:2005av,Suzuki:2004dw,Suzuki:2004uz} e referências.
Revisaremos o método LCO para este operador, e a respectiva geração
dinâmica de massa,
\cite{Verschelde:2001ia,Knecht:2001cc,Dudal:2002pq,Dudal:2003vv,Dudal:2003pe},
no calibre de Landau.

Finalmente, apresentaremos os operadores locais compostos por campos
fantasmas, que também têm chamado a atenção nos últimos anos
\cite{Schaden:1999ew,Kondo:2000ey,Schaden:2000fv,Schaden:2001xu,Dudal:2002xe,Sawayanagi:2003dc,Kondo:2001nq,Dudal:2004rx,Capri:2005vw}.
Obviamente tais operadores não são invariantes de calibre, uma vez
que os campos de Faddeev-Popov aparecem devido a fixação de calibre.
Contudo, tais operadores proporcionam interessantes consequências
físicas nas teorias de Yang-Mills. Faremos um resumo dos principais
efeitos relacionados a condensação destes operadores, no calibre de
Landau. Em particular, abordaremos a questão através do método LCO.

\section{Gauge invariant operators}

\subsection{Abelian case}

Consideremos o limite Abeliano das teorias de Yang-Mills, ou seja, a
ação da QED, \cite{Feynman:1949zx,Feynman:1950ir,Feynman:1958ty},
\begin{equation}
S_{QED}=\frac{1}{4}\int{d^4}xF_{\mu\nu}F_{\mu\nu}\;,\label{3max0}
\end{equation}
onde o tensor eletromagnético é dado por
\begin{equation}
F_{\mu\nu}=\partial_\mu{A}_\nu-\partial_\nu{A}_\mu\;.\label{3eletro1}
\end{equation}
Lembramos que a ação da QED é invariante sob transformações de
calibre na forma
\begin{equation}
\delta{A}_\mu=-\partial_\mu\omega\;,\label{3transf0}
\end{equation}
onde $\omega$ é o parâmetro do grupo de calibre $U(1)$.

O objeto invariante de calibre mais simples que podemos considerar
com o campo $A_\mu$ é a componente transversa deste campo, $A_{\mu
}^{T}$,
\begin{equation}
A_\mu^T=\left(\delta_{\mu\nu}-\frac{\partial_\mu\partial_\nu}{\partial^2}\right)A_\nu\;.\label{3trans0}
\end{equation}
Desta forma, um operador de dimensão dois associado seria
\begin{equation}
\mathcal{O}_1^{Abel}(A)=\int{d^4}xA_\mu^TA_\mu^T\;.\label{3op1}
\end{equation}

A segunda possibilidade é considerar a quantidade
$\int{d^4}xA_\mu{A}_\mu$ e minimizá-la com respeito a transformações
de calibre, veja \cite{Gubarev:2000nz,Gubarev:2000eu},
\begin{equation}
\mathcal{O}_2^{Abel}(A)=A_{\min}^2=\mathrm{min}\int{d^4}xA_\mu{A}_\mu\;.\label{3op2}
\end{equation}
É importante ressaltar que foi provado em \cite{Gubarev:2000eu} que
o funcional $A_{\min}^2$ é um parâmetro de ordem para o estudo da
transição de fase da QED compacta em três dimensões.

Uma terceira possibilidade para um operador de massa invariante de
calibre no caso Abeliano seria considerar o termo de Stückelberg,
dado por, \cite{Ruegg:2003ps}
\begin{equation}
\mathcal{O}_3^{Abel}(A)=\int{d^4}x\left(A_\mu+\partial_\mu\phi\right)^2\;,\label{3op3}
\end{equation}
onde $\phi$ é um campo escalar sem dimensão. O operador
$\mathcal{O}_3^{Abel}(A)$ é invariante sob a transformação
(\ref{3transf0}) juntamente com
\begin{equation}
\delta\phi=\omega\;.\label{3gauge1}
\end{equation}
Curiosamente, o termo de Stückelberg (\ref{3op3}) pode ser reescrito
na forma de um modelo sigma de calibre com simetria $U(1)$,
\begin{equation}
\mathcal{O}^{Abel}_3(A)=\int{d^4}x\left(A_\mu-\frac{i}{e}G^{-1}\partial_\mu{G}\right)^2\;.\label{3op3a}
\end{equation}
com
\begin{equation}
G=e^{ie\phi}\;.\label{3quant0}
\end{equation}
De forma que as transformações de calibre (\ref{3transf0}) e
(\ref{3gauge1}) são agora dadas por
\begin{eqnarray}
A_\mu&\rightarrow&A_\mu+\frac{i}{e}V^{-1}\partial_{\mu}V\;,\nonumber\\
G&\rightarrow&GV\;,\label{3gauge2}
\end{eqnarray}
onde
\begin{equation}
V=e^{ie\omega}\;.\label{3quant1}
\end{equation}
Esta propriedade será importante para discutirmos a generalização do
termo de Stückelberg no caso não Abeliano.

A quarta e última possibilidade de um operador invariante de calibre
é dada pelo objeto não local
\begin{equation}
\mathcal{O}_4^{Abel}(A)=-\frac{1}{2}\int{d^4}xF_{\mu\nu}\frac{1}{\partial^2}F_{\mu\nu}\;.\label{3op4}
\end{equation}

Apesar da forma aparentemente diferente de cada um destes
operadores, pode-se mostrar facilmente, veja detalhes em
\cite{Capri:2005dy}, que estes operadores são equivalentes ao nível
clássico,
\begin{equation}
\mathcal{O}_1^{Abel}(A)\equiv\mathcal{O}_2^{Abel}(A)\equiv\mathcal{O}_3^{Abel}(A)\equiv\mathcal{O}_4^{Abel}(A)\;,\label{3equiv1}
\end{equation}
Ainda, no caso Abeliano, estes operadores são renormalizáveis quando
acoplados a um parâmetro de massa, \cite{Capri:2005dy}. Por exemplo,
ação
\begin{equation}
S=S_{YM}+\frac{m}{2}\mathcal{O}_4^{Abel}(A)\;,\label{3actionm1}
\end{equation}
é renormalizável. Para isso, o termo não não local pode ser colocado
numa forma local com a ajuda de campos auxiliares. Mais detalhes
podem ser encontrados em \cite{Capri:2005dy}.

\subsection{Non-Abelian case}

No caso Abeliano todos os operadores invariantes de calibre
discutidos se mostraram equivalentes entre si, (\ref{3equiv1}).
Consideremos agora o caso mais interessante, ou seja, a ação de
Yang-Mills (\ref{1ym1}), que descreve o caso não Abeliano. Veremos
agora como estes operadores se generalizam ao caso não Abeliano e
até onde a relação de equivalência (\ref{3equiv1}) permanece válida.

\subsubsection{Operator $A_{\min}^2$}

O operator $\mathcal{O}_2^{Abel}$ da expresão (\ref{3op2}) pode ser
generalizado ao caso não Abeliano considerando a minimização do
funcional $\mathrm{Tr}\int{d^4}xA_\mu^UA_\mu^U$ ao longo da órbita
de calibre de $A_\mu^a$,
\cite{Semenov,Zwanziger:1990tn,Dell'Antonio:1989jn,Dell'Antonio:1991xt,vanBaal:1991zw,
Gubarev:2000nz,Gubarev:2000eu}, de acordo com
\begin{eqnarray}
A_{\min}^2&\equiv&\min_{\{U\}}\mathrm{Tr}\int{d^4}xA_\mu^UA_\mu^U\;,\nonumber\\
A_\mu^U&=&U^{\dagger}A_\mu{U}+\frac{i}{g}U^\dagger\partial_\mu{U}\;.\label{3Amin0}
\end{eqnarray}
A segunda das (\ref{3Amin0}) é tida como a definição da órbita de
calibre.

É importante ter em mente que, apesar de o processo de minimização
acima descrito fazer com que o operador $A_{\min}^2$ se torne
invariante de calibre, devemos ressaltar que encontrar a forma
explícita do mínimo absoluto alcançado pelo funcional
$\mathrm{Tr}\int{d^4}xA_\mu^{U}A_\mu^U$ consiste em um passo
altamente não trivial. Na prática, encontrar o mínimo absoluto deste
operador requer a solução do problema das ambigüidades de
Gribov\footnote{Esta questão será discutida na terceira parte desta
tese.}. Contudo, foi demonstrado em
\cite{Semenov,Zwanziger:1990tn,Dell'Antonio:1989jn,Dell'Antonio:1991xt,vanBaal:1991zw}
que o operador $\mathrm{Tr}\int{d^4}xA_\mu^UA_\mu^U$ atinge seu
mínimo absoluto ao longo da órbita de $A_\mu^a$. Como discutido em
\cite{Capri:2005dy}, uma configuração de mínimo relativo de
$\mathrm{Tr}\int{d^4}xA_\mu^UA_\mu^U$ é obtida quando, por exemplo,
$U=h$ de forma que $A_\mu^h$ é um campo transverso,
$\partial_{\mu}A_\mu^h=0$. A forma explícita de $A_\mu^h$ é dada por
uma série de potências em $A_\mu^a$, \cite{Lavelle:1995ty},
\begin{equation}
A_\mu^h=\left(\delta_{\mu\nu}-\frac{\partial_\mu\partial_\nu}{\partial^2}\right)\left\{A_{\nu
}-ig\left[ \frac{1}{\partial ^{2}}\partial A,A_\nu\right]
+\frac{ig}{2}\left[ \frac{1}{\partial ^{2}}\partial A,\partial _{\nu
}\frac{1}{\partial ^{2}}\partial A\right]
+O(A^{3})\right\}\;.\label{3min0}
\end{equation}
Substituindo a expressão (\ref{3min0}) em (\ref{3Amin0}) deduzimos
uma expressão em séries de potências em $A_\mu^a$ para o operador
$A^2_{\min}$,
\begin{eqnarray}
A_{\min}^2&=&\frac{1}{2}\int{d^4}x\left[A_\mu^a\left(\delta_{\mu\nu}-\frac{\partial_\mu\partial_\nu}{\partial^2}\right)A_\nu^a-
gf^{abc}\left(\frac{\partial_\nu}{\partial^2}\partial{A}^a\right)\left(\frac{1}{\partial^2}\partial{A}^b\right)A_\nu^c\right]+O(A^4)\;.\nonumber\\
& & \label{3min1}
\end{eqnarray}

Ainda, a configuração de mínimo relativo $A_\mu^h$ é tal que, além
de transversa é, ordem a ordem em $g$, invariante de calibre, veja
em \cite{Capri:2005dy}. Esta simples propriedade mostra que
$A_{\min}^2$ também generaliza o operador $\mathcal{O}_1^{Abel}$
definido em (\ref{3op1}).

\subsubsection{Stückelberg term}

O termo de Stückelberg, $\mathcal{O}^{Abel}_3$, dado em
(\ref{3op3}), pode ser generalizado ao caso não Abeliano,
\cite{Ruegg:2003ps}, de acordo com a expresão
\begin{equation}
\mathcal{O}_S=\mathrm{Tr}\int{d^4}x\left(A_\mu-\frac{i}{g}G^{-1}\partial_\mu{G}\right)^2\;,\label{3stueck0}
\end{equation}
onde
\begin{equation}
G=e^{ig\phi^aT^a}\;,\label{3U0}
\end{equation}
com $\phi^a$ sendo um campo escalar sem dimensão. A expressão
(\ref{3stueck0}) é invariante sob transformações de calibre na forma
\begin{eqnarray}
A_\mu&\rightarrow&V^{-1}A_\mu{V}+\frac{i}{g}V^{-1}\partial_\mu{V}\;,\nonumber\\
G&\rightarrow&GV\;.\label{3gauge4}
\end{eqnarray}

De acordo com \cite{Esole:2004rx,Capri:2005dy}, o termo de
Stückelberg é classicamente equivalente ao ope-rador $A_{\min}^2$. A
prova é feita, \cite{Esole:2004rx,Capri:2005dy}, extraíndo as
equações de movimento de $\phi^a$ a partir de (\ref{3stueck0}) e
substituindo-as no mesmo, encontrando a relação de equivalência
\begin{equation}
\mathcal{O}_S\equiv{A}_{\min}^2\;.\label{3equiv2}
\end{equation}

\subsubsection{Operator $\mathrm{tr}\int{d^4}xF_{\mu\nu}\frac{1}{D^2}F_{\mu\nu}$}

O último operador Abeliano a ser generalizado seria
$\int{d^4}xF_{\mu\nu}\frac{1}{\partial^2}F_{\mu\nu}$, exposto na
expressão (\ref{3op4}). De acordo com \cite{Jackiw:1997jg} tal
generalização é feita facilmente através da subs-tituição da
derivada ordinária pela derivada covariante,
$\partial\rightarrow{D}$, ou seja,
\begin{equation}
\mathcal{O}_o=\mathrm{Tr}\int{d^4}xF_{\mu\nu}\frac{1}{D^2}F_{\mu\nu}=
\frac{1}{2}\int{d^4}xF_{\mu\nu}^a\left[(D^2)^{-1}\right]^{ab}F_{\mu\nu}^b\;.\label{3op4a}
\end{equation}
No caso Abeliano, todos os quatro operadores considerados eram, de
fato, equivalentes entre si. Aqui, no caso não Abeliano, vimos que a
generalização dos operadores $\mathcal{O}^{Abel}_1$,
$\mathcal{O}^{Abel}_2$ e $\mathcal{O}^{Abel}_3$ são equivalentes
entre si. Contudo, o operador $\mathcal{O}_o$ se mostra diferente de
$A^2_{\min}$. Para vermos isso, basta observarmos a expressão de
$A^2_{\min}$ em termos do tensor intensidade de campo,
\cite{Zwanziger:1990tn},
\begin{eqnarray}
A_{\min}^2&=&-\frac{1}{2}\mathrm{Tr}\int{d^4}x\left(F_{\mu\nu}\frac{1}{D^2}F_{\mu\nu}+
2i\frac{1}{D^2}F_{\lambda\mu}\left[\frac{1}{D^2}D_\kappa{F}_{\kappa\lambda},\frac{1}{D^2}D_\nu{F}_{\nu\mu}\right]\right.\nonumber\\
&&-2i\left.\frac{1}{D^2}F_{\lambda\mu}\left[\frac{1}{D^2}D_{\kappa}F_{\kappa\nu},\frac{1}{D^2}D_{\nu}F_{\lambda\mu}\right]\right)+O(F^4)\;.\label{3zzw}
\end{eqnarray}
Portanto, no caso não Abeliano temos duas classes de operadores de
massa invariantes de calibre.

\subsection{Discussion}

Discutimos, nas teorias de Yang-Mills, quatro operadores de dimensão
dois invariantes de calibre que podem estar relacionados com o
fenômeno da geração dinâmica de massa. Mostramos que, de fato,
existem duas classes de operadores, uma dada pelo operador
$A_{\min}^2$ e suas representações e outra dada pelo operador
$\mathcal{O}_o$. Nos resta agora saber o que fazer com estes
operadores, ou seja, como estudar os efeitos que estes causam na
teoria e como associá-los a uma massa dinâmica.

Como proposto em \cite{Capri:2005dy}, podemos pensar em considerar
modelos massivos na forma
\begin{equation}
S_m=S_{YM}+\frac{m^2}{2}\mathcal{O}^i\;,\label{3actionM1}
\end{equation}
onde $\mathcal{O}^i\in\left\{A_{\min}^2,\mathcal{O}_o\right\}$ e o
parâmetro de massa $m^2$, ao invés de ser um parâmetro livre, seria
fixado através de uma equação de gap. No caso de $A_{\min}^2$, a
ação (\ref{3actionM1}) se torna difícil de se lidar devido ao fato
de $A_{\min}^2$ ser dado por uma expansão em $A_\mu^a$ com termos
altamente não locais, (\ref{3Amin1}). Um conhecimento da solução do
problema de Gribov se mostra necessário para lidar com tal
operador\footnote{Como discutiremos na terceira parte desta tese,
tal solução ainda não está a nossa disposição.}. Poderíamos ainda
ter a esperança de nos depararmos com uma teoria renormalizável se
utilizarmos a representação, local, de Stückelberg (\ref{3stueck0}).
Mas, infelizmente, esta representação não é polinomial, o que
invalida automaticamente a renormalizabilidade do modelo devido ao
aparecimento de infinitos vértices.

Por outro lado, $\mathcal{O}_o$ possui uma aparência mais amigável.
Primeiramente, este operador pode ser introduzido em qualquer
calibre, sem termos que solucionar nenhum problema não trivial do
tipo Gribov. Ainda, $\mathcal{O}_o$ pode ser colocado numa
representação local e polinomial com a ajuda de um conjunto de
campos auxiliares apropriados, veja os detalhes em
\cite{Capri:2005dy}. Nesta representação a teoria proposta
(\ref{3actionM1}) é, pelo menos, renormalizável por contagem de
potências. Mas, infelizmente, nem mesmo nesta representação a ação
(\ref{3actionM1}) se mostra renormalizável. Ao invés disso, uma
teoria alternativa às teorias de Yang-Mills foi identificada, uma
teoria de calibre não Abeliana massiva renormalizável, veja os
detalhes em \cite{Capri:2005dy,Capri:2006ne}.

Como discutido no início deste capítulo, inspirados nos resultados
da rede, procuramos uma massa dinâmica dentro das teorias de
Yang-Mills, e não modelos alternativos. Devido a dificuldade de
lidar com operadores não locais, vamos então simplificar tais
operadores para o caso específico do calibre de Landau (calibre de
interesse desta tese). Veremos que poderemos lidar com operadores
locais e ainda relacioná-los ao caso geral dos operadores não locais
aqui discutidos.

\section{Local operators}

Nos últimos anos tem crescido o interesse por operadores de dimensão
dois, em particular o condensado de gluons
$\left\langle{A}_\mu^aA_\mu^a\right\rangle$ no calibre de Landau.
Estudos têm sido efetuados através de várias técnicas não
perturbativas teóricas, fenomenológicas e computacionais na rede,
veja, por exemplo,
\cite{Gubarev:2000eu,Gubarev:2000nz,Boucaud:2001st,Boucaud:2005rm,Kondo:2001nq,Verschelde:2001ia,Dudal:2002pq,
Dudal:2003vv,Dudal:2003by,Dudal:2004rx,Browne:2004mk,Gracey:2004bk,Dudal:2005na,Furui:2005bu,Gubarev:2005it,Slavnov:2005av,Suzuki:2004dw,Suzuki:2004uz}
e referências contidas. Não vamos entrar nos detalhes dos muitos
métodos utilizados para tratar este condensado. Nos ateremos ao
método aqui utilizado, o formalismo LCO.

No método LCO, o operador ${A}_\mu^aA_\mu^a$ e suas generalizações a
outros calibres foram intensamente estudados,
\cite{Verschelde:2001ia,Dudal:2002pq,Dudal:2003by,Dudal:2003np,Dudal:2003by,Dudal:2004rx,Dudal:2005na}.
De acordo com esta construção uma massa para o glúon é gerada
dinamicamente no nível árvore, devido a um condensado $\left\langle
A_\mu^aA_\mu^a\right\rangle\neq0$.

Outra classe de condensados, talvez menos conhecidas são os chamados
condensados de campos fantasmas BCS e Overhauser,
$\left\langle{f}^{abc}\oc^bc^c\right\rangle$,
$\left\langle{f}^{abc}c^bc^c\right\rangle$ e
$\left\langle{f}^{abc}\oc^b\oc^c\right\rangle$. Estes foram
estudados primeiramente no calibre Abeliano máximo,
\cite{Schaden:1999ew,Kondo:2000ey,Schaden:2000fv,Schaden:2001xu} no
caso $SU(2)$. Em \cite{Dudal:2002xe,Sawayanagi:2003dc}, foi mostrado
que a massa gerada para os gluons não diagonais é, na verdade
taquiônica, a um laço.

No caso do calibre de Landau, tais condensados foram estudados no
formalismo LCO em \cite{Lemes:2002rc,Dudal:2003dp}. Um ação efetiva
foi construída para os operadores $f^{abc}\oc^b c^c$, $f^{abc}c^b
c^c$ e $f^{abc}\oc^b \oc^c$ simultaneamente, preservando a simetria
the $SL(2,\mathbb{R})$.

O estudo conjunto entre $A_\mu^aA_\mu^a$ e $gf^{abc}\bar{c}^bc^c$
foi efetuado em \cite{Capri:2005vw} e compõe o tema principal desta
parte.

\subsection{Operator $A_\mu^aA_\mu^a$ and the Landau gauge}

Comecemos por discutir as propriedades do operador de massa
$A_\mu^aA_\mu^a$. Este operador, apesar de não ser invariante de
calibre, pode nos dizer muito sobre os operadores discutidos na
seção anterior. Para entendermos isso, consideremos a forma
explícita de $A_{\min}^2$, (\ref{3min0}). Este operador, no calibre
de Landau, se simplifica enormemente devido ao fato de o calibre ser
composto por configurações transversas exclusivamente. De fato, no
calibre de Landau, a expressão (\ref{3min0}) se reduz a
\begin{equation}
A_{\min}^2\bigg|_{\mathrm{Landau}}=\widetilde{\mathcal{O}}(A)=\frac{1}{2}\int{d^4}xA_\mu^aA_\mu^a\;.\label{3Amin1}
\end{equation}
Ainda, é trivial checar que, ao nível clássico,
$\widetilde{\mathcal{O}}$ é invariante de calibre. Desta forma temos
um operador local exatamente associado a $A^2_{\min}$ no caso de um
calibre específico (Landau).

Podemos ainda, associar $\widetilde{\mathcal{O}}$ à segunda classe
de operadores invariantes de calibre, ou seja, $\mathcal{O}_o$
definido em (\ref{3op4a}). Para tal, basta tomarmos a ordem em $g^0$
da expresão (\ref{3op4a}), e verificar que
\begin{equation}
\mathcal{O}_o=-\mathrm{Tr}\int{d^4}xA_\mu^TA_\mu^T+O(g)\;,\label{3equiv3}
\end{equation}
onde $A_\mu^T$ é a componente transversa do glúon $A_\mu$. Desta
forma, no calibre de Landau,
\begin{equation}
\mathcal{O}_o=-2\widetilde{\mathcal{O}}+O(g)\;.\label{3equiv4}
\end{equation}

As relações (\ref{3Amin1}) e (\ref{3equiv3}) significam que o estudo
do operador local $A_\mu^aA_\mu^a$ no calibre de Landau pode vir a
fornecer pistas da condensação dos operadores invariantes de
calibre. E, consequentemente, evidências da geração dinâmica de
massa. É importante ressaltar que o modelo massivo descrito pela
ação
\begin{equation}
S_{m2}=S_{YM}+m^2\widetilde{\mathcal{O}}\;,\label{3actionM2}
\end{equation}
é renormalizável a todas as ordens em teoria de perturbações,
\cite{Dudal:2002pq}.

Daqui para frente, a menos que o contrário seja dito, trabalharemos
esclusivamente no calibre de Landau.

\subsection{Operator $A_\mu^aA_\mu^a$ and the LCO technique}

Vamos discutir brevemente as propriedades do operador
$A_\mu^aA_\mu^a$ quando introduzido através do método LCO. Como
veremos mais detalhadamente o método LCO para o caso mais geral no
próximo capítulo e na parte seguinte desta tese, passaremos
rapidamente pelos principais pontos físicos deste tópico. Os
detalhes técnicos podem ser encontrados em
\cite{Verschelde:2001ia,Knecht:2001cc,Dudal:2002pq,Dudal:2003vv,Dudal:2003pe}.

Esta seção é incluída a título de exemplo inicial do método LCO, bem
como para comparação dos resultados que obteremos na terceira parte
desta tese, quando tratarmos das ambigüidades de Gribov. O mesmo não
ocorrerá com os condensados fantasmas.

\subsubsection{Renormalizability}

Para estudarmos as propriedades do operador $A_\mu^aA_\mu^a$ nas
teorias de Yang-Mills, consideramos a ação
\begin{equation}
\Sigma=S_{YM}+S_{gf}+S_{LCO}+S_{ext}\;.\label{4action1}
\end{equation}
onde a ação de Yang-Mills está disposta na expresão (\ref{1ym1}) e o
termos de fixação de calibre, impondo o calibre de Landau, está dado
em (\ref{1landau1}). A ação LCO no caso do operador
$A_\mu^aA_\mu^a$, de acordo com o Ap. \ref{ap_ferram}, é na forma
\begin{eqnarray}
S_{LCO}&=&s\int{d^4x}\left(\frac{\lambda}{2}A_\mu^aA_\mu^a-\frac{\zeta}{2}\lambda{J}\right)\nonumber\\
 &=&\int{d^4x}\left(\frac{J}{2}A_\mu^aA_\mu^a+\lambda{A_\mu^a\partial_\mu{c}^a}-\frac{\zeta}{2}J^2\right)\;,\label{4lco1}
\end{eqnarray}
onde as fontes (campos clássicos) $\lambda$ e $J$ são introduzidas
para manter a simetria BRST existente neste calibre, (\ref{1brs1}).
De fato, estas fontes formam um dubleto BRST, sendo portanto
inofensivas ao setor físico das teorias de Yang-Mills,
\begin{eqnarray}
s\lambda&=&J\;,\nonumber\\
sJ&=&0\;.\label{4brs3}
\end{eqnarray}
O parâmetro LCO, $\zeta$, é introduzido para absorver as
divergências associadas à função de correlação
$\left<A^2(x)A^2(y)\right>$. É fácil entender a necessidade deste
termo, pois é permitido por contagem de potências, veja
tabela\footnote{No caso do operador $A_\mu^aA_\mu^a$, de número
quântico fantasma nulo, as fontes $\lambda$ e $J$ possuem número
quântico fantasma dado respectivamente por $-1$ e $0$.}
\ref{aptable2} localizada no Ap. \ref{ap_ferram}. Ademais, este
parâmetro possui papel fundamental para se estabelecer a invariância
do potencial efetivo sob o grupo de renormalização. Finalmente,
temos a ação de fontes invariantes BRST,
\begin{eqnarray}
S_{ext}&=&s\int{d^4x}\left(-\Omega^a_{\mu}A^a_\mu+L^ac^a\right)\nonumber\\
 &=&\int{d^4x}\left(-\Omega^a_{\mu}D_\mu^{ab}c^b+\frac{g}{2}f^{abc}L^ac^bc^c\right)\;,\label{apext}
\end{eqnarray}
onde as fontes $\Omega_\mu^a$ e $L^a$ são invariantes sob
transformações BRST,
\begin{equation}
s\Omega_\mu^a=sL^a=0\;.\label{4brs2}
\end{equation}
A ação de fontes externas $S_{ext}$ é introduzida de forma a sermos
consistentes com o princípio de ação quântica e podermos escrever
identidades de Ward para as simetrias da ação (\ref{4action1}).
Lembramos que os números quânticos das fontes BRST $\Omega_\mu^a$ e
$L^a$, bem como dos campos $A_\mu^a$, $c^a$, $\bar{c}^a$ e $b^a$,
podem ser encontrados na tabela \ref{aptable1}.

Para provar a renormalizabilidade da ação (\ref{4action1}) é
conveniente utilizar a teoria de renormalização algébrica
\cite{book}, veja também Ap. \ref{ap_ferram}. Tecnicamente, este
método consiste em encontrarmos o contratermo mais geral possível,
compatível com as simetrias da ação (\ref{4action1}) e mostrar que
este contratermo pode ser reabsorvido na ação (\ref{4action1})
através da redefinição multiplicativa dos campos e parâmetros da
mesma ação. Fisicamente, devemos mostrar que a ação (\ref{4action1})
permanece estável sob correções quânticas. Vamos então enumerar as
simetrias do modelo através de identidades de Ward
\begin{itemize}
\item Identidade de Slavnov-Taylor associada a simetria BRST
\begin{equation}
\mathcal{S}(\Sigma)=\int{d^4x}\left(\frac{\delta\Sigma}{\delta\Omega_\mu^a}\frac{\delta\Sigma}{\delta{A}_\mu^a}+
\frac{\delta\Sigma}{\delta{L}^a}\frac{\delta\Sigma}{\delta{c}^a}+
b^a\frac{\delta\Sigma}{\delta\bar{c}^a}+
J\frac{\delta\Sigma}{\delta\lambda}\right)=0\;,\label{4slav1}
\end{equation}

\item Fixação de calibre e a equação dos campos anti-fantasmas
\begin{eqnarray}
\frac{\delta\Sigma}{\delta{b}^a}&=&\partial_\mu{A}_\mu^a\;,\nonumber\\
\frac{\delta\Sigma}{\delta\bar{c}^a}+\partial_\mu\frac{\delta\Sigma}{\delta\Omega_\mu^a}&=&0\;.\label{4gauge1}
\end{eqnarray}

\item Equação dos campos fantasmas
\begin{equation}
\int{d^4x}\left(\frac{\delta\Sigma}{\delta{c}^a}+gf^{abc}\bar{c}^b\frac{\delta\Sigma}{\delta{b}^c}\right)=
gf^{abc}\int{d^4x}\left(\Omega_\mu^bA_\mu^c+L^bc^c\right)\;,\label{4ghost1}
\end{equation}

\item Equação de inserção
\begin{equation}
\int{d^4x}\left(\frac{\delta\Sigma}{\delta\lambda}+c^a\frac{\delta\Sigma}{\delta{b}^a}\right)=0\;.\label{4lambda1}
\end{equation}
\end{itemize}
É importante ter em mente que a equação de inserção (\ref{4lambda1})
só é possível devido ao fato de o operador
$\widetilde{\mathcal{O}}$, em (\ref{3Amin1}), ser classicamente
invariante de calibre. Como consequência a equação de inserção
(\ref{4lambda1}) gera a simetria $SL(2,\mathbb{R})$,
\begin{equation}
\int{d^4x}\left(\bar{c}^a\frac{\delta\Sigma}{\delta{c^a}}+\frac{\delta\Sigma}{\delta{L}^a}\frac{\delta\Sigma}{\delta{b}^a}\right)=0\;.\label{4sl1}
\end{equation}
pertencente à álgebra de Nakanishi-Ojima (álgebra NO),
\cite{Nakanishi:1980dc}. A simetria $SL(2,\mathbb{R})$, assim como a
ágebra NO são conhecidas propriedades do calibre de Landau.

O contratermo mais geral possível, de acordo com a teoria de
renormalização algébrica \cite{book}, respeitando às identidades de
Ward (\ref{4slav1}-\ref{4lambda1}), é dado por
\begin{equation}
S^c=a_0S_{YM}+\int{d^4}x\left\{a_1\left[A_\mu^a\frac{\delta{S}_{YM}}{\delta{A}_\mu^a}+\left(\Omega_\mu^a+\partial_\mu\bar{c}^a\right)\partial_\mu{c}^a+
\frac{1}{2}JA_\mu^aA_\mu^a\right]+\frac{a_8}{2}\zeta{J}^2\right\}\;,\label{4count1}
\end{equation}
onde $a_i$ são parâmetros de renormalização independentes. Note que
o coeficiente do termo $JA^2$ é o parâmetro associado à
renormalização do glúon. Esta propriedade vem da equação de inserção
(\ref{4lambda1}). Ainda, vemos que o termo LCO $\zeta{J}^2$ se
renormaliza independentemente dos demais termos da teoria. Estas
divergências são as acima mencionadas associadas à função de
correlação $\left<A^2(x)A^2(y)\right>$, justificando assim a
introdução do parâmetro LCO $\zeta$.

Resta testar a renormalizabilidade multiplicativa da teoria na
presença do operador $A_\mu^aA_\mu^a$. De fato, o contratermo
(\ref{4count1}) pode ser reabsorvido na ação clássica
(\ref{4action1}) através da redefinição multiplicativa dos campos,
fontes e parâmetros da teoria, de acordo com as expressões
(\ref{apren1}-\ref{apren3}) no Ap. \ref{ap_ferram}. Juntamente com
\begin{eqnarray}
J_0&=&Z_JJ\;,\nonumber\\
\lambda_0&=&Z_\lambda\lambda\;,\nonumber\\
\zeta_0&=&Z_\zeta\zeta\;.
\end{eqnarray}
Os fatores de renormalização independentes são dados por
$Z_A^{1/2}$, $Z_g$ e $Z_\zeta$. Para o glúon e a constante de
acoplamento temos
\begin{eqnarray}
Z_A^{1/2}&=&1+\epsilon\left(a_1-\frac{a_0}{2}\right)\;,\nonumber\\
Z_g&=&1+\epsilon\frac{a_0}{2}\;,\label{4ren1}
\end{eqnarray}
enquanto que para o parâmetro LCO temos
\begin{equation}
Z_\zeta=1+\epsilon(-a_8+2a_1-2a_0)\;.\label{4ren2}
\end{equation}
Ainda, como resultado principal desta análise temos a renormalização
das fontes LCO
\begin{eqnarray}
Z_\lambda&=&Z_c^{-1/2}Z_A^{-1/2}\;,\nonumber\\
Z_J&=&Z_\lambda^2=Z_gZ_A^{-1/2}\;,\label{4ren3}
\end{eqnarray}
onde a segunda das (\ref{4ren3}) não é obtida diretamente como um
teorema de não renormali-zação, mas sim obtida devido ao fato de o
termo $JA^2$ não possuir contratermo independente. Ressaltamos que
este teorema pode ser obtido de maneira formal considerando um
sistema de fontes mais sofisticado, veja \cite{Dudal:2002pq}.

Ainda, o teorema de não renormalização do vértice
glúon-fantasma-antifantasma \cite{Blasi:1990xz} permanece válido,
devido a equação fantasma (\ref{4ghost1}),
\begin{equation}
Z_c=Z_{\bar{c}}=Z_A^{-1/2}Z_g^{-1}\;.\label{4ren4}
\end{equation}
Os demais objetos se renormalizam da mesma forma que o caso sem o
operador $A_\mu^aA_\mu^a$,
\begin{eqnarray}
Z_L&=&Z_b^{-1/2}=Z_A^{1/2}\nonumber\;,\\
Z_\Omega&=&Z_g^{-1/2}Z_A^{-1/4}\nonumber\;,\label{4ren4a}
\end{eqnarray}

Com isso terminamos a análize das propriedades de renormalização do
operador $A_\mu^aA_\mu^a$.

\subsubsection{Effective action and the renormalization group}

O próximo passo consiste em contruir um potencial efetivo associado
ao operador $A_\mu^aA_\mu^a$. A ação efetiva é definida por,
\cite{Peskin}\footnote{Apesar de não estar explícito, a expressão
(\ref{4eff1}) é assumida renormalizada. Utilizaremos essa hipótese
sempre que falarmos de quantidades quânticas. Contudo, não
utilizaremos a notação renormalizada padrão para não deixar a
leitura carregada.},
\begin{equation}
e^{-\Gamma}(J)=\int{DAD\bar{c}DcDb}\exp\left\{-S_{YM}-S_{gf}-
\int{d^4x}\left(\frac{J}{2}A_\mu^aA_\mu^a-\frac{\zeta}{2}J^2\right)\right\}\;,\label{4eff1}
\end{equation}
Um problema com a ação efetiva é o fato de $\zeta$ ser, até então,
um parâmetro livre da teoria introduzido para absorver divergências
apenas. Temos então um problema com a unicidade da energia do vácuo.
Um segundo problema a ser encarado é o termo em $J^2$, que sugere
que a interpretação da ação efetiva como energia de vácuo está
perdida.

O primeiro problema é resolvido escolhendo $\zeta$ apropriadamente,
de forma a existir apenas uma solução para a ação efetiva. Isto é
feito através das equações do grupo de renormalização, veja, por
exemplo, \cite{Verschelde:2001ia} e o apêndice \ref{ap_ferram}. De
fato, escolhendo $\zeta$ como uma série de Laurent em $g^2$
\begin{equation}
\zeta=\frac{\zeta_0}{g^2}+\zeta_1+g^2\zeta_2+g^4\zeta_3+\ldots\;,\label{4zeta1}
\end{equation}
a ação efetiva (\ref{4eff1}) obedece a seguinte EGR homogênea
\begin{equation}
\left[\mu\frac{\p}{\p\mu}+\beta(g^2)\frac{\p}{\p{g^2}}-\gamma_{A^2}(g^2)\int{d^4x}\;J\frac{\delta}{\delta{J}}\right]\Gamma=0\;,\label{4rge2}
\end{equation}
onde
\begin{eqnarray}
\beta(g^2)&=&\mu\frac{\p{g^2}}{\p\mu}\;,\nonumber\\
\gamma_{A^2}(g^2)&=&\mu\frac{\p}{\p\mu}\ln{Z}_J\;.\label{4diman2}
\end{eqnarray}
Em particular, de acordo com a eq. (\ref{4ren3}), segue que a
dimensão anômala do operador $A_\mu^aA_\mu^a$ é dada por
\begin{equation}
\gamma_{A^2}(g^2)=-\left[\frac{\beta(g^2)}{2g^2}+\gamma_A(g^2)\right]\;.\label{4diman1}
\end{equation}
Vemos que, devido ao teorema de não renormalização (\ref{4ren3}), a
dimensão anômala do operador $A_\mu^aA_\mu^a$ não é uma quantidade
independente da teoria, sendo dependente da função beta e da
dimensão anômala do campo de calibre. Este resultado pode ser
generalizado a outros calibres como o MAG e o Curci-Ferrari, devido
a existência da simetria $SL(2,\mathbb{R})$, veja
\cite{Dudal:2003pe}. Uma importante observação a respeito do
parâmetro LCO $\zeta$ é o fato de sua solução influenciar ordens
diferentes na expansão do grupo de renormalização. Desta forma, para
resolvermos as equações na ordem $n$, necessitamos conhecer o valor
de $\zeta$ a ordem $n+1$.

O segundo problema, associado ao termo quadrático em $J$, pode
facilmente ser resolvido através da introdução do campo auxiliar de
Hubbard-Stratonovich, $\sigma$. Este truque é efetuado através da
introdução da identidade, na integral de caminho (\ref{4eff1}),
escrita como
\begin{equation}
1=\mathcal{N}\int{D\sigma}e^{-\frac{1}{2\zeta}\int{d^4x}\left[\frac{\sigma}{g}+\frac{1}{2}A_\mu^aA_\mu^a-\zeta{J}\right]^2}\;,\label{4hub1}
\end{equation}
onde $\mathcal{N}$ é um fator de normalização. Desta forma a
expressão (\ref{4eff1}) passa a ser escrita como
\begin{equation}
e^{-\Gamma(J)}=\int{DAD\bar{c}DcDb}\exp\left\{-S_{YM}-S_{gf}-S_\sigma+\int{d^4x}\frac{J\sigma}{g}\right\}\;,\label{4eff1x}
\end{equation}
onde
\begin{equation}
S_\sigma=\int{d^4x}\left[\frac{\sigma^2}{2g^2\zeta}+\frac{1}{2}\frac{\sigma}{g\zeta}A_\mu^aA_\mu^a+
\frac{1}{8\zeta}\left(A_\mu^aA_\mu^a\right)^2\right]\;.\label{4sigma1}
\end{equation}
Vemos que a identidade (\ref{4hub1}) elimina o termo quadrático em
$J$. Ademais, é fácil ver que o campo $\sigma$ se relaciona com o
operador composto $A_\mu^aA_\mu^a$ através da relação
\begin{equation}
\left\langle{A_\mu^aA_\mu^a}\right\rangle=-\frac{2}{g}\left\langle\sigma\right\rangle\;.\label{4rel1}
\end{equation}
Assim, se $\sigma$ desenvolver um valor esperado de vácuo não
trivial, teremos um valor não trivial para  condensado
$A_\mu^aA_\mu^a$. Tal condensado gera um termo de massa para o
glúon, através da relação
\begin{equation}
m^2=-\frac{\left\langle{A_\mu^aA_\mu^a}\right\rangle}{2\zeta}
\end{equation}

Não vamos entrar nos detalhes técnicos do cálculo do potencial
efetivo, uma vez que as questões físicas do mesmo foram abordadas,
vamos apenas apresentar os resultados. Os detalhes podem ser
encontrados em \cite{Verschelde:2001ia}.

\subsubsection{Vacuum energy and dynamical mass}

De acordo com
\cite{Verschelde:2001ia,Knecht:2001cc,Dudal:2002pq,Browne:2003uv} é
possível encontrar uma solução da equação do gap a um laço com
parâmetro de expansão relativamente pequeno,
$\frac{g^2N}{16\pi^2}=\approx0.193$. Tal solução é caracterizada, no
esquema de renormalização $\MSbar$ e regularização dimensional, por
\begin{eqnarray}
m^2&\approx&(2.031\lms)^2\;,\nonumber\\
E_\mathrm{vac}&\approx&-0.323\lms^4\;,\label{4res1}
\end{eqnarray}
onde a escala de renormalização foi escolhida de modo a eliminar os
logarítmos, ou seja $\omu^2=m^2$.

\subsection{Ghost operators}

Por fim, vamos apresentar as possibilidades de operadores de massa
compostos por campos fantasmas. Existem quatro possibilidades,
$gf^{abc}c^bc^c$, $gf^{abc}\bar{c}^b\bar{c}^c$,
$gf^{abc}c^b\bar{c}^c$ e $\bar{c}^ac^a$. Por analogia ao
supercondutor Abeliano, os dois primeiros pertencem ao chamado canal
BCS, em analogia à condensação dos pares de Cooper, enquanto o
terceiro pertence ao canal Overhauser, em analogia ao efeito
Overhauser. A quarta possibilidade será deixada de lado, pois no
calibre de Landau este operador não se faz necessário. Isto segue
das identidades de Ward que excluem a formação do condensado
$\left<\bar{c}^ac^a\right>$.

O estudo dos operadores $gf^{abc}c^bc^c$,
$gf^{abc}\bar{c}^b\bar{c}^c$ e $gf^{abc}c^b\bar{c}^c$ foi feito em
\cite{Dudal:2003dp} através do formalismo LCO. Neste trabalho foi
provada a renormalizabilidade da ação LCO descrevendo tais
operadores. Ainda, estes condensados se relacionam através da
simetria $SL(2,\mathbb{R})$. Tal simetria caracteriza as transições
entre os vácuos BCS e Overhauser, mostrando que são equivalentes
entre si. Esta propriedade permitirá que estudemos apenas o setor
Overhauser nesta tese.

Ainda, tais condensados carregam índice de cor, implicando na quebra
da simetria de cor. Contudo, foi mostrado também em
\cite{Dudal:2003dp}, que esta quebra ocorre no setor não físico da
teoria. Outra forma de entender isso, é através das excitações de
Goldstone associadas a quebra da simetria $SL(2,\mathbb{R})$, que
são não físicas \cite{Schaden:1999ew,Dudal:2003dp}. Este resultado é
uma consequência direta da invariância BRST. Outro efeito que estes
condensados acarretam é o fato de gerar massa para o glúon. Contudo,
esta massa se mostra taquiônica a um laço.

Ao contrário do caso $A_\mu^aA_\mu^a$, não vamos entrar em muitos
detalhes neste capítulo uma vez que os efeitos associados ao
condensado Overhauser serão todos discutidos no capítulo que se
segue. Para mais detalhes sobre os condensados fantasmas nos
referimos a \cite{Lemes:2002rc,Dudal:2003dp}.

\chapter{Combined analysis of the condensates $\left<A_\mu^aA_\mu^a\right>$ and $\left<f^{abc}\bar{c}^bc^c\right>$}

Neste capítulo vamos discutir as consequências dos condensados de
vácuo não triviais $\left<A_\mu^aA_\mu^a\right>$ e
$\left<f^{abc}\bar{c}^bc^c\right>$ quando seus efeitos são levados
em consideração simultaneamente. Mostraremos a renormalizabilidade
do método LCO para estes condensados para o caso geral $SU(N)$. Para
o resto do capítulo faremos cálculos para o caso mais simples
$SU(2)$. Em particular, faremos o cálculo a um laço da ação quântica
efetiva, da energia do vácuo e dos condensados propriamente ditos.
Faremos também um estudo das consequências físicas deste
condensados. Os detalhes técnicos deste capítulo podem ser
encontrados em \cite{Capri:2005vw}.

\section{Renormalizability}

Nesta seção demostraremos que as teorias de Yang-Mills permanecem
renormalizáveis quando introduzimos, simultaneamente, os operadores
compostos $A_\mu^aA^a_\mu$ e $gf^{abc}\bar{c}^bc^c$, no calibre de
Landau, através do método LCO.

Para acoplar os operadores compostos na ação de Yang-Mills no
calibre de Landau consideramos a ação completa
\begin{equation}
\Sigma=S_{YM}+S_{gf}+S_{LCO}+S'_{LCO}+S_{ext}+S'_{ext}\;.\label{5action1}
\end{equation}
Nesta ação $S_{YM}$ e $S_{gf}$ são dadas, respectivamente, em
(\ref{1ym1}-\ref{1landau1}). O termo de fontes externas acopladas às
variações BRST não lineares, $S_{ext}$, foi descrita em
(\ref{1ext1}). A ação $S_{LCO}$ descreve o termo LCO associado ao
operador gluônico $A_\mu^aA_\mu^a$, disponível em (\ref{4lco1}),
enquanto a ação $S'_{LCO}$ descreve o operador
$gf^{abc}\bar{c}^bc^c$ e é dada por
\begin{eqnarray}
S'_{LCO}&=&s\int{d^4x}\biggl(gf^{abc}\tau^a\bar{c}^bc^c-\frac{\rho}{2}\omega^a\tau^a\biggr)\nonumber\\
&=&\int{d^4x}\biggl(gf^{abc}\omega^a\bar{c}^bc^c-gf^{abc}\tau^ab^cc^c+\frac{g^2}{2}f^{abc}f^{cde}\tau^a\bar{c}^bc^dc^e-
\frac{\rho}{2}\omega^a\omega^a\biggr)\label{4lco3}
\end{eqnarray}
onde as fontes LCO formam dubletos BRST\footnote{Os números
quânticos das fontes LCO do operador fantasma podem ser encontrados
na tabela \ref{aptable2}, onde os números fantasmas das fontes
$\tau$ e $\omega$ são, respectivamente, $-1$ e $0$.} de acordo com
\begin{eqnarray}
s\tau^a&=&\omega^a\;,\nonumber\\
s\omega^a&=&0\;.\label{4doublets1}
\end{eqnarray}
Por fim, temos a ação extra $S'_{ext}$, dada por
\begin{eqnarray}
S'_{ext}&=&s\int{d^4x}\biggl(\beta\frac{g}{2}f^{abc}\tau^a\tau^bc^s+\gamma\tau^a\partial_{\mu}A^a_\mu\biggr)\nonumber\\
&=&\int{d^4x}\biggl(\beta{g}f^{abc}\omega^a\tau^bc^c+\beta\frac{g^2}{4}f^{abc}f^{cde}\tau^a\tau^bc^dc^e+\gamma\omega^a\partial_\mu{A}^a_\mu+
\gamma\tau^a\partial_{\mu}D_\mu^{ab}c^b\biggr)\;,\nonumber\\
&&\label{5supplement1}
\end{eqnarray}
onde $\beta$ e $\gamma$ são novos parâmetros independentes sem
dimensão\footnote{Estes parâmetros aparecem apenas neste capítulo e
não existirá confusão caso reutilizemos tais letras gregas em outros
capítulos para descrever outras quantidades.}. A ação $S'_{ext}$ é
introduzida por motivos de renormalizabilidade.

A ação completa (\ref{5action1}), está na forma apropriada para que
suas simetrias sejam des-critas através de identidades de Ward, veja
apêndice \ref{ap_ferram}. Identidades estas que são listadas a
seguir:
\begin{itemize}
\item  Identidade de Slavnov-Taylor
\begin{equation}
\mathcal{S}(\Sigma)=\int{d^4}x\biggl(\frac{\delta\Sigma}{\delta\Omega_\mu^a}\frac{\delta\Sigma}{\delta{A}_\mu^a}+
\frac{\delta\Sigma}{\delta{L}^a}\frac{\delta\Sigma}{\delta{c}^a}+b^a\frac{\delta\Sigma}{\delta\bar{c}^a}+
\omega^a\frac{\delta\Sigma}{\delta\tau^a}+J\frac{\delta\Sigma}{\delta\lambda}\biggr)=0\;.\label{5ST1}
\end{equation}

\item  Condição de Landau modificada
\begin{equation}
\frac{\delta\Sigma}{\delta{b}^a}=\partial_\mu{A}_\mu^a+gf^{abc}\tau^bc^c\;.\label{5gauge-fixing1}
\end{equation}

\item  Equação dos campos antifantasmas modificada
\begin{equation}
\frac{\delta\Sigma}{\delta\bar{c}^a}+\partial_\mu\frac{\delta\Sigma}{\delta\Omega_\mu^a}-gf^{abc}\tau^b\frac{\delta\Sigma}{\delta{L}^c}=
-gf^{abc}\omega^bc^c\;.\label{5anti-ghost1}
\end{equation}

\item  Equação dos campos fantasmas modificada
\begin{equation}
\mathcal{G}^a(\Sigma)=\Delta_{class}^a\;,\label{5ghosteq1}
\end{equation}
com
\begin{eqnarray}
\mathcal{G}^a&=&\int{d^4}x\biggl(\frac{\delta}{\delta{c}^a}+gf^{abc}\bar{c}^b\frac{\delta}{\delta{b}^c}+
gf^{abc}\tau^b\frac{\delta}{\delta\omega^c}\biggr)\;,\nonumber\\
\Delta_{class}^a&=&\int{d^4}x\Bigl[gf^{abc}\Bigl(\Omega_\mu^bA_\mu^c-L^bc^c-\omega^b\bar{c}^c+(\beta-\rho)\tau^b\omega^c-\tau^bb^c\Bigl)\Bigr]
\nonumber\;.
\end{eqnarray}

\item  Equação de $\lambda$ modificada
\begin{equation}
\int{d^4}x\biggl(\frac{\delta\Sigma}{\delta\lambda}+c^a\frac{\delta\Sigma}{\delta{b}^a}-2\tau^a\frac{\delta\Sigma}{\delta{L}^a}\biggr)=0\;.
\label{5tau-equation1}
\end{equation}
Esta identidade expressa a invariância BRST, ao nível clássico, do
operador $A_\mu^aA_\mu^a$.
\end{itemize}

Lembrando que, como discutido no apêndice \ref{ap_ferram}, de acordo
com a teoria de renormalização algébrica, os termos nos lados
direitos das equações (\ref{5gauge-fixing1})-(\ref{5ghosteq1}),
sendo lineares nos campos quânticos, representam uma quebra que
permanece clássica, não evoluindo às correções quânticas,
\cite{book}. Ainda, de acordo com a teoria de renormalização
algébrica, pode-se mostrar que, o contratermo mais geral possível a
ser adicionado a ação (\ref{5action1}), respeitando as identidades
(\ref{5ST1}-\ref{5ghosteq1}), pode ser escrito como
\begin{equation}
\Sigma^{(1)}=a_0S_{YM}+\mathcal{B}_{\Sigma}\Delta^{-1},\label{5general-CT1}
\end{equation}
onde $\mathcal{B}_{\Sigma}$ é o operador de Slavnov-Taylor
linearizado, nilpotente, dado por
\begin{equation}
\mathcal{B}_{\Sigma}=\int{d^4}x\biggl(\frac{\delta\Sigma}{\delta\Omega_\mu^a}\frac{\delta}{\delta{A}_\mu^a}+
\frac{\delta\Sigma}{\delta{A}_\mu^a}\frac{\delta}{\delta\Omega_\mu^a}+\frac{\delta\Sigma}{\delta{L}^a}\frac{\delta}{\delta{c}^a}+
\frac{\delta\Sigma}{\delta{c}^a}\frac{\delta}{\delta{L}^a}+b^a\frac{\delta}{\delta\bar{c}^a}+\omega^a\frac{\delta}{\delta\tau^a}+
J\frac{\delta}{\delta\lambda}\biggr)\;,\label{5linearized1}
\end{equation}
e $\Delta ^{-1}$ escrito como
\begin{eqnarray}
\Delta^{-1}&=&\int{d^4}x\left\{a_1\left[A_\mu^a\left(\Omega_\mu^a+\partial_\mu\bar{c}^a\right)-\frac{\lambda}{2}A_\mu^aA_\mu^a\right]\right.\nonumber\\
&+&\left.\frac{a_8}{2}\zeta\lambda{J}+a_9\gamma\tau^a\partial_\mu{A}_\mu^a+\frac{a_{14}}{2}\rho\tau^a\left(g{f}^{abc}\tau^bc^c-
\omega^a\right)\right\}\;.\label{5general-delta-1}
\end{eqnarray}
O passo final para demostrar a renormalizabilidade da ação
(\ref{5action1}) é checar a estabilidade da mesma sob correções
quânticas. Este passo é feito reabsorvendo o contratermo
(\ref{5general-CT1}) na ação (\ref{5action1}) através da redefinição
multiplicativa dos campos, fontes e parâmetros da teoria. A
definição da redefinição multiplicativa dos campos e fontes e
parâmetros está escrita em (\ref{apren1}), com
\begin{eqnarray}
\Phi&\in&\left\{A,b,c,\bar{c}\right\}\;,\nonumber\\
\mathcal{J}&\in&\left\{\Omega,L,\lambda,J,\tau,\omega\right\}\;,\nonumber\\
\xi&\in&\left\{g,\zeta,\rho,\beta,\gamma\right\}\;.\label{5ren1}
\end{eqnarray}
De fato, temos a renormalizabilidade multiplicativa. No caso da
constante de acoplamento e do campo de calibre os fatores são dados
de acordo com a expressão (\ref{4ren1}), como no caso sem operadores
compostos. Da mesma forma os campos de Faddeev-Popov, o campo de
Lautrup-Nakanishi e as fontes BRST se renormalizam de acordo com
(\ref{4ren4}-\ref{4ren4a}). As quantidades LCO associadas ao
operador $A_\mu^aA_\mu^a$ também não sofrem alteração, e continuam
na forma (\ref{4ren2}-\ref{4ren3}). Para os demais objetos,
\begin{eqnarray}
Z_\omega&=&Z_A^{1/2}\;,\nonumber\\
Z_\tau&=&Z_g^{-1/2}Z_A^{3/4}\nonumber\;,\\
Z_\beta&=&1+\epsilon\Bigl(\frac{\rho{a}_{14}}{\beta}-a_0-2a_1\Bigr)\nonumber\;,\\
Z_\rho&=&1+\epsilon(a_{14}-a_0-2a_1)\nonumber\;,\\
Z_\zeta&=&1+\epsilon(a_8+2a_0+2a_1)\nonumber\;,\\
Z_\gamma&=&1+\epsilon(a_9-a_0-a_1)\;.\label{5Zs1}
\end{eqnarray}

Uma importante observação está no fato de $Z_J$ e $Z_{\omega}$ serem
escritos como combinação dos fatores de renormalização do glúon e da
constante de acoplamento. Este teroema de não renormalização implica
no fato de as dimensões anômalas dos operadores compostos
$A_\mu^aA_\mu^a$ e $gf^{abc}\bar{c}^bc^c$ serem expressas como
combinações lineares da dimensão anômala do glúon e da função beta,
\cite{Dudal:2002pq,Dudal:2003dp}.

\section{One-loop quantum effective action in $\MSbar$ scheme}

\subsection{Initial remarks}

Vamos agora calcular a ação efetiva para os condensados
$\left\langle A_\mu^aA_\mu^a\right\rangle$ e
$\left\langle{f}^{abc}\bar{c}^bc^c\right\rangle$.

Uma vez provada a renormalizabilidade, podemos colocar a zero as
fontes desnecessárias ao cálculo da ação quântica,
$\tau^a=\lambda=\Omega_\mu^a=L^a=0$. Note que nenhum dos termos
dependentes nestas fontes são necessários para o cálculo do
potencial efetivo. É importante ressaltar, ainda, que podemos
esquecer o termo $\omega^a\partial_\mu{A}_\mu^a$ na equação
(\ref{5supplement1}), pois no calibre de Landau $\p_\mu{A}_\mu^a=0$.
Contudo, podemos um pouco mais formais e efetuar a transformação de
Jacobiano unitário $b^\prime=b+\gamma\omega$ no funcional gerador de
diagramas conexos $\mathcal{W}(\omega,J)$, o que leva a eliminação
dos parâmetros $\beta$ e $\gamma$.

Temos ainda dois parâmetros LCO livres, $\rho$ e $\zeta$. Como
demonstrado em \cite{Verschelde:2001ia,Dudal:2003by,Dudal:2003dp},
tais parâmetros podem ser fixados através das equações do grupo de
renormalização, veja também apêndice \ref{ap_ferram}. Lembramos
ainda que o valor explícito do contratermo proporcinal a $\omega^a$
não é afetado pela presença de $J$ e vice versa. Desta forma, os
valores previamente determinados para $\rho$ e $\zeta$ permanecem os
mesmos que os casos isolados estudados no capítulo anterior,
\cite{Verschelde:2001ia,Dudal:2003by,Dudal:2003dp,Capri:2005vw}.
Tais valores são dados por\footnote{Neste capítulo estaremos sempre
utilizando regularização dimensional com a convesão $d=4-\epsilon$ e
esquema de renormalização $\MSbar$.}
\begin{eqnarray}
\rho&=&\rho_0+\rho_1g^2+\ldots\;,\nonumber\\
\zeta&=&\frac{\zeta_0}{g^2}+\zeta_1+\ldots\;,\label{5a2}
\end{eqnarray}
onde
\begin{eqnarray}
\rho_0&=&-\frac{6}{13}\;,\;\;\;\;\;\;\rho_1=-\frac{95}{312\pi^2}\;,\nonumber\\
\zeta_0&=&\frac{27}{26}\;,\;\;\;\;\;\;\;\;\;\zeta_1=\frac{161}{52}\frac{3}{16\pi^2}\;,\label{5a2bis}
\end{eqnarray}
para o caso $SU(2)$.

A ação relevante para o cálculo d ação efetiva é, portanto,
\begin{eqnarray}
S&=&S_{YM}+S_{gf}+\int{d^4x}\left(gf^{abc}\omega^a\bar{c}^bc^c+\frac{1}{2}JA_\mu^aA^a_\mu-\frac{\rho}{2}\omega^a\omega^a-\frac{\zeta}{2}J^2\right)\;.
\label{5Snodig}
\end{eqnarray}
Os termos quadráticos nas fontes, que não são compatíveis com a
interpretação da ação quântica como energia do vácuo, podem ser
eliminados através de uma transformação de Hubbard-Stratonovich,
desta forma
\cite{Verschelde:2001ia,Dudal:2003dp,Dudal:2003by,Capri:2005vw},
\begin{eqnarray}
S_{\sigma\phi}(J,\omega)&=&S_{YM}+S_{gf}+\int{d^4}x\left[\frac{\phi^a\phi^a}{2g^2\rho}+\frac{1}{\rho}\phi^a{f}^{abc}\bar{c}^bc^c+
\frac{g^2}{2\rho}\left(f^{abc}\bar{c}^bc^c\right)^2\right]\nonumber\\
&+&\int{d^4}x\left[\frac{\sigma^2}{2g^2\zeta}+\frac{\sigma}{2g\zeta}A_\mu^aA_\mu^a+\frac{1}{8\zeta}\left(A_\mu^aA_\mu^a\right)^2-
\omega^a\frac{\phi^a}{g}-J\frac{\sigma}{g}\right]\;,\label{5a1}
\end{eqnarray}
onde as fontes se acoplam linearmente aos campos auxiliares $\sigma$
e $\phi^a$. Ainda, valem as seguintes identificações
\cite{Verschelde:2001ia,Dudal:2003dp,Dudal:2003by,Capri:2005vw}
\begin{eqnarray}
\left\langle\phi^a \right\rangle &=& -g^2\left\langle f^{abc}\oc^a c^b\right\rangle\;,\nonumber \\
  \left\langle \sigma \right\rangle &=&-\frac{g}{2}\left\langle
  A_\mu^2\right\rangle\;,\label{5ident1}
\end{eqnarray}
Note que a ação (\ref{5a1}) será multiplicativamente renormalizável
e obedecerá a uma equação homogênea do grupo de renormalização.

Assim, a ação efetiva poderá ser calculada através da definição
usual da energia do vácuo,
\begin{equation}
e^{-\Gamma}=\int{[D\Phi]}e^{-S_{\sigma\phi}(J=\omega=0)}\;,\label{5eff1}
\end{equation}
onde a medida $[D\Phi]=DADbD\bar{c}DcD\phi{D}\sigma$ é tomada com
relação a todos os campos relevantes.

\subsection{Quantum action}

Para o cálculo explícito do potencial efetivo vamos considerar a
aproximação a um laço e vamos nos concentrar, pelo resto deste
capítulo, ao caso $SU(2)$. Mostraremos que valores não triviais para
os condensados favorecem uma energia de vácuo negativa,
independentemente da escala.

Para a ação efetiva a um laço
$\Gamma^{(1)}\left(\sigma,\phi\right)$, basta considerarmos os
termos quadráticos na ação $S_{\sigma\phi}$, (\ref{5a1}), tomando os
campos de Hubbard-Stratonovich como configurações de vácuo estáveis.
Assim, a ação quântica a um laço se reduz ao cálculo de
\begin{eqnarray}
e^{-\Gamma^{(1)}(\sigma,\phi)}&=&\int{DAD\bar{c}Dc}\exp\left\{\frac{1}{2}\int{d^4x}A^a_\mu\left[\delta_{\mu\nu}\left(\partial^2-\frac{\sigma}{g\zeta}\right)-
\left(1-\frac{1}{\alpha}\right)\partial_\mu\partial_\nu\right]A_\nu^a\right.\nonumber\\
&-&\left.\int{d^4}x\bar{c}^a\left(\delta^{ac}\partial^2-\frac{1}{\rho}\phi^b\varepsilon^{abc}\right)c^c-
\int{d^4}x\left(\frac{\phi^a\phi^a}{2g^2\rho}+\frac{\sigma^2}{2g^2\zeta}\right)\right\}\;,\label{5eff1x}
\end{eqnarray}
Com regularização dimensional e no esquema $\MSbar$ deduzimos que
\begin{equation}
\Gamma^{(1)}(\sigma,\phi)=\Gamma_{A^2}(\sigma)+\Gamma_{gh}\left(\phi\right)\;,\label{5a4}
\end{equation}
com \cite{Verschelde:2001ia,Dudal:2003by,Capri:2005vw}
\begin{equation}
\Gamma_{A^2}(\sigma)=\frac{\sigma^2}{2\zeta_0}\left(1-\frac{\zeta_1}{\zeta_0}g^2\right)+
\frac{3\left(N^2-1\right)}{64\pi^2}\frac{g^2\sigma^2}{\zeta_0^2}\left(\ln\frac{g\sigma}{\zeta_0\overline{\mu}^2}-\frac{5}{6}\right)\;,\label{5a5}
\end{equation}
enquanto \cite{Dudal:2003dp,Capri:2005vw}
\begin{equation}
\Gamma_{gh}\left(\phi\right)=\frac{\phi^2}{2g^2\rho_0}\left(1-\frac{\rho_1}{\rho_0}g^2\right)+
\frac{1}{32\pi^2}\frac{\phi^2}{\rho_0^2}\left(\ln\frac{\phi^2}{\rho_0^2\overline{\mu}^4}-3\right)\;,\label{5a6}
\end{equation}
onde $\phi=\phi^3$, $\phi^a=\phi\delta^{a3}$. Tal decomposição
equivale a escolher a configuração de vácuo na direção Abeliana do
espaço de cor, correspondente ao subgrupo de Cartan de $SU(2)$, o
qual é gerado pela matriz de Pauli diagonal $\sigma^3$.

As configuraçòes de mínimo, descrevendo o vácuo, são encontradas
resolvendo as equações de gap
\begin{equation}
\frac{\partial \Gamma^{(1)}\left( \sigma ,\phi \right) }{\partial \sigma }=\frac{%
\partial \Gamma^{(1)}\left( \sigma ,\phi \right) }{\partial \phi }=0\;,\label{5a7}
\end{equation}
que resultam em
\begin{eqnarray}
\frac{1}{\zeta _{0}}\left( 1-\frac{\zeta _{1}}{\zeta _{0}}%
g^{2}\right) +2\frac{9}{64\pi ^{2}}\frac{g^{2}}{%
\zeta _{0}^{2}}\left( \log \frac{g\sigma _{*}}{\zeta _{0}\overline{%
\mu }^{2}}-\frac{5}{6}\right) +\frac{9 }{64\pi ^{2}}%
\frac{g^{2}}{\zeta _{0}^{2}} &=&0\;,  \nonumber \\
\frac{1}{g^{2}\rho _{0}}\left( 1-\frac{\rho _{1}}{\rho _{0}}g^{2}\right) +2%
\frac{1}{32\pi ^{2}}\frac{1}{\rho _{0}^{2}}\left( \log \frac{\phi _{*}^{2}}{%
\rho _{0}^{2}\overline{\mu }^{4}}-3\right) +2\frac{1}{32\pi ^{2}}\frac{1}{%
\rho _{0}^{2}} &=&0\;,  \label{5a8}
\end{eqnarray}
onde $\left( \sigma _{*},\phi _{*}\right)$ denotam os valores de
soluções não triviais. Substituindo as (\ref{5a8}) na ação efetiva
(\ref{5eff1x}), encontramos, para a energia do vácuo,
\begin{equation}
\Evac =\Gamma^{(1)}\left( \sigma
_{*},\phi _{*}\right) =-\frac{9}{128\pi ^{2}}\frac{%
g^{2}}{\zeta _{0}^{2}}\sigma _{*}^{2}-\frac{1}{32\pi ^{2}}\frac{\phi
_{*}^{2}}{\rho _{0}^{2}}\;.  \label{5a9}
\end{equation}
Concluímos que valores não triviais para os condensados são
dinamicamente favoráveis, uma vez que a energia do vácuo a um laço é
diminuída por estas configurações. Este resultado é independente da
escala escolhida.

\subsection{Renormalization group invariance}

Para checar a invariância da ação efetiva a um laço (\ref{5a4}),
primeiro redefinimos algumas quantidades de acordo com
\begin{eqnarray}
m^2&=&\frac{g\sigma}{\zeta_0}\;,\nonumber\\
\omega&=&\frac{\phi}{\left|\rho_0\right|}\;.\label{5a15b}
\end{eqnarray}
de forma que a ação efetiva se escreva como
\begin{eqnarray}
\Gamma^{(1)}(m^2,\omega)&=&\zeta_0\frac{m^4}{2g^2}\left(1-\frac{\zeta_1}{\zeta_0}g^2\right)+
\frac{3\left(N^2-1\right)}{64\pi^2}m^4\left(\ln\frac{m^2}{\overline{\mu}^2}-\frac{5}{6}\right)-
\rho_0\frac{\omega^2}{2g^2}\left(1-\frac{\rho_1}{\rho_0}g^2\right)\nonumber\\
&+&\frac{\omega^2}{32\pi^2}\left(\ln\frac{\omega^2}{\overline{\mu}^4}-3\right)\;,\label{5ll2}
\end{eqnarray}
De acordo com a técnica LCO, a ação quântica deve obedecer a
seguinte EGR
\begin{equation}
\overline{\mu}\frac{d}{d\overline{\mu}}\Gamma^{(1)}(m^2,\omega)=\left[\omu\frac{\p}{\p\omu}+\beta(g^2)\frac{\p}{\p
g^2}+\gamma_{\omega}(g^2)\omega\frac{\p}{\p\omega}+\gamma_{m^2}(g^2)m^2\frac{\p}{\p{m^2}}\right]\Gamma^{(1)}(m^2,\omega)=0\;.\label{5rg1}
\end{equation}
onde
\begin{eqnarray}
\omu\frac{\p m^2}{\p\omu}&=&\gamma_{m^2}(g^2)m^2\;,\nonumber\\
\overline{\mu}\frac{\p\omega}{\p\overline{\mu}}&=&\gamma_\omega(g^2)\omega\;,\label{5diman1}
\end{eqnarray}
Das expressões (\ref{5ren1}) e das definições (\ref{5a15b}) é fácil
deduzir que
\begin{eqnarray}
\gamma_{m^2}(g^2)&=&\frac{\beta(g^2)}{2g^2}-\gamma_A(g^2)\;,\nonumber\\
\gamma_\omega(g^2)&=&\frac{\beta(g^2)}{2g^2}+\gamma_A(g^2)\;.\label{5diman2}
\end{eqnarray}
Utilizando as expressões (\ref{5a2bis}), juntamente com os valores a
um laço da função beta e da dimensão anômala do glúon, (veja, por
exemplo, \cite{Gracey:2002yt})
\begin{eqnarray}
\beta^{(1)}(g^2)&=&-\frac{22}{3}\frac{g^4N}{16\pi^2}\;,\nonumber\\
\gamma_A^{(1)}(g^2)&=&-\frac{13}{6}\frac{g^2N}{16\pi^2}\;,\label{5beta1}
\end{eqnarray}
é trivial checar a validade da equação (\ref{5ll2}).

\subsection{Numerical results}

Vimos na expressão (\ref{5a9}) que a formação de ambos os
condensados é favorecida, devido ao valor negativo da energia do
vácuo, independentemente da escala escolhida. Contudo, devemos ser
capazes de encontrar uma solução consistente das equações de gap
(\ref{5a8}) para confirmar este resultado formal. Para tal, devemos
lidar com o problema de termos agora duas escalas de massa e,
consequentemente, dois tipos de logaritimos potencialmente grandes.
Este problema foi resolvido tecnicamente em \cite{Capri:2005vw},
onde os deta-lhes podem ser encontrados. Fundamentalmente,
utilizamos a invariância sob o grupo de renormalização para ressomar
os logaritimos dominantes na ação efetiva. Ainda, este procedimento
pode ser feito separadamente para $m^2$ e $\omega$ pois estes
setores não se misturam na aproximação de um laço. Esta propriedade
pode ser utilizada para definir cada escala separadamente pois
teremos duas expansões totalmente independentes. Desta forma, cada
setor obdece a uma EGR independente. Assim, de acordo com
\cite{Capri:2005vw}, podemos trabalhar com dois parâmetros de
expansão, cujos valores são dados pela solução das equações de gap
após a ressoma dos logaritimos dominantes
\begin{eqnarray}
\left.\frac{\og^2N}{16\pi^2}\right|_{N=2}&=&\frac{9}{37}\approx0.243\;,\nonumber\\
\left.\frac{\widetilde{g}^2N}{16\pi^2}\right|_{N=2}&=&\frac{36}{385}\approx0.094\;,\label{5ll14}
\end{eqnarray}
onde as quantidades $\bar{q}$ são calculadas na escala $\omu=m^2$
enquanto quantidades $\widetilde{q}$ são calculadas na escala
$\omu=\omega$. Assim, da expressão a um laço do esquema $\MSbar$
\begin{equation}
    g^2(\omu)=\frac{1}{\beta_0\ln\frac{\omu^2}{\lms^2}}\;,\label{5ll15}
\end{equation}
juntamente com as soluções das equações de gap (\ref{5ll14}),
obtemos as estimativas
\begin{eqnarray}
\overline{m}^2=e^{\frac{37}{33}}\lms^2&\approx&\nonumber3.07\lms^2\;,\\
\widetilde{\omega}=e^{\frac{35}{12}}\lms^2&\approx&18.48\lms^2\;,\label{5ll16}
\end{eqnarray}
Para a energia do vácuo, (\ref{5a9}), obtemos
\begin{equation}
    \Evac\approx-1.15\lms^4\;.\label{5ll17}
\end{equation}

Os parâmetros de expansão (\ref{5ll14}), sendo relativamente
pequenos, nos fornecem resultados bastante confiáveis. Contudo, uma
análise a mais ordens na expansão em laços é necessária para
confirmar e melhorar nossos resultados.

\section{Consequences of non-trivial condensates}

Vamos agora analizar algumas consequências físicas que ocorrem
devido a presença dos condensados
$\left\langle{A}_\mu^aA_\mu^a\right\rangle$ e
$\left\langle\varepsilon^{abc}\bar{c}^bc^c\right\rangle$. Em
particular, vamos discutir a transversalidade do propagador do glúon
e da polarização do vácuo. Faremos a análise através de identidades
de Ward. Veremos também que há uma quebra de degenerescência no
espectro de massa do glúon. Para tal é conveniente fazermos algumas
observações a cerca da ação efetiva.

De fato, devido ao fato de a transformação de Hubbard-Stratonovich
ser um mapeamento exato, a ação
\begin{equation}
S_{\sigma\phi}=S_{YM}+S_{gf}+S_\sigma+S_\phi\;,\label{5b1}
\end{equation}
onde $S_\sigma$ é dado por (\ref{4sigma1}) e $S_\phi$ por
\begin{equation}
S_\phi=\int{d^4}x\left[\frac{\phi^a\phi^a}{2g^2\rho}+\frac{1}{\rho}\varepsilon^{abc}\phi^a\bar{c}^bc^c+
\frac{g^2}{2\rho}\left(\varepsilon^{abc}\bar{c}^bc^c\right)^2\right]\;,\label{4action2}
\end{equation}
possui invariância BRST. Esta propriedade é facilmente estabelecida
decompondo os campos de Hubbard Stratonovich em termos de flutuações
quânticas em torno do condensado de vácuo, isto é
\begin{eqnarray}
\sigma&=&\sigma_*+\widetilde{\sigma}\;,\nonumber\\
\phi^a(x)&=&\delta^{a3}\phi_*+\widetilde{\phi}^a(x)\;,\nonumber\\
\left\langle\widetilde{\phi}^a(x)\right\rangle&=&0\;.\label{5b2}
\end{eqnarray}
Para as transformações BRST obtemos
\begin{eqnarray}
s\sigma&=&s\widetilde{\sigma}=gA_\mu^a\p_\mu{c}^a\;,\nonumber\\
s\sigma_*&=&0\;,\nonumber\\
s\phi^a&=&s\widetilde{\phi}^a=-g^2s(\varepsilon^{abc}\bar{c}^bc^c)=-g^2\left(\varepsilon^{abc}b^bc^c+
\frac{g}{2}\varepsilon^{abc}\varepsilon^{cmn}\bar{c}^bc^mc^n\right)\;,\nonumber\\
s\phi_*&=&0\;,\label{5b4}
\end{eqnarray}
que, juntamente com as eqs. (\ref{1brs1}), definem uma invariância
exata da ação $S_{\sigma\phi}$
\begin{equation}
sS_{\sigma\phi}=0\;.\label{5b5}
\end{equation}

Outra importante observação está na escolha do condensado e suas
flutuações na forma (\ref{5b2}). Esta escolha sugere um
desacoplamento nos setores Abeliano e não-Abeliano do grupo $SU(2)$.
De fato, a expressão (\ref{5b2}) sugere que façamos a decomposição
do índice de cor como $a=\{3,A\}$ com $A\in\{1,2\}$. E o campo
$\phi^a$ se decompõe como
$\phi^a=\{\phi^3,\phi^A\}=\{\phi,\phi^A\}$. Desta forma
\begin{eqnarray}
\phi(x)&=&\phi_*+\widetilde{\phi}(x)\;,\nonumber\\
\phi^A(x)&=&\widetilde{\phi}^a(x)\;,\;.\label{5b2a}
\end{eqnarray}
Substituindo as definições (\ref{5b2}) e (\ref{5b2a}) na ação
(\ref{5b1}), temos que
\begin{eqnarray}
S_{\sigma\phi}&=&S_{YM}+S_{gf}+\int{d^4}x\left[\frac{\sigma_*^2}{2g^2\zeta}+
\frac{\phi_*^2}{2g^2\rho}+
\frac{\sigma_*\widetilde{\sigma}}{g^2\zeta}+
\frac{\phi_*\widetilde{\phi}}{g^2\rho}+
\frac{\widetilde{\sigma}^2}{2g^2\zeta}+
\frac{\widetilde{\phi}^A\widetilde{\phi}^A}{2g^2\rho}+ \frac{1}{2}\frac{\sigma_*}{g\zeta}A_\mu^aA_\mu^a\right.\nonumber\\
&+&\left.\frac{1}{\rho}\phi_*\varepsilon^{AB}\bar{c}^Ac^B+
\frac{1}{2}\frac{\widetilde{\sigma}}{g\zeta}A_\mu^aA_\mu^a+
\frac{1}{\rho}\widetilde{\phi}^a\varepsilon^{abc}\bar{c}^bc^c+
\frac{g^2}{2\rho}\left(\varepsilon^{abc}\bar{c}^bc^c\right)^2+\frac{1}{8\zeta}\left(A_\mu^aA_\mu^a\right)^2\right]\;,\nonumber\\
&&\label{5b1a}
\end{eqnarray}
Vemos assim, que no nível árvore o propagador fantasma já apresenta
tal decomposição, devido a presença de um condensado Overhauser não
trivial. Para ficar evidente, vamos escrever os propagadores a ordem
zero, no vácuo não trivial. No caso do propagador do glúon temos,
\cite{Verschelde:2001ia,Dudal:2003by},
\begin{equation}
D_{\mu\nu}^{ab}(k)=\delta^{ab}\frac{1}{k^2+m^2}\left(\delta_{\mu\nu}-\frac{k_\mu{k}_\nu}{k^2}\right)\;.\label{5c1}
\end{equation}
Vemos que o parâmetro $m^2$, definido em (\ref{5a15b}), corresponde
a uma massa gerada dinamicamente para o glúon, no nível árvore. Para
o propagador dos campos fantasmas temos, \cite{Dudal:2003dp},
\begin{eqnarray}
G^{33}(k)&=&\frac{1}{k^2\;}\;,\nonumber\\
G^{AB}(k)&=&\frac{\delta^{AB}k^2-\omega\varepsilon^{AB}}{k^4+\omega^2}\;,\label{5c2}
\end{eqnarray}

\subsection{Transversality analysis}

\subsubsection{Gluon propagator}

Para mostrar que o propagador do glúon permanece transverso
consideramos o gerador funcional das funções de Green conexas $Z^c$,
obtido a partir da ação quântica $\Gamma$ através de uma
transformação de Legendre. Temos então a inclusão das fontes de
Schwinger $I_b^a$, $J_\mu^a$ e $K_c^a$, para os campos $b^a$,
$A_\mu^a$ e $c^a$, respectivamente. A ação efetiva $\Gamma$ obedece
a seguinte identidade de Ward
\begin{equation}
\frac{\delta\Gamma}{\delta{b}^a}=\partial_\mu{A}_\mu^a+g^2\varepsilon^{abc}F^bc^c\;,\label{5ident2}
\end{equation}
que se transforma, para o funcional $Z^c$, em
\begin{equation}
I_b^a=\p_\mu\frac{\delta{Z}^c}{\delta{J}_\mu^a}+g^2\varepsilon^{ade}F^d\frac{\delta{Z}^c}{\delta{K}_c^e}\;,\label{5gp2}
\end{equation}
de onde derivamos facilmente a relação de transversalidade do
propagador do glúon, a todas as ordens em teoria de perturbações
\begin{equation}
0=\p_\mu^x\frac{\delta^2Z^c}{\delta{J}_\mu^a(x)\delta{J}_\mu^b(y)}\bigg|_{F,I,J,K=0}\;.\label{5gp3}
\end{equation}

\subsubsection{Vacuum polarization}

Para o estudo da polarização do vácuo vamos utilizar a identidade de
Slavnov-Taylor para a ação quântica $\Gamma$. Assim, de forma a
escrevermos uma identidade de Slavnov-Taylor consistente com o PAQ,
veja \cite{book} e apêndice \ref{ap_ferram}, adicionamos à ação
(\ref{5b1}) um termo de fontes BRST invariantes acopladas às
variações BRST não lineares,
\begin{equation}
S''_{ext}=\int
d^{4}x\left(\Omega_\mu^asA_\mu^a+L^asc^a+Rs\sigma+F^as\widetilde{\phi}^a\right)\;,\label{5b7}
\end{equation}
A ação completa
\begin{equation}
\Sigma_{\sigma\phi}=S_{\sigma\phi}+S''_{ext}\;,\label{5b8}
\end{equation}
obedece a uma identidade de Slavnov-Taylor, também válida no nível
quântico, de forma que
\begin{equation}
\mathcal{S}(\Gamma)=\int{d^4}x\left(\frac{\delta\Gamma}{\delta\Omega_\mu^a}\frac{\delta\Gamma}{\delta{A}_\mu^a}+
\frac{\delta\Gamma}{\delta{L}^a}\frac{\delta\Gamma}{\delta{c}^a}+\frac{\delta\Gamma}{\delta{R}}\frac{\delta\Gamma}{\delta\widetilde{\sigma}}
+\frac{\delta\Gamma}{\delta{F}^a}\frac{\delta\Gamma}{\delta\widetilde{\phi}^a}+
b^a\frac{\delta\Gamma}{\delta\bar{c}^a}\right)=0\;,\label{5b11}
\end{equation}
onde
\begin{equation}
\Gamma=\Sigma_{\sigma\phi}+\epsilon\Sigma_{\sigma\phi}^{(1)}+\epsilon^2\Sigma_{\sigma\phi}^{(2)}+\ldots\label{5b13}
\end{equation}
Considerando a aproximação a primeira ordem em $\epsilon$ e atuando
em (\ref{5b11}) com o operador teste
\begin{equation}
\frac{\delta^2}{\delta{c}^a(x)\delta{A}_\nu^b(y)}\label{5b16}
\end{equation}
e colocando todos os campos e fontes a zero,
$A_\mu^a=\Omega_\mu^a=c^a=L^a=\widetilde{\phi}^a=R=F^a=b^a=\bar{c}^a=0$,
obtemos a seguinte identidade de Ward para a polarização do vácuo
\begin{eqnarray}
\partial_\mu^x\frac{\delta^2\Gamma^{(1)}}{\delta{A}_\mu^a(x)\delta{A}_\nu^{b}(y)}&=&
\frac{\phi_*}{\rho_0}\frac{\delta^2\left[\int{d^4}z\left(\varepsilon^{3np}b^nc^p\right)_z\;
\cdot\Gamma\right]^{(1)}}{\delta{c}^a(x)\delta{A}_\nu^b(y)}\nonumber\\
&+&\frac{\sigma_*}{g\zeta_0}\frac{\delta^2\left[\int{d^4}z\left(A_\alpha^a\partial_\alpha{c}^a\right)_z\;
\cdot\Gamma\right]^{(1)}}{\delta{c}^a(x)\delta{A}_\nu^b(y)}\;.\label{5b17a0}
\end{eqnarray}
O lado direito da expressão (\ref{5b17a0}) sugere a quebra da
trasnversalidade devido a presença dos condensados Overhauser e de
gluons. Contudo o último termo, associado ao condensado de gluons, é
nulo, devido a transversalidade do propagador do glúon no nível
árvore. Para argumentar sobre isso consideremos a transversalidade
do propagador do glúon (\ref{5gp3}), que segue da condição
$\partial_\mu{A}_\mu^a=0$. Portanto o segundo termo em
(\ref{5b17a0}), proporcional ao condensado $\sigma_*$, é, de fato,
nulo. Isto porque temos a presença de gluons internos se propagando
na expansão em laços, sempre acompanhados de quadridivergências. Um
cálculo explícito a um laço conduzirá à confirmação deste resultado.
Desta forma, temos
\begin{equation}
\partial_\mu^x\frac{\delta^2\Gamma^{(1)}}{\delta{A}_\mu^a(x)\delta{A}_\nu^{b}(y)}=
\frac{\phi_*}{\rho_0}\frac{\delta^2\left[\int{d^4}z\left(\varepsilon^{3np}b^nc^p\right)_z\;
\cdot\Gamma\right]^{(1)}}{\delta{c}^a(x)\delta{A}_\nu^b(y)}=\phi_*\mathcal{K}_\nu^{ab}(x,y)\;.\label{5b17}
\end{equation}
O lado esquerdo da expressão (\ref{5b17}) representa a polarização
do vácuo enquanto que $\mathcal{K}_\nu^{ab}(x,y)$ representa a
função de Green $1PI$ com inserção do operador composto
$\int{d^4}z\left(\varepsilon^{3np}b^nc^p\right)_z$, com um glúon e
um fantasma amputados externamente.

Devido a invariância de Lorentz, podemos escrever a transformada de
Fourier da identidade de Ward (\ref{5b17}) como\footnote{Lembramos
que $\omega\propto\phi_*$.}
\begin{equation}
p_\mu\Pi_{\mu\nu}^{ab}(p,\omega)=\omega{a}^{ab}(p,\omega)p_\nu\;,\label{5b22}
\end{equation}
onde $\Pi_{\mu\nu}^{ab}(p,\omega)$ é a polarização do vácuo. Esta
identidade de Ward mostra formalmente que a polarização do vácuo
deixa de ser transversa devido a presença do condensado Overhauser,
a um laço. Note ainda que o condensado $\sigma_*$ não quebra a
transversalidade da polarização do vácuo. Veremos mais adiante a
confirmação deste resultado formal quando calcularmos a massa
dinâmica do glúon explicitamente, veja \cite{Capri:2005vw}, a um
laço.

\subsubsection{Poles of the gluon propagator}

Para analizar os pólos do propagador do glúon, reescrevemos a
identidade de Ward (\ref{5b22}) na forma
\begin{equation}
p_\mu\left[\Pi_{\mu\nu}^{ab}(p,\omega)-a^{ab}(p,\omega)\delta_{\mu\nu}\right]=0\;,\label{5b23}
\end{equation}
Utilizando esta relação para o cálculo do fator de forma do glúon a
um laço pode-se deduzir que
\begin{equation}
\frac{1}{9}\left\langle{A}_\mu^a(p)A_\mu^a(-p)\right\rangle_{(1)}=\frac{1}{p^2+m^2+\frac{\Pi^{aa}(p,\omega)+a^{aa}(p,\omega)}{3}}\;,\label{5prop1}
\end{equation}
onde foi utilizada a decomposição
\begin{equation}
\Pi_{\mu\nu}^{ab}(p,\omega)=\left(\delta_{\mu\nu}-\frac{p_\mu{p}_\nu}{p^2}\right)\Pi^{ab}(p,\omega)+a^{ab}(p,\omega)\delta_{\mu\nu}\;.\label{5w4}
\end{equation}
e a soma sobre os índices de cor é assumida. Da expressão
(\ref{5prop1}) concluímos que ambos condensados afetam o polo do
propagador do glúon. Além disso, o polo é afetado pela quebra de
transversalidade da polarização do vácuo. Contudo, essa não é toda a
história. Da expressão dos propagadores fantasmas (\ref{5c2}) vemos
que a um laço, teremos diferentes contribuições nos setores Abeliano
e não Abeliano. Devido a isso, é conveniente decompor a identidade
(\ref{5b23}) da mesma forma. Portanto, escrevemos a polarização do
vácuo como,
\begin{eqnarray}
\Pi_{\mu\nu}^{33}(p,\omega)&=&\left(\delta_{\mu\nu}-\frac{p_\mu{p}_\nu}{p^2}\right)\Pi(p,\omega)+a(p,\omega)\delta_{\mu\nu}\;,\nonumber\\
\Pi_{\mu\nu}^{AB}(p,\omega)&=&\delta^{AB}\left[\left(\delta_{\mu\nu}-\frac{p_\mu{p}_\nu}{p^2}\right)\tilde{\Pi}(p,\omega)+
\tilde{a}(p,\omega)\delta_{\mu\nu} \right]\;,\label{5b23x}
\end{eqnarray}
veja os detalhes em \cite{Capri:2005vw}. Desta forma a componente
Abeliana do propagador do glúon se escreve como
\begin{equation}
\left\langle{A}_\mu^3(p)A_\nu^3(-p)\right\rangle_{(1)}=\frac{1}{p^2+m^2+\Pi(p,\omega)+a(p,\omega)}\left(\delta_{\mu\nu}-\frac{p_\mu{p}_\nu}{p^2}\right)
\;,\label{5prop1a}
\end{equation}
enquanto que as componentes não Abelianas ficam na forma
\begin{equation}
\left\langle{A}_\mu^A(p)A_\mu^A(-p)\right\rangle_{(1)}=
\frac{\delta^{AB}}{p^2+m^2+\tilde{\Pi}(p,\omega)+\tilde{a}(p,\omega)}\left(\delta_{\mu\nu}-\frac{p_\mu{p}_\nu}{p^2}\right)\;,\label{5prop1b}
\end{equation}
Lembrando que, evidentemente,
\begin{equation}
\left\langle{A}_\mu^3(p)A_\nu^A(-p)\right\rangle_{(1)}=0\;,\label{5prop1c}
\end{equation}
É evidente que $\Pi\ne\tilde{\Pi}$. Concluímos assim que a quebra da
transversalidade não só afeta o polo dos propagadores como, devido a
escolha da direção do condensado no espaço de cor, existe uma
diferença nestes pólos, entre os setores Abeliano e não Abeliano,
indicando uma quebra da degenerescência da massa dinâmica do glúon.
Veremos a seguir como esta diferença gera efeitos físicos não
triviais. Note ainda que, como demonstrado formalmente
anteriormente, existe a confirmação de que o propagador é
transverso.

\section{Dynamical mass computation}

Vamos agora apresentar os resultados explícitos através de um
roteiro que facilitará o entendimento do cálculo da contribuição dos
condensados para a massa do glúon. Os detalhes deste cálculo pode
ser encontrado em \cite{Capri:2005vw}.

É importante ter em mente que a definição de massa aqui considerada
é a massa efetiva obtida através da polarização do vácuo a momento
nulo. Esta escolha é feita pois a massa como polo do propagador
exige o conhecimento da polarização do vácuo a momento $p$, que de
fato, é muito difícil de se obter no presente caso, veja
\cite{Capri:2005vw} para uma discusão mais detalhada. Lembramos que
o estudo da massa como polo do propagador foi feito
\cite{Browne:2004mk,Gracey:2004bk}, com o método LCO, contudo, este
estudo está além da ambição desta tese.

\subsection{Contributions}

Para a contribuição de um laço para a massa teremos que calcular a
contribuição dos campos de Faddeev-Popov,
$\left[\Pi_{\mu\nu}^{ab}(0)\right]_{\mathrm{gh}}$, bem como a
contribuição do laço de gluons,
$\left[\Pi_{\mu\nu}^{ab}(0,\omega)\right]_{\mathrm{gl}}$. Neste
último, temos ainda a contribuição das flutuações do condensado
$\sigma_*$, através do vértice $\widetilde{\sigma}A^2$. Note que
estes diagramas deixam de ser nulos no caso em que o propagador do
glúon é massivo.

Comecemos pela componente de gluons. Tal cálculo pode ser encontrado
em \cite{Browne:2004mk,Gracey:2004bk}, para o caso geral $N$.
Adaptando este resultao para o caso $N=2$ e tomando o limite
$p^2=0$, chegamos a
\begin{equation}
\left[\Pi_{\mu\nu}^{ab}(0)\right]_{\mathrm{gl}}=\frac{g^2}{16\pi^2}m^2\left(-\frac{7}{48}+\frac{17}{8}\ln\frac{m^2}{\omu^2}\right)\delta^{ab}\delta
_{\mu\nu}\;.\label{5pvbis}
\end{equation}

No caso da contribuição da componente fantasma, vimos que a
decomposição Abeliana é necessária. De acordo com
\cite{Capri:2005vw}, o setor não Abeliano corresponde a
\begin{equation}
\left[\Pi_{\rho\rho}^{AB}(0)\right]_{\mathrm{gh}}=-g^2\varepsilon^{Amn}\varepsilon^{Bpq}\int\frac{d^dk}{\left(2\pi\right)^d}k^2\left\langle
\bar{c}^mc^q\right\rangle_k\left\langle\bar{c}^pc^n\right\rangle_k=2g^2\delta^{AB}\int\frac{d^dk}{\left(2\pi\right)^d}\frac{k^4}{k^2\left(k^4+\omega^2
\right)}\;,\label{5v8}
\end{equation}
o que resulta em
\begin{eqnarray}
\left[\Pi_{\rho\rho}^{AB}(0)\right]_{\mathrm{gh}}=-\delta^{AB}\frac{\omega{g}^2}{16\pi}\;.\label{5v9}
\end{eqnarray}

Finalmente, o setor Abeliano, é obtido através de, veja
\cite{Capri:2005vw},
\begin{equation}
\left[\Pi_{\rho\rho}^{33}(0)\right]_{\mathrm{gh}}=-g^2\int\frac{d^dk}{\left(2\pi\right)^d}\frac{k^2}{\left(k^4+
\omega^2\right)^2}\left(-2k^4+2\omega^2\right)\;.
\end{equation}
De onde é fácil deduzir que
\begin{equation}
\left[\Pi_{\rho\rho}^{33}(0)\right]_{\mathrm{gh}}-\frac{g^2\omega}{8\pi}\;.\label{5v6}
\end{equation}

Chamamos a atenção para o fato de
$\left[\Pi_{\rho\rho}^{33}(0)\right]_{\mathrm{gh}}\neq\left[\Pi_{\rho\rho}^{AB}(0)\right]_{\mathrm{gh}}$,
confirmando o argumento apresentado anterioprmente que o condensado
$\left\langle\varepsilon^{abc}\bar{c}^bc^c\right\rangle$ quebra a
degenerescência da massa do glúon. Note ainda que, este resultado
está de acordo com o resultado obtido no calibre de Curci-Ferrari
\cite{Sawayanagi:2003dc}, onde a componente Abeliana é duas vezes
maior que a componente não Abeliana.

\subsection{Results and interpretations}

Como oservado em \cite{Dudal:2002xe,Sawayanagi:2003dc}, a
contribuição dos condensados fantasmas para a massa do glúon é
negativa. Desta forma, se levarmos em consideração apenas os
condensados fantasmas, teremos, a um laço, uma massa dinâmica
taquiônica para o glúon. Contudo, a contribuição devido ao
condensado $A_\mu^aA_\mu^a$ é positiva. Estes efeitos, de fato, se
compensam e resutam em massas positivas,
\begin{eqnarray}
m_{\mathrm{Ab}}^2&=&m^2+\delta{m}^2-\frac{g^2\omega}{32\pi}\;,\nonumber\\
m_{\mathrm{nAb}}^2&=&m^2+\delta{m}^2-\frac{g^2\omega}{64\pi}\;,\label{5f1}
\end{eqnarray}
onde $m_{\mathrm{Ab}}$ e $m_{\mathrm{Ab}}$ representam,
respectivamente, as massas de gluons Abelianos e nào Abelianos. A
quantidade $\delta{m}^2$ representa a contribuição proveniente de
diagramas de gluons, extraído da expressão (\ref{5pvbis}),
\begin{eqnarray}
\delta{m}^2=\frac{g^2}{16\pi^2}m^2\left(-\frac{7}{48}+\frac{17}{8}\ln\frac{m^2}{\omu^2}\right)\;.\label{5deltam}
\end{eqnarray}
Observe que independentemente da escala escolhida temos a relação
\begin{equation}
m_{\mathrm{nAb}}^2>m_{\mathrm{Ab}}^2\;,\label{5f1x}
\end{equation}

Como discutido no cálculo da energia do vácuo, teremos que lidar com
o problema da existência de duas escalas de massa. O mesmo
tratamento pode ser feito para o presente cálculo, de forma que os
valores explítos são calculados através de
\begin{eqnarray}
m_{\mathrm{Ab}}^2&=&\overline{m}^2+\frac{\overline{g}^2}{16\pi^2}\overline{m}^2\left(-\frac{7}{48}\right)-\frac{\widetilde{g}^2\widetilde{\omega}}{32\pi}
\;,\nonumber\\
m_{\mathrm{nAb}}^2&=&\overline{m}^2+\frac{\overline{g}^2}{16\pi^2}\overline{m}^2\left(-\frac{7}{48}\right)-\frac{\widetilde{g}^2\widetilde{\omega}}{64\pi}
\;,\label{5resultaten}
\end{eqnarray}

Substituindo os valores (\ref{5ll16}) em (\ref{5resultaten}),
chegamos a
\begin{eqnarray}
m_{\mathrm{Ab}}^2&\approx&1.66\lms^2\;,\nonumber\\
m_{\mathrm{nAb}}^2&\approx&2.34\lms^2\;,\label{5f1bis}
\end{eqnarray}

A relação (\ref{5f1x}) é, usualmente, interpretada como uma
evidência da dominância Abeliana. Por exemplo, no caso do MAG, este
efeito é observado,
\cite{Amemiya:1998jz,Bornyakov:2003ee,Kondo:2000ey,Dudal:2004rx,Capri:2005tj,Capri:2006vv,Dudal:2006ib,Capri:2006cz}.
Analogamente, podemos interpretar a relação (\ref{5f1x}) juntamente
com os valores numéricos consistentes com o grupo de renormalização
(\ref{5f1bis}) como uma indicação da dominância Abeliana no calibre
de Landau\footnote{Infelizmente, até nosso conhecimento, um valor
numérico explícito para $\lms^{N=2;N_f=0}$ está ainda indisponível
na literatura. Isso impossibilita uma estimativa exata das massas
(\ref{5f1bis}).}. Ressaltamos que no caso do MAG, a decomposição nos
setores Abeliano e não Abeliano é efetuada diretamente na fixação de
calibre. No caso do calibre de Landau, não existe esta quebra do
grupo \emph{a priori}, o torna o efeito da quebra de degenerescência
um resultado extremamente forte.

\part{GRIBOV AMBIGUITIES}

\chapter{Preliminary notions}

O trabalho original de Gribov, \cite{Gribov:1977wm}, chama a atenção
para o fato de que o processo de quantização das teorias de
Yang-Mills falha num determinado momento, veja
\cite{Sobreiro:2005ec} para uma introdução pedagógica sobre o
assunto. Gribov mostrou que o método de Faddeev-Popov,
\cite{Faddeev:1967fc}, funciona muito bem para tratar problemas
quânticos perturbativos. Contudo, ao ir para o regime
não-perturbativo, o método de Faddeev-Popov resulta ser incompleto.
De fato, o problema de Gribov é caracterizado pelo fato de que o
procedimento de fixação de calibre para eliminar os graus de
liberadade não físicos da teoria não é suficiente para se obter uma
quantização consistente. De fato, ao sair do regime perturbativo
detecta-se ainda uma simetria de calibre residual identificada
através das chamadas cópias de Gribov.

Gribov, ainda em seu trabalho original \cite{Gribov:1977wm},
aperfeiçoou o método de quantização de Faddeev-Popov
\cite{Faddeev:1967fc} para o caso específico dos calibres de Landau
e Coulomb. Sua solução consiste em restringir o domínio de
integração do espaço funcional dos campos de calibre para uma região
onde exista um número menor de cópias de Gribov, a chamada primeira
região de Gribov ou simplesmente região de Gribov. O efeito físico
desta restrição se reflete nos propagadores da teoria. O propagador
do glúon mostra-se suprimido na região de baixas energias enquanto o
propagador dos campos de Faddeev-Popov se torna mais singular, nesta
região de energia, quando comparada à predição perturbativa. Esta
singularidade caracteriza a existência de forças de longo alcance
quando no regime de baixas energias, imprencidíveis para o fenômeno
do confinamento da cor. Ainda, os pólos do propagador do glúon
deixam de ser reais, quando se leva em conta a restrição de Gribov,
evidenciando o caráter não observacional dos campos de calibre,
eliminando-os do espectro físico da teoria, como deve ser para
objetos confinados.

Com isso Gribov mostrou que o problema do confinamento está
diretamente relacionado com o processo de quantização das teorias de
Yang-Mills, ou seja, as cópias de Gribov estão presentes exatamente
na região de baixas energias, onde o regime é caracterizado pelo
acoplamento forte e a teoria de perturbação não é mais válida. Desta
forma, uma compreensão mais apurada dos aspectos da quantização das
teorias de Yang-Mills são uma necessidade para a compreensão do
confinamento.

Apesar de o tratamento de Gribov ter sido feito essencialmente nos
calibres de Landau e Coulomb, a existência das ambigüidades de
Gribov é uma patologia presente em qualquer calibre e qualquer
teoria de calibre topologicamente não trivial. De fato, Singer
mostrou formalmente, \cite{Singer:1978dk}, que as ambigüidades de
Gribov ocorrem devido ao fato de as teorias de Yang-Mills serem não
triviais sob o ponto de vista topológico\footnote{Veja
\cite{Bertlmann,Nakahara,Nash} para a explicação da geometria por
trás das teorias de Yang-Mills.}. Ou seja, fixar o calibre não é um
passo possível globalmente, no sentido de englobar todo o espaço
funcional, sendo possível apenas localmente. Matematicamente, para
se quantizar as teorias de Yang-Mills, devemos definir uma seção no
fibrado principal. Isto só é possível se o fibrado principal for
trivial, o que não é verdade no caso das teorias de Yang-Mills.

Após a descoberta das ambigüidades de Gribov, uma série de trabalhos
\cite{Semenov,Zwanziger:1981nu,Zwanziger:1982na,Zwanziger:1988jt,Dell'Antonio:1989jn,
Zwanziger:1989mf,Zwanziger:1990tn,Dell'Antonio:1991xt,Zwanziger:1992qr,Zwanziger:2002sh}
fez com que um grande avanço no entendimento da região de Gribov
fosse desenvolvido. A região de Gribov, no calibre de Landau, ainda
possui cópias dentro dela
\cite{Semenov,Dell'Antonio:1989jn,Dell'Antonio:1991xt,vanBaal:1991zw}.
Mas a restrição não é inconsistente pois, de fato, toda órbita de
calibre pasa pelo menos uma vez na região de Gribov. Esta importante
propriedade, demonstrada em
\cite{Zwanziger:1981nu,Zwanziger:1982na,Zwanziger:1988jt}, diz que,
mesmo que a restrição à primeira região de Gribov não elimine todas
as cópias, uma grande parte delas é eliminada, tudo fora da região
de Gribov são cópias de Gribov, portanto, configurações de campo não
físicas. Descobriu-se também a existência de uma região que seria
mais fundamental que a região de Gribov, a chamada região modular
fundamental (RMF),
\cite{Semenov,Dell'Antonio:1989jn,Dell'Antonio:1991xt,vanBaal:1991zw}.
Esta região seria então livre de cópias, exceto por sua borda. As
cópias da borda da RMF seriam então identificadas e a topologia
desta nova região seria por demais complexa, mas, enfim, uma região
livre de cópias. Contudo, trabalhar com essa região é algo não
trivial, e sua implementação à teoria de modo eficaz ainda é um
problema a ser resolvido. Uma tentativa de implementar a restrição a
RMF foi feita, com relativo sucesso, em \cite{vanBaal:1991zw}, na
formulação Hamiltoniana.

Em particular, o estudo da primeira região de Gribov permitiu que a
restrição a esta região fosse implementada na integral de caminho de
forma local e renormalizável, no calibre de Landau/Coulomb,
\cite{Zwanziger:1989mf,Zwanziger:1992qr,Maggiore:1993wq}. Isto
significa que, no calibre de Landau, existe uma Lagrangiana que
descreve as teorias de Yang-Mills onde a integral de caminho possui
domínio de integração igual a região de Gribov, eliminando,
portanto, muitas das cópias de Gribov. Esta Lagrangiana possui as
propriedades básicas essenciais para se efetuar cálculos, isto é,
local, renormalizável e obedece ao grupo de renormalização. Contudo,
tentativas de cálculos reais,
\cite{Dudal:2005na,Gracey:2005cx,Gracey:2006dr} a um e dois laços
não se mostraram tão simples assim, reforçando a idéia de que o
problema de Gribov e sua solução são de caráter essencialmente não
perturbativo. Contudo, o caráter qualitativo dos resultados são
consistentes com os cálculos numéricos feitos na rede
\cite{Cucchieri:1999sz,Bonnet:2001uh,Langfeld:2001cz,Cucchieri:2003di,Bloch:2003sk,Furui:2003jr,Furui:2004cx}
bem como cálculos utilizando as equações de Scwhinger-Dyson
\cite{Fischer:2006ub}.

Ainda no calibre de Landau, foi estudada a compatibilidade da ação
de Gribov-Zwanziger com a geração dinâmica de massa,
\cite{Sobreiro:2004us,Sobreiro:2004yj,Dudal:2005na}. Nestes
trabalhos, os resultados são compatíveis com as predições da rede
\cite{Cucchieri:1999sz,Bonnet:2001uh,Langfeld:2001cz,Cucchieri:2003di,Bloch:2003sk,Furui:2003jr,Furui:2004cx}
e das equações de Schwinger-Dyson \cite{Fischer:2006ub}. Ou seja, o
propagador do glúon permanece suprimido na região infravermelha,
sendo esta supressão ainda mais forte quando se leva em conta a
geração dinâmica de massa. Ao mesmo tempo, o propagador do ghost
continua mais singular que a predição perturbativa.

É importante ressaltar que pouco foi feito com relação às
ambigüidades de Gribov em outros calibres que não os calibres de
Landau e Coulomb. Recentemente, estudos destes problemas foram
feitos nos calibres lineares covariantes \cite{Sobreiro:2005vn}
encontrando supressão do progador do glúon transverso devido ao
horizonte de Gribov e supressão do propagador longitudinal do glúon
devido à geração dinâmica de massa. Ainda, uma quantidade associada
a restrição do domínio de integração se mostra mais singular que a
predição perturbativa. Esta quantidade possui o mesmo papel que o
propagador do ghost no calibre de Landau. Ainda, este resultado,
obtido analiticamente, está em acordo qualitativo com a predição
numérica calculada utilizando-se os métodos da QCD na rede,
\cite{Giusti:1996kf,Giusti:1999im,Giusti:2000yc}. Neste capítulo
dedicaremos uma seção a este tópico, devido ao fato de os calibres
lineares covariantes serem a generalição linear natural do calibre
de Landau.

Outro calibre explorado com relação às ambigüidades de Gribov é o
calibre máximo Abeliano, o chamado MAG \cite{'tHooft:1981ht}, no
caso de $SU(2)$. Em \cite{Quandt:1997rg,Bruckmann:2000xd} as
propriedades básicas das cópias de Gribov no MAG são discutidas em
detalhe. Em \cite{Capri:2005tj}, seguindo a pres-crissão do trabalho
original de Gribov, \cite{Gribov:1977wm}, a influência do horizonte
nos propagadores dos campos fundamentais das teorias de Yang-Mills
são discutidos no nível árvore, encontrando resultados em completo
acordo qualitativo com as predições da rede, vide
\cite{Kronfeld:1987vd,Kronfeld:1987ri,Amemiya:1998jz,Bornyakov:2003ee}.
Em particular, o propagador do glúon diagonal apresenta uma
supressão na região infravermelha devido ao horizonte de Gribov
enquanto os propagadores não diagonais são suprimidos devido apenas
à geração dinâmica de massa.

Os resultados expostos até agora neste capítulo são todos obtidos a
temperatura zero. Somente muito recentemente, o horizonte de Gribov
e sua influência nas teorias de Yang-Mills foram estudados
analiticamente à temperatura finita. Como resultado, obtém-se um
equação de estado para o plasma de quarks e gluons
\cite{Zwanziger:2004np,Zwanziger:2005nk}. Este resultado não só está
de acordo com os resultados anteriores
\cite{Linde:1980ts,Gross:1980br,Feynman:1981ss}, como os
aperfeiçoam.

É importante ter em mente que um resultado final ainda não está
estabelecido. Até o presente momento, não sabemos como se quantizar
as teorias de Yang-Mills de maneira completa e eficiente. Da mesma
forma, e diretamente relacionado, está o fato de que um método não
perturbativo completo não está ainda estabelecido.

Vamos rever a problemática de Gribov pelo resto deste capítulo.
Discutiremos os principais pontos relativos às ambigüidades na
quantização de Faddeev-Popov \cite{Faddeev:1967fc} e como esse
problema se reflete na teoria. Revisaremos a solução proposta por
Gribov \cite{Gribov:1977wm,Sobreiro:2005ec} e em seguida o
aperfeiçoamento desta desenvolvido por Zwanziger, obtendo uma
formulação local e renormalizável das teorias de Yang-Mills livre de
um grande número de cópias de calibre. Como na parte anterior vamos
nos ater ao calibre de Landau a temperatura nula.

\section{Gribov problem}

\subsection{Gribov copies}

A quantização das teorias de Yang-Mills pode ser efetuada, por
exemplo, através do método de Faddeev-Popov \cite{Faddeev:1967fc}.
Neste procedimento a integral de caminho apropriada para descrever
quanticamente a teoria, ou seja, com o vínculo garantindo a
imposição do calibre de Landau,
\begin{equation}
\partial_\mu{A}_\mu^a=0\;,\label{6land}
\end{equation}
é
\begin{equation}
Z=\int{DA}\delta(\partial_\mu{A}_\mu^a)\det(\mathcal{M}^{ab})e^{-S_{YM}}\;,\label{6fadpop1}
\end{equation}
onde $\mathcal{M}^{ab}$ é o operador de Faddeev-Popov, dado por,
\begin{equation}
\mathcal{M}^{ab}=-\partial_\mu{D}_\mu^{ab}\;.\label{6fadpopop}
\end{equation}
Gribov mostrou, em \cite{Gribov:1977wm}, que a condição
(\ref{6land}) não fixa o calibre univocamente. Ou seja, para uma
dada configuração $A_\mu$ obedecendo ao calibre de Landau, existe
uma configuração equivalente $\tilde{A}_\mu$, que também obedece ao
calibre de Landau, ou seja,
\begin{eqnarray}
\tilde{A}_\mu&=&A_\mu+U^\dagger{D}_\mu{U}\;,\nonumber\\
\partial_\mu\tilde{A}_\mu&=&\partial_{\mu}A_\mu\;=\;0\;.\label{6copias1}
\end{eqnarray}
A configuração $\tilde{A}_\mu$ é chamada de {\it cópia de Gribov}
associada ao campo $A_\mu$. Substituindo a primeira das
(\ref{6copias1}) na segunda, chegamos à relação
\begin{equation}
\partial_\mu\left(U^\dagger{D}_\mu{U}\right)=0\;.\label{6copias2}
\end{equation}
A equação (\ref{6copias2}) é a {\it equação das cópias}, ou seja, as
soluções desta equação para os elementos do grupo $U$ definem a
existência de cópias para uma dada configuração $A_\mu$. Exemplos de
cópias de Gribov podem ser encontradas, por exemplo, no trabalho
original de Gribov \cite{Gribov:1977wm} e em \cite{Henyey:1978qd}.

A equação (\ref{6copias2}) define todas as possíveis cópias de
Gribov associadas à configuração $A_\mu$. No caso de cópias muito
próximas, nos restringimos a transformações infinitesimais. Neste
caso, é trivial mostrar que, à primeira ordem, a equação das cópias
(\ref{6copias1}) se resume a
\begin{equation}
\mathcal{M}^{ab}\omega^b=0\;,\label{6copias3}
\end{equation}
onde $\mathcal{M}^{ab}$ é o operador de Faddeev-Popov
(\ref{6fadpopop}) e $\omega^a$ o parâmetro da transformação de
calibre\footnote{Veja o apêndice \ref{ap_nota} para as notações e
convensões utilizadas.}.

\subsection{Gribov region and Gribov horizon}

\begin{figure}[ht]
\centering \epsfig{file=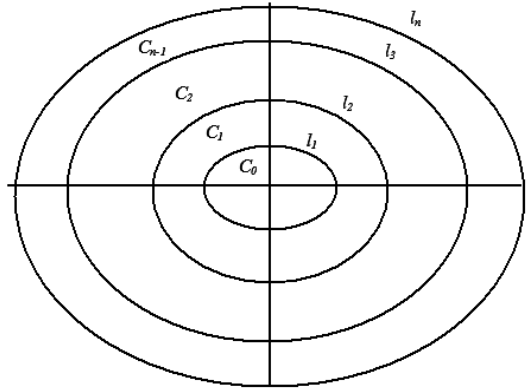,width=8cm} \caption{Regiões e
horizontes de Gribov.}\label{6fig1}
\end{figure}

A equação de cópias infinitesimais (\ref{6copias3}) fornece
importantes informações sobre como as cópias de Gribov se arrumam no
espaço funcional dos campos de calibre. Primeiramente, vemos que as
cópias infinitesimais aparecem onde o operador de Faddeev-Popov
possui autovalores nulos. Vamos então escrever a equação de
autovalores para o operador de Faddeev-Popov
\begin{equation}
\mathcal{M}^{ab}\omega^b=\epsilon\omega^a\;,\label{6eigen1}
\end{equation}
onde os autovalores dependem da configuração do campo de calibre
$\epsilon=\epsilon(A)$. Esta equação nos permite definir os chamados
horizontes de Gribov. Para tal, entendemos esta equação como um tipo
de equação de Schrödinger, com $A_\mu^a$ fazendo o papel do
potencial. Ainda, é fácil ver que o operador de Faddeev-Popov, no
calibre de Landau, é um operador hermitiano\footnote{Note que as
equações de cópias (\ref{6copias2}) e (\ref{6copias3}) são válidas
também nos calibres lineares covariantes. Contudo, o fato de os
campos não serem transversos destrói a hermiticidade do operador de
Faddeev-Popov, e, consequentemente, as propriedades que se seguem
neste capítulo.}, possuindo assim autovalores reais. Para pequenos
valores de $A_\mu^a$ o termo cinético $-\partial^2$, em
(\ref{6eigen1}), é dominante, onde $k$ representa o momento. Assim,
denotando o conjunto de autovalores de uma dada configuração
$A_\mu^a$ por
$\left\{\epsilon_1(A),\epsilon_2(A),\epsilon_3(A),\ldots\right\}$,
temos que, para $A_\mu^a$ pequeno, todos autovalores $\epsilon_i(A)$
são positivos, {\it i.e.} $\epsilon_i(A)>0$. Contudo, conforme o
campo $A_\mu^a$ cresce em magnitude, um dos autovalores, digamos
$\epsilon_1(A)$, se torna nulo. E, conforme a magnitude de $A_\mu^a$
aumenta mais, este autovalor se torna negativo. Para um valor ainda
maior de $A_\mu^a$, um segundo autovalor, $\epsilon_2(A)$, se anula,
se tornando negativo conforme a configuração de calibre aumenta
ainda mais, e assim por diante. Analisando este comportamento,
dividimos o domínio do espaço funcional dos campos de calibre em
regiões $\left\{C_{0},C_{1},C_{2},\ldots,C_{n}\right\}$ nas quais o
operador de Faddeev-Popov, (\ref{6fadpopop}), possui
$\left\{0,1,2,\ldots,n\right\}$ autovalores negativos, conforme
representado na Fig. \ref{6fig1}. As regiões $C_i$ são as chamadas
regiões de Gribov. As curvas que separam tais regiões, $l_i$, são os
horizontes de Gribov, definidos como as curvas sobre as quais o
operador de Faddeev-Popov possui autovalores nulos.

A primeira região de Gribov, $C_0$, será chamada simplesmente de
{\it região de Gribov}, enquanto que o primeiro horizonte, $l_0$,
será chamado de {\it horizonte de Gribov}. Ainda, denotaremos o
horizonte por $l_0\equiv\partial{C}_0$.

Note que a configuração $A_\mu^a=0$ é o ponto mais ``{\it distante}"
do horizonte de Gribov, pois induz o maior autovalor do operador de
Faddeev-Popov. Desta forma, podemos concluir que o vácuo
perturbativo reside no centro da região de Gribov.

\subsection{Properties of the Gribov region}

A região de Gribov possui, estritamente, autovalores positivos do
operador de Faddeev-Popov. Esta propriedade nos permite definir a
região de Gribov como
\begin{equation}
C_0\equiv\left\{A_\mu^a\;\big|\;\partial_\mu{A}_\mu^a=0,\;\mathcal{M}^{ab}>0\right\}\;.\label{6reg1}
\end{equation}
Próximo ao horizonte, os autovalores se aproximam de zero e as
cópias, cada vez mais, são próximas umas das outras. Ou seja, as
cópias infinitesimais estão próximas ao horizonte.

A região $C_0$ possui algumas importantes propriedades, que, podem
ser usadas em favor de eliminar as cópias de Gribov. Vamos enunciar
tais propriedades. A demonstração destas propriedades pode ser
encontrada na literatura de referência especificada em cada uma das
propriedades, a seguir.
\begin{itemize}
\item Propriedade 1: {\it Para qualquer campo de calibre localizado na região $C_i$, próximo ao
horizonte $\partial{C}_i$, existe uma configuração equivalente na
região $C_{i+1}$, também próximo ao horizonte $\partial{C}_i$.}

A demonstração desta propriedade, feita em \cite{Gribov:1977wm}, é
desenvolvida mostrando que tais cópias possuem autovalores iguais
com sinais opostos. Ou seja, estão em lados diferentes do horizonte.
Esta propriedade sugere a restrição do domínio do espaço funcional
de $A_\mu^a$ à região de Gribov. De fato, isso foi sugerido pelo
próprio Gribov, \cite{Gribov:1977wm}. Contudo, antes de mostrarmos
os efeitos da restrição, vamos desenvolver outras propriedades que
generalizam e aperfeiçoam a propriedade de Gribov. Veja Fig.
\ref{6fig2a}.
\begin{figure}[ht]
\centering \epsfig{file=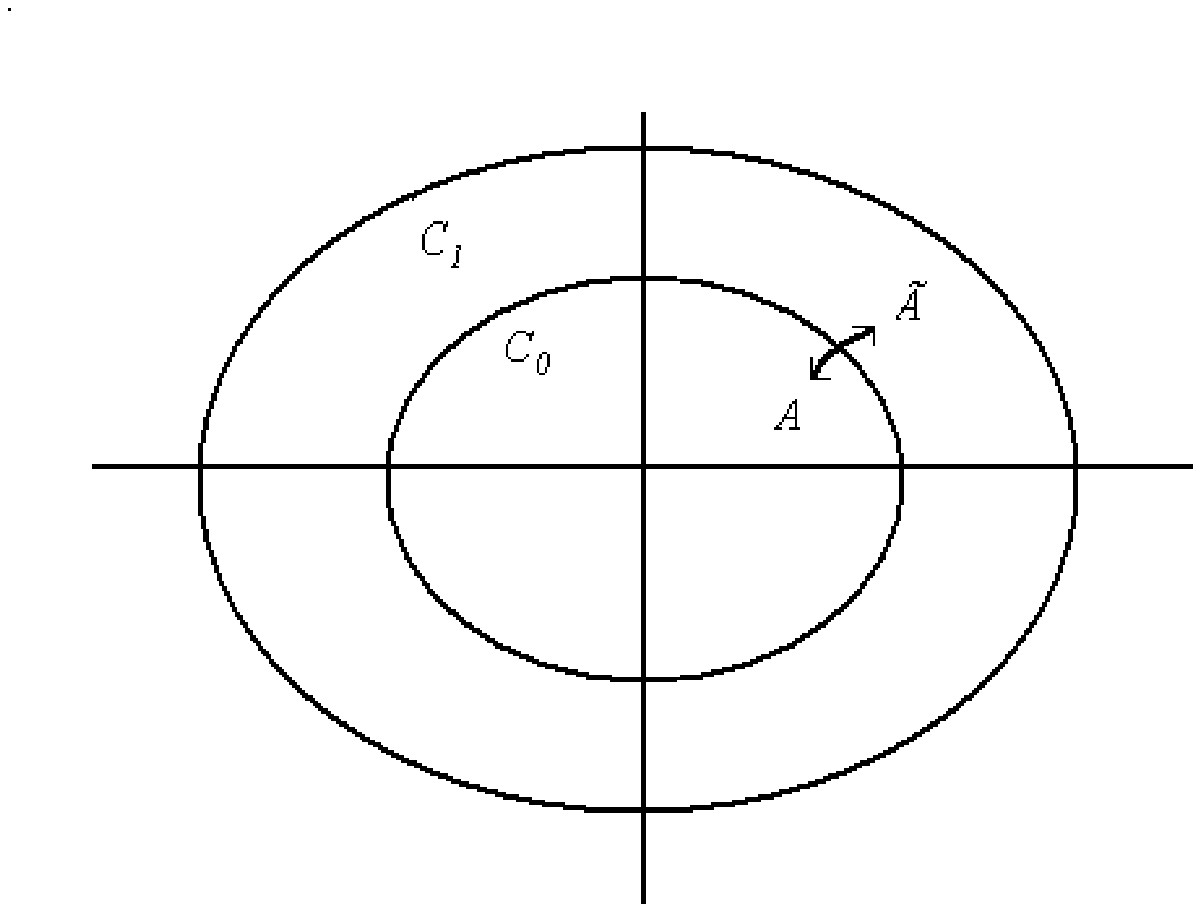,width=8cm} \caption{Cópias de
Gribov próximas aos horizontes.}\label{6fig2a}
\end{figure}

\item Propriedade 2: {\it A região $C_0$ é convexa.}

Esta propriedade, demonstrada em \cite{Zwanziger:1982na} utilizando
o método de campo de fundo, nos diz que se um campo pode ser escrito
como a interpolação de dois outros campos que pertençam a região de
Gribov, então este campo também pertence a região de Gribov. Ou
seja, se $\{A_1,A_2\}$ são campos pertencentes a região $C_0$,
$\{A_1,A_2\}\in{C}_0$, e se o campo $A$ pode ser escrito como
$A=\alpha{A}_1+(1-\alpha)A_2$, onde o parâmetro de interpolação é
dado por $0\le\alpha\le1$, então $A$ também habita a região $C_0$,
$A\in{C}_0$.

\item Propriedade 3: {\it A região $C_0$ possui fronteira em todas as direções.}

A demostração desta propriedade está feita em
\cite{Zwanziger:1982na}. Esta simples propriedade nos diz que sempre
podemos afirmar se uma dada configuração finita está dentro ou fora
da região $C_0$. Ou seja, a região $C_0$ é finita e bem definida em
todas as direções, de forma que sempre podemos atingir sua fronteira
(o horizonte de Gribov) com um limite apropriado. Outra forma de
entendermos esta propriedade é dizendo que toda configuração
existente dentro da região de Gribov está a uma ``{\it distância}''
finita do vácuo perturbativo $A_\mu^a=0$, que nada mais é que o
centro da região de Gribov.

\item Propriedade 4: {\it Toda órbita de calibre passa pela região
$C_0$.}

Esta propriedade se encontra demonstrada em
\cite{Zwanziger:1981nu,Zwanziger:1982na}. Por órbita de calibre
entendemos a quantidade definida na segunda das (\ref{3Amin0}), ou
seja, a variação do campo $A_\mu$ ao longo dos elementos $U$ do
grupo. Portanto, para uma configuração geral $A_\mu$, não
pertencente à $C_0$, existe uma configuração equivalente dentro de
$C_0$. Assim, o resultado de Gribov (Propriedade 1) para cópias
próximas se generaliza dizendo que todas as configurações que não
pertencem à região $C_0$ são cópias de Gribov de alguma configuração
que existe dentro da região $C_0$.

\item Propriedade 5: {\it A região $C_0$ está contida dentro do Elipsoide de
Landau, $\mathcal{E}_L$, $C_0\supseteq{\mathcal{E}_L}$.}

Este resultado, discutido em
\cite{Zwanziger:1988jt,Dell'Antonio:1989jn,Zwanziger:1989mf}, pode
ser entendido como uma apri-moração da propriedade 3, estabelecendo
um limite geométrico para a região de Gribov. O chamado elipsóide de
Landau é definido pelo hiperelipsóide
\begin{equation}
\int\frac{d^4q}{(2\pi)^2}\frac{a_\mu^a(q)a_\mu^{a\dagger}(-q)}{q^2}=\mathcal{C}\;,\label{6elip1}
\end{equation}
onde $\mathcal{C}$ é uma constante dependente da dimensão do grupo
de calibre. As quantidades $a_\mu^{a\dagger}$ e $a_\mu^a$ são
coeficientes de expansão de $A_\mu^a$. Este resultado é extremamente
importante pois, tomando-se o valor esperado da equação
(\ref{6elip1}) encontra-se que o propagador do glúon
$D_{\mu\nu}^{ab}(q)$ obedece a relação
\begin{equation}
\int\frac{d^4q}{(2\pi)^2}\frac{D_{\mu\mu}^{aa}(q)}{q^2}\le\frac{\mathcal{C}}{4}\;.\label{6elip2}
\end{equation}
O resultado (\ref{6elip2}) implica numa contradição com relação a
predição perturbativa do grupo de renormalização
\cite{Dell'Antonio:1989jn}.
\end{itemize}

Apesar de estas propriedades mostrarem a grande importância da
região de Gribov, esta ainda não é livre de cópias
\cite{Semenov,Zwanziger:1990tn,Dell'Antonio:1989jn,Dell'Antonio:1991xt,vanBaal:1991zw,
Gubarev:2000nz,Gubarev:2000eu}, pois não se sabe quantas vezes cada
órbita cruza a fronteira de $C_0$. De fato estas cópias existem e
estão associadas a cópias sem carga topológica \cite{Henyey:1978qd}.
Contudo, ainda podemos definir uma região mais refinada, chamada
região modular fundamental. Apesar de sua implementação na teoria
ser altamente não trivial e não ter sido feita na integral de
caminho até hoje, esta região seria livre de cópias. Vamos
desenvolver uma breve discussão sobre a região modular fundamental
na próxima seção. Contudo, vamos nos ater a região de Gribov pelo
resto desta tese.

\subsection{Fundamental modular region and the functional $A_{\min}^2$}

É muito simples entender porque existe a possibilidade de ainda
haver cópias dentro da região de Gribov. De acordo com a propriedade
4 discutida anteriormente, cada órbita de calibre passa pela região
de Gribov pelo menos uma vez, mas nada garante que seja uma única
vez. Outra forma de vermos isso é através da definição da região de
Gribov $C_0$ em (\ref{6reg1}). Pensemos novamente no funcional
$A_{\min}^2$ descrito em (\ref{3Amin0}). Podemos definir a região de
Gribov $C_0$ como o conjunto de campos $A_\mu^a$ que estabilizam o
funcional $A_{\min}^2$. Impondo a nulidade da variação primeira em
(\ref{3Amin0}) selecionamos as configurações que obedecem ao calibre
de Landau (\ref{6land}), estas são as configurações que tornam o
funcional $A_{\min}^2$ estacionário. Requerendo agora que a variação
segunda de (\ref{3Amin0}) seja positiva, encontrando assim a equação
de cópias infinitesimais (\ref{6copias3}), ou seja, as configurações
que impõem a estabilidade do funcional $A_{\min}^2$. Contudo, este
procedimento encontra todas as configurações que tornam o funcional
$A_{\min}^2$ estável, ou seja, todos os mínimos, absolutos ou não.
Obviamente, os mínimos de uma mesma órbita são cópias de Gribov e
devem ser filtrados de maneira mais refinada. Assim, devemos
selecionar os mínimos absolutos de cada órbita para eliminar as
cópias de Gribov. Definimos então, a chamada \emph{região modular
fundamental} (RMF), denotada por $\Lambda$, como o conjunto de
configurações que tornam o funcional $A_{\min}^2$ um mínimo
absoluto,
\begin{equation}
\Lambda\equiv\left\{\mathcal{A}_\mu^a\;\big|\;\partial_\mu\mathcal{A}_\mu^a=0,\;
\mathcal{M}^{ab}(\mathcal{A})>0\Leftrightarrow{\mathcal{A}_{\min}^2}<A_{\min}^2\;\forall\;A\ne\mathcal{A}\right\}
\;.\label{6reg2}
\end{equation}
Apesar de mais requintada, a RMF pode ainda posuir cópias de Gribov.
Essas cópias são caracterizadas por diferentes configurações
existindo na mesma órbita e gerando o mesmo mínimo absoluto para
$A_{\min}^2$. De fato essas cópias existem e vivem na fronteira de
$\Lambda$ e seriam eliminadas através da identificação topológica
dessas cópias. Esse procedimento torna a topologia de $\Lambda$
altamente não trivial, dificultando ainda mais sua operacionalidade.

Vale ressaltar que, recentemente, foi mostrado formalmente em
\cite{Zwanziger:2003cf}, que as cópias dentro da primeira região de
Gribov não afetam os valores esperados da teoria. A prova é feita
argumentando que a região de Gribov e a RMF possuem uma fronteira em
comum, e a maior contribuição para a integral de caminho viria desta
fronteira. Contudo, o fato de a RMF ser de difícil acesso,
impossibilita cálculos explícitos que comprovem este argumento.
Mesmo assim, esse resultado torna muito promissor e estimulante o
estudo da região de Gribov no lugar da RMF, e, de fato, é isso que
faremos ao longo desta tese.

\section{Restriction to the Gribov region}

Fazendo uso das propriedades discutidas na seção anterior, em
particular a quarta propriedade e o fato de que existem indícios de
que as cópias dentro da região de Gribov não afetam os valores
esperados da teoria \cite{Zwanziger:2003cf}, podemos restringir o
domínio de integração do espaço funcional dos campos de calibre à
primeira região de Gribov, eliminando assim, um grande número de
cópias. A integral de caminho (\ref{6fadpop1}) deve, então, ser
modificada para
\begin{equation}
Z=\int{DA}\delta(\partial_\mu{A}_\mu^a)\det(\mathcal{M}^{ab})e^{-S_{YM}}\mathcal{V}(C_0)\;,\label{6fadpop2}
\end{equation}
onde o funcional $\mathcal{V}(C_0)$ garante que a integração será
feita apenas na região de Gribov. A determinação do funcional
$\mathcal{V}(C_0)$ pode ser feita com o auxílio do propagador dos
campos de Faddev-Popov. Antes de determinar $\mathcal{V}(C_0)$ vamos
ver o que o este funcional e o propagador fantasma nos dizem sobre o
vácuo da QCD. Após esta análise discutiremos brevemente como Gribov
fez a restrição considerando o propagador dos campos de
Faddeev-Popov \cite{Gribov:1977wm}. Depois exporemos o método de
Zwanziger\footnote{Os dois prcedimentos são extremamente técnicos e
não entraremos nestes detalhes aqui de forma a não deixar a leitura
do texto muito pesada.} de ressomar todos os termos que contribuem
para o propagador dos campos fantasmas \cite{Zwanziger:1989mf}.

\subsection{Gribov horizon and the QCD vacuum}

Vamos analizar a equação das cópias (\ref{6copias3}). Concluímos
anteriormente que o vácuo perturbativo, $A_\mu^a=0$, reside no
centro da região de Gribov. Ademais, esta equação nos permite
definir o horizonte de Gribov como o conjunto de campos $A_\mu^a$
que geram os primeiros autovalores nulos do operador de
Faddeev-Popov. Por outro lado, o operador de Faddeev-Popov é o
operador que aparece atuando nos fantasmas de Faddeev-Popov. Assim,
conforme nos aproximamos do horizonte, o inverso deste operador se
torna cada vez maior, até que, no horizonte, ele torna-se singular.
Ainda, como na região de Gribov o operador de Faddeev-Popov é
positivo-definido, veja (\ref{6reg1}), o propagador de Faddev-Popov
também o será.

Vamos analisar o propagador dos campos de Fadeev-Popov. A teoria de
perturbações aplicada ao cálculo do propagador fantasma utilizando a
integral de caminho de Faddeev-Popov (\ref{6fadpop1}) nos leva a
\begin{equation}
G^{ab}(k)=\frac{\delta^{ab}}{k^2}
\frac{1}{\left(1-\frac{11g^2N}{48\pi^2}\log\frac{\Lambda^2}{k^2}\right)^{\frac{9}{44}}}\;,
\label{6ghost1}
\end{equation}
onde $\Lambda$ é o corte ultravioleta e $N$ é a dimensão do Grupo,
dada pelo número de Casimir $f^{acd}f^{bcd}=N\delta^{ab}$. O
propagador (\ref{6ghost1}) possui dois pólos, um em $k_1^2=0$ e
outro em $k_2^2=\Lambda^2e^{-\frac{48\pi^2}{11Ng^2}}$. É fácil ver
que quando $k^2$ é grande o termo que gera o pólo $k_2$ desaparece e
o termo que gera $k_1$ deixa de ser singular. Nesta escala de
energia sabemos que a teoria de perturbações é válida e portanto,
estamos lidando com o vácuo perturbativo $A_\mu^a=0$, ou seja, longe
do horizonte. Analizando o termo que gera o pólo $k_2$ vemos que,
para energias acima do pólo $k_2^2$, o fator $1$ domina e o
propagador (\ref{6ghost1}) é positivo, indicando que ainda estamos
dentro da região de Gribov. Para energias abaixo do pólo $k^2_2$ o
propagador (\ref{6ghost1}) se torna complexo, significando que, em
algum momento, cruzamos o horizonte de Gribov para fora da região de
Gribov. Desta forma, $\mathcal{V}(C_0)$ impediria a existência deste
pólo. Ainda, $\mathcal{V}(C_0)$ impediria a existência de um pólo
finito pois este acarretaria no mesmo problema do pólo $k_2^2$. Por
outro lado, o pólo $k_1^2=0$ está longe da origem, $A_\mu^a=0$, e
não possui o problema de cruzar o horizonte pois não podemos ir a
valores menores que o zero. Assim, este pólo é a única possibilidade
de descrever o horizonte. Podemos então esperar que o vácuo real da
QCD seja descrito pelas configurações que estão próximos ao
horizonte.

Note que esta é uma análise heurística e, até mesmo, ingênua do
vácuo da QCD. Contudo, em um recente resultado, \cite{Maas:2005qt},
foram encontradas evidências explícitas de que esta suposição seja
verdadeira mostrando que configurações topológicas de vácuo tendem a
existir próximo ao horizonte. Veremos a seguir como isso pode ser
evidênciado apenas a partir da restrição na integral de caminho.

\subsection{Gribov approximation}

\begin{figure}[ht]
\centering \epsfig{file=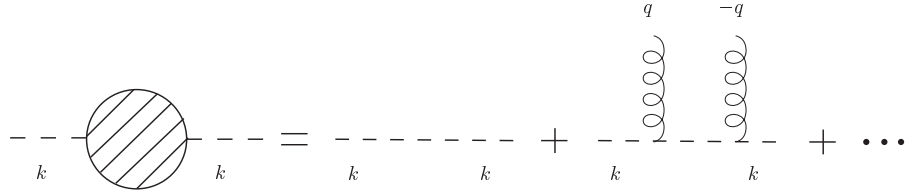,width=12cm} \caption{Propagador
fantasma com gluons externos.}\label{6fig3}
\end{figure}

Após esta análise do propagador perturbativo, concluímos que para
garantirmos que a integração seja feita apenas dentro da região de
Gribov é suficiente impormos que o propagador fantasma não possua
pólos finitos. O contrário implicaria que estamos fora da região de
Gribov. De acordo com \cite{Gribov:1977wm} a condição de não
existência de pólos finitos no propagador fantasma pode ser
desenvolvida através do cálculo perturbativo do mesmo, a segunda
ordem, considerando os gluons como campos externos e assumindo a
conservação da energia, conforme a Fig. \ref{6fig3}.

Esta aproximação, como feito em detalhe em
\cite{Gribov:1977wm,Sobreiro:2005ec}, resulta na condição
\begin{equation}
\frac{Ng^2}{4(N^2-1)}\int\frac{d^4k}{(2\pi)^4}\frac{A_\mu^a(k)A_\mu^a(-k)}{k^2}<1\;.\label{6nopole1}
\end{equation}
Esta condição implica no seguinte funcional de restrição
\begin{equation}
\mathcal{V}(C_0)=\exp\left\{-Ng^2\gamma^4\int\frac{d^4k}{(2\pi)^4}\frac{A_\mu^a(k)A_\mu^a(-k)}{k^2}+
4(N^2-1)\gamma^4\right\}\;. \label{6rest1}
\end{equation}
Na expressão (\ref{6rest1}) o parâmetro $\gamma$, chamado parâmetro
de Gribov, é determinado através do requerimento de que a energia do
vácuo dependa minimamente deste parâmetro. Este requerimento, à
primeira ordem, é traduzido através da equação do gap
\begin{equation}
\frac{3}{4}Ng^2\int\frac{d^4k}{(2\pi)^4}\frac{1}{k^4+2Ng^2\gamma^4}=1\;.\label{6gap1}
\end{equation}
De fato, a equação (\ref{6gap1}) fixa o valor de $\gamma$, ou seja,
o parâmetro de Gribov não é um parâmetro livre, mas determinado de
tal forma que a condição (\ref{6nopole1}) seja satisfeita. Note
ainda que o parâmetro de Gribov possui dimensão de massa.

O efeito físico da restrição se reflete na alteração da ação
determinando o peso pro-babilístico na integral de caminho. O termo
extra, proporcional ao parâmetro de Gribov, juntamente com a equação
do gap (\ref{6gap1}), garante que a integração ocorra apenas para
configurações que estejam dentro da primeira região de Gribov. Este
termo, sendo quadrático no campo de calibre, altera o propagador do
glúon mesmo no nível árvore. Um cálculo simples do propagador do
glúon, utilizando a integral de caminho (\ref{6fadpop2}) e levando
em conta o funcional (\ref{6rest1}), produz
\begin{equation}
D_{\mu\nu}^{ab}(k)=\delta^{ab}\frac{k^2}{k^4+2Ng^2\gamma^4}\left(\delta_{\mu\nu}-\frac{k_\mu{k}_\nu}{k^2}\right)\;.
\label{6gluon1}
\end{equation}
Note que este propagador é finito no limite infravermelho. Outro
efeito físico associado a este propagador é o fato de os pólos serem
imaginários, indicando que o glúon não pertence ao espectro físico
de Yang-Mills como deve ser para objetos confinados. Outra forma de
ver isso é através da violação da positividade deste propagador
\cite{Alkofer:2003jj,Cucchieri:2004mf,Dudal:2005na}.
\begin{figure}[ht]
\centering \epsfig{file=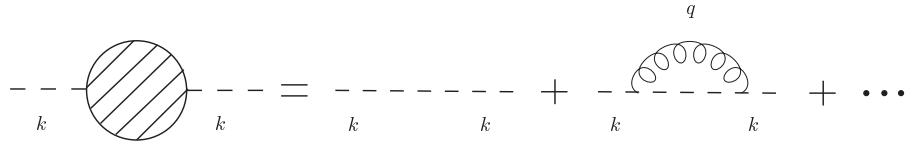,width=12cm} \caption{Propagador
fantasma a um laço.}\label{6fig4}
\end{figure}

O propagador dos campos de Faddeev-Popov se altera conforme
correções quânticas são levadas em consideração, isso devido ao
propagador do glúon (\ref{6gluon1}). De fato, podemos calcular o
propagador dos fantasmas a um laço em teoria de pertubações,
considerando o glúon agora um campo quântico, conforme a Fig.
\ref{6fig4}. O limite infravermelho deste propagador se comporta
como
\begin{equation}
G(k)\bigg|_{k\rightarrow0}=\frac{1}{N^2-1}G^{aa}(k)\bigg|_{k\rightarrow0}\propto\frac{1}{k^4}\;.\label{6ghost2}
\end{equation}
Vemos que o propagador dos fantasmas é mais singular que a predição
perturbativa, indicando a existência de forças de longo alcance na
teoria. É importante ressaltar que a equação do gap (\ref{6gap1}) é
fundamental para estabelecer este resultado.

Retornaremos a questão dos propagadores no fim desta parte. Vamos
agora discutir o método de Zwanziger de aprimoração da restrição de
Gribov.

\subsection{Horizon function}

A restrição feita à maneira de Gribov (seção anterior) consiste numa
aproximação da condição de ausência de pólos no propagador dos
fantasmas e a utilização desta condição para truncar o domínio de
integração da integral de caminho (\ref{6fadpop1}). A condição de
ausência de pólos pode ser ressomada a todas as ordens de forma a
encontramos uma forma mais completa para o funcional de restrição
$\mathcal{V}(C_0)$ definido em (\ref{6fadpop2}). Os detalhes do
método podem ser encontrados em \cite{Zwanziger:1989mf}. Esse
procedimento é feito impondo diretamente que o ope-rador de
Faddeev-Popopov possua apenas autovalores positivos. Entendendo a
equação de autovalores (\ref{6eigen1}) como um tipo de equação de
Schrödinger e tratando o termo dependente de $A_\mu^a$ como um
potencial perturbativo pode-se calcular, com boa aproximação, os
autovalores, a todas as ordens. Esse procedimento não tem ligação
alguma com as técnicas de segunda quantização, sendo um problema
puro de Mecânica Quântica, ou seja, primeira quantização. É o
problema de encontrar o espectro do operador de Faddeev-Popov para
um potencial $A_\mu$ genérico e em seguida impor que este potencial
não produza estados li-gados. Os passos para se encontrar os
autovalores do operador de Faddeev-Popov podem ainda omitir algumas
importantes considerações que levam o resultado final a ser, ainda,
aproximado, embora muito mais preciso que o resultado de Gribov
(\ref{6rest1}). A condição encontrada é dada por
\begin{equation}
8(N^2-1)-4NH>0\;,\label{6nopole2}
\end{equation}
onde $H$ é a chamada {\it função horizonte},
\begin{equation}
H=H[A]=g^2\int{d^4x}f^{abc}A_\mu^b\left(\mathcal{M}^{-1}\right)^{ad}f^{dec}A_\mu^e\;.\label{6hor1}
\end{equation}
O lado esquerdo da expressão (\ref{6nopole2}) está relacionada com
os autovalores do operador de Faddeev-Popov. A expressão
(\ref{6nopole2}) diz que o funcional de restrição pode ser escrito,
por exemplo, como
\begin{equation}
\mathcal{V}(C_0)=\exp\left\{-\gamma^4H+4(N^2-1)\gamma^4\right\}\;,\label{6rest2}
\end{equation}
O parâmetro de Gribov é determinado através da equação do gap,
obtida requerendo que o parâmtero de Gribov seja tal que a energia
do vácuo dependa minimamente de $\gamma$,
\begin{equation}
\left<H[A]\right>=4(N^2-1)\;.\label{6gap2}
\end{equation}
A equação do gap (\ref{6gap2}) também é conhecida como {\it condição
de horizonte}.

A restrição (\ref{6rest2}), juntamente com a condição de horizonte
(\ref{6gap2}), garante que o domínio de integração no espaço
funcional seja restrito à região de Gribov. Ademais, não é difícil
entender que a maior contribuição à integral de caminho
(\ref{6fadpop2}) vem das configurações próximas ao horizonte. Para
tal, basta observar que, próximo ao horizonte o funcional $H$ começa
a divergir, devido à presença do operador $\mathcal{M}^{-1}$. Assim,
na região infravermelha, o novo termo domina a integral de caminho.
Devido ao fato de a função exponencial poder ser usada como
representação integral da função delta, o funcional de restrição
pode ser descrito por
\begin{equation}
\mathcal{V}(C_0)=\delta[4(N^2-1)-H]\;.\label{6rest3}
\end{equation}
Esta expressão não só mostra que a equação do gap é satisfeita como
diz que a maior contribuição da integral de caminho vem de
configurações que vivem próximas ao horizonte. Note que podemos
representar a exponencial por uma função delta pois esta
representação é consistente com a equação do gap. Outra forma de
entender isso é observar que o funcional de restrição, na forma em
que aparece em (\ref{6rest2}), pode ser entendido como um termo de
Boltzmann no conjunto estatístico canônico. Assim, assumindo a
equivalência entre o conjunto estatístico canônico e o
microcanônico\footnote{De fato, esta equivalência existe neste caso,
como mostrado em \cite{Zwanziger:1989mf}.}, podemos mudar a
exponencial pela delta.

Vimos anteriormente, analisando o propagador perturbativo dos campos
fantasmas (\ref{6ghost1}), que o verdadeiro vácuo da QCD é composto
por configurações que vivem na fronteira da região de Gribov. Agora,
não só se confirma esta hipótese, como também vimos que estas
configurações são as que mais contribuem para a integral de caminho
(\ref{6fadpop2}). Obviamente, estudando a região ultravioleta, estas
configurações deixam de ser dominantes, contribuindo pouco aos
efeitos perturbativos.

Note que, tomando o limite apropriado
$\mathcal{M}\approx\partial^2$, os resultados aqui presentes se
reduzem aos da aproximação de Gribov (seção anterior).

\section{Stability of Gribov-Zwanziger action}

A integral de caminho (\ref{6fadpop2}), explicitamente escrita, é na
forma
\begin{equation}
Z=\int{DAD\bar{c}DcDb}\exp\left\{-S_{nl}\right\}\;,\label{6fadpop3}
\end{equation}
onde a ação não local $S_{nl}$ é conhecida como {\it ação de
Gribov-Zwanziger}, dada por
\begin{equation}
S_{nl}=S_{YM}+S_{gf}+\gamma^4H-4(N^2-1)\gamma^4\;.\label{6action1}
\end{equation}
A ação de Yang-Mills é dada por (\ref{1ym1}), a ação de fixação de
calibre, impondo o calibre de Landau, é dada por (\ref{1landau1}) e,
obviamente, a função horizonte é definida em (\ref{6hor1}). Uma
propriedade necessária para a ação (\ref{6action1}) é a estabilidade
sob correções quânticas. Contudo, a consistência da teoria ao nível
quântico requer como condições necesárias que a ação de partida seja
local e renormalizável por contagem de potências. Notavelmente, a
ação (\ref{6action1}) pode ser escrita numa representação local.
Ainda, de forma a tratar da renormalizabilidade da ação de
Gribov-Zwanziger, sob o ponto de vista da renormalização algébrica
\cite{book}, é necesária a existência de uma simetria BRST. A
simetria BRST pode ser definida se colocarmos a teoria de
Gribov-Zwanziger imersa numa classe de teorias mais gerais e, no
final das contas, toma-se o limite apropriado\footnote{Esta seção
será um pouco mais técnica pois devemos apresentar as propriedades e
simetrias da ação local (\ref{6action2})}. Discutiremos as
implicações físicas da localização, da existência de uma simetria
BRST nilpotente e outras simetrias e, por fim, da
renormalizabilidade da ação (\ref{6action1}). Os detalhes dos
resultados aqui discutidos podem ser encontrados em
\cite{Zwanziger:1989mf,Zwanziger:1992qr,Maggiore:1993wq}.

\subsection{Localization of the horizon function}

A função horizonte na integral de caminho (\ref{6fadpop3}) pode ser
escrita numa representação local com o auxílio de um conjunto de
campos auxiliares bosônicos e fermiônicos, veja
\cite{Zwanziger:1989mf,Zwanziger:1992qr}. Esta relação entre as
representações não-local e local é exata, dada por
\begin{eqnarray}
\exp\left\{-\gamma^4H\right\}&=&\int{D\varphi{D}\bar{\varphi}D\omega{D}\bar{\omega}}
\exp\left\{\int{d^4x}\left[-\bar{\varphi}^{ac}_\mu\mathcal{M}^{ab}\varphi^{bc}_\mu+
\bar{\omega}^{ac}_\mu\mathcal{M}^{ab}\omega^{bc}_\mu+\right.\right.\nonumber\\
&+&\left.\left.\gamma^2gf^{abc}A_\mu^a\left(\varphi^{bc}_\mu+\bar{\varphi}^{bc}_\mu\right)\right]+
4(N^2-1)\gamma^4\right\}\;,
\end{eqnarray}
provendo a ação local de partida, chamada {\it ação física}
\begin{equation}
S_{phys}=S_{YM}+S_{gf}+S_{aux}+S_{\gamma}\;,\label{6action2}
\end{equation}
onde, novamente, $S_{YM}$ é dada por (\ref{1ym1}) e $S_{gf}$ por
(\ref{1landau1}). Os demais termos são, o que chamaremos de {\it
termo de campos auxiliares},
\begin{equation}
S_{aux}=\int{d^4x}\left(\bar{\varphi}^{ac}_\mu\mathcal{M}^{ab}\varphi^{bc}_\mu-
\bar{\omega}^{ac}_\mu\mathcal{M}^{ab}\omega^{bc}_\mu\right)\;,\label{6aux1}
\end{equation}
e o {\it termo de mistura},
\begin{equation}
S_{\gamma}=-\gamma^2gf^{abc}\int{d^4x}\;A_\mu^a\left(\varphi^{bc}_\mu+\bar{\varphi}^{bc}_\mu\right)-4(N^2-1)\gamma^4\;.
\label{6gamma1}
\end{equation}

Os campos auxiliares $\{\varphi,\bar{\varphi}\}$ são de natureza
bosônica enquanto os campos $\{\omega,\bar{\omega}\}$ possuem
natureza fermiônica. Os últimos são  na verdade, fantasmas, pois são
campos com spin inteiro e estatística de Fermi. Obviamente, assim
como os campos de Faddeev-Popov, estes campos não são campos
físicos, no sentido de serem observáveis. Esta propriedade vem do
fato de os campos auxiliares poderem ser integrados e eliminados da
integral de caminho. Contudo, a presença de campos auxiliares torna
a ação de Gribov-Zwanziger passível de um estudo de sua
renormalizabilidade.

\subsection{Symmetries and Ward identities}

Uma vez que temos uma ação local que leva em conta a presença do
horizonte de Gribov, (\ref{6action2}), devemos estabelecer uma
simetria BRST de forma a tratar o problema da renormalizabilidade da
ação (\ref{6action2}). Ainda, simetrias extras são sempre bem
vindas.

De fato, a ação física (\ref{6action2}) possui muitas simetrias
escondidas. Uma dessas simetrias é a $U(f)$, $f=4(N^2-1)$, global.
Esta simetria nos permite, no caso geral, fazer uso do índice
composto $(a,\mu)\rightarrow{i}$, de forma que os campos auxiliares
são escritos como
$\left\{\varphi^{ab}_\mu,\bar{\varphi}^{ab}_\mu,\omega^{ab}_\mu,\bar{\omega}^{ab}_\mu\right\}=
\left\{\varphi^a_i,\bar{\varphi}^a_i,\omega^a_i,\bar{\omega}^a_i\right\}$.
Esta simetria possibilita a definição de um número quântico
adicional para os campos auxiliares, a {\it carga $Q_f$}. Os números
quânticos e dimensões ultravioletas dos campos auxiliares estão
dispostos na tabela \ref{6table1}, lembrando que apenas estes campos
e as fontes extras, que ainda vamos introduzir, possuem carga $Q_f$
não nula.
\begin{table}[t]
\centering
\begin{tabular}{|c|c|c|c|c|}
\hline
campos & $\varphi$ & $\bar{\varphi}$ & $\omega$ & $\bar{\omega}$ \\
\hline
dimensão & $1$ & $1$ & $1$ & $1$\\
número fantasma & $0$ & $0$ & $1$ & $-1$\\
carga $Q_f$ & $1$ & $-1$ & $1$ & $-1$\\
\hline
\end{tabular}
\caption{Dimensão e números quânticos dos campos auxiliares de
Zwanziger.} \label{6table1}
\end{table}

Podemos ainda definir uma simetria BRST nilpotente, onde os campos
auxiliares formam dubletos BRST, pertencendo assim à parte trivial
da cohomologia de BRST. As transformações BRST dos campos usuais das
teorias de Yang-Mills são dadas pela simetria BRST usual,
(\ref{1brs1}). Para os campos auxiliares temos
\begin{eqnarray}
s\varphi^a_i&=&\omega^a_i\;,\nonumber\\
s\omega^a_i&=&0\;,\nonumber\\
s\bar{\omega}^a_i&=&\bar{\varphi}^a_i\;,\nonumber\\
s\bar{\varphi}^a_i&=&0\;.\label{6brs1}
\end{eqnarray}

Para evidenciar estas, e outras, simetrias a ação física
(\ref{6action2}) é descrita através de uma teoria mais geral,
envolvendo vértices e fontes (campos clássicos) extras,
\begin{equation}
S=S_{YM}+S_{gf}+S'_{aux}+S_s\;,\label{6action3}
\end{equation}
onde a ação de campos auxiliares é alterada para
\begin{eqnarray}
S'_{aux}&=&s\int{d^4x}\;\bar{\omega}^a_i\mathcal{M}^{ab}\varphi^b_i\;,\nonumber\\
 &=&S_{aux}-gf^{abc}\int{d^4x}\;\partial_\mu\bar{\omega}_i^aD_\mu^{bd}c^d\varphi^c_i\;,\label{6aux2}
\end{eqnarray}
o que inclui um termo com vértices extras. Estes vértices, na
verdade, não têm influência nenhuma na teoria e podem ser eliminados
através de uma mudança de variáveis de Jacobiano unitário. Veremos
este ponto dentro de alguns parágrafos. A ação de fontes, que
substitui o termo de mistura, é dada por
\begin{eqnarray}
S_s&=&s\int{d^4x}\left(-U^a_{i\mu}D_\mu^{ab}\varphi^b_i-V^a_{i\mu}D_\mu^{ab}\bar{\omega}^b_i-
U^a_{i\mu}V^a_{i\mu}\right)\nonumber\\
&=&\int{d^4x}\left(-M^a_{i\mu}D_\mu^{ab}\varphi^b_i-V^a_{i\mu}D_\mu^{ab}\bar{\varphi}^b_i+
U^a_{i\mu}D_\mu^{ab}\omega^b_i-N^a_{i\mu}D_\mu^{ab}\bar{\omega}^b_i+\right.\nonumber\\
&-&\left.gf^{abc}U^a_{i\mu}D_\mu^{bd}c^d\varphi^c_i+
gf^{abc}V^a_{i\mu}D_\mu^{bd}c^d\bar{\omega}^c_i+U^a_{i\mu}N^a_{i\mu}-
M^a_{i\mu}V^a_{i\mu}\right)\;,\label{6gamma2}
\end{eqnarray}
onde as transformações BRST das fontes são dadas por
\begin{eqnarray}
sU^a_{i\mu}&=&M^a_{i\mu}\;,\nonumber\\
sM^a_{i\mu}&=&0\;,\nonumber\\
sV^a_{i\mu}&=&N^a_{i\mu}\;,\nonumber\\
sN^a_{i\mu}&=&0\;.\label{6brs2}
\end{eqnarray}
Os números quânticos das fontes estão dispostos na tabela
\ref{6table2}. Note que estas fontes são introduzidas para garantir
a invariância por transformações de BRST. De outra forma, a ação de
mistura, (\ref{6gamma1}), quebraria a simetria BRST.
\begin{table}[t]
\centering
\begin{tabular}{|c|c|c|c|c|}
\hline
fontes & $M$ & $V$ & $N$ & $U$ \\
\hline
dimensão & $2$ & $2$ & $2$ & $2$\\
número fantasma & $0$ & $0$ & $1$ & $-1$\\
carga $Q_f$ & $-1$ & $1$ & $1$ & $-1$\\
\hline
\end{tabular}
\caption{Dimensão e números quânticos das fontes de Zwanziger.}
\label{6table2}
\end{table}

A ação (\ref{6action3}) descreve uma teoria geral que engloba a ação
de Gribov-Zwanziger, (\ref{6action2}). Esta ação, utilizada de modo
a efetuar a renormalização da teoria, deve, no fim das contas, fluir
para a teoria de Gribov-Zwanziger, através de um limite apropriado.
Tal limite é atingido requerendo que as fontes auxiliares de
Zwanziger atinjam seus valores físicos
\begin{eqnarray}
M^{ab}_{\mu\nu}&=&V^{ab}_{\mu\nu}\;=\;\delta^{ab}\delta_{\mu\nu}\gamma^2\;,\nonumber\\
U^{ab}_{\mu\nu}&=&N^{ab}_{\mu\nu}\;=\;0\;.\label{6phys1}
\end{eqnarray}
Com esse limite todos os termos originais são recuperados, e alguns
vértices extras também. Isso porque, nesse limite, temos,
\begin{equation}
\lim_{\{M,V,U,N\}\rightarrow\gamma^2,0}S=S_{phys}-
gf^{abc}\int{d^4x}\;\partial_\mu\bar{\omega}_i^aD_\mu^{bd}c^d\varphi^c_i\;.\label{6extra1}
\end{equation}
Contudo, este termo não afeta a teoria pois pode ser eliminado
através de uma mudança de variável de Jacobiano unitário,
\begin{equation}
\omega^a_i\rightarrow\omega^a_i+gf^{abc}\mathcal{M}^{-1}\partial_\mu\left(D_\mu^{bd}c^d\varphi^c_i\right)\;.
\label{6transf1}
\end{equation}

Uma vez entendida a ação local (\ref{6action3}), podemos enumerar as
simetrias por ela gozada, em termos de identidades de Ward. Para
tal, como já explicado antes, devemos considerar a ação completa com
termos de fontes para controlar as transformações BRST não lineares
(Veja \cite{book} e Ap. \ref{ap_ferram}). Adicionando o termo de
fontes externas usuais, (\ref{1ext1}), a ação completa é dada por
\begin{equation}
\Sigma=S+S_{ext}\;.\label{6action4}
\end{equation}
A ação (\ref{6action4}) possui um rico conjunto de simetrias
descritas aqui através das seguintes identidades de Ward:
\begin{itemize}
\item Identidade de Slavnov-Taylor
\begin{eqnarray}
\mathcal{S}(\Sigma)&=&\int{d^4x}\left(\frac{\delta\Sigma}{\delta\Omega^a_\mu}\frac{\delta\Sigma}{\delta{A}^a_\mu}+
\frac{\delta\Sigma}{\delta{L}^a}\frac{\delta\Sigma}{\delta{c}^a}+b^a\frac{\delta\Sigma}{\delta\bar{c}^a}+
\omega^a_i\frac{\delta\Sigma}{\delta\varphi^a_i}+\bar{\varphi}^a_i\frac{\delta\Sigma}{\delta\bar{\omega}^a_i}+
\right.\nonumber\\
&+&\left.M^a_{i\mu}\frac{\delta\Sigma}{\delta{U}^a_{i\mu}}+
N^a_{i\mu}\frac{\delta\Sigma}{\delta{V}^a_{i\mu}}\right)=0\;.\label{6st1}
\end{eqnarray}

\item O calibre de Landau e a equação de movimento dos campos antifantasmas
\begin{eqnarray}
\frac{\delta\Sigma}{\delta{b}^a}&=&\partial_\mu{A}_\mu^a\;,\nonumber\\
\frac{\delta\Sigma}{\delta\bar{c}^a}+\partial_\mu\frac{\delta\Sigma}{\delta\Omega^a_\mu}&=&0\;.\label{6b1}
\end{eqnarray}

\item Equação de movimento dos campos fantasmas

\begin{equation}
\mathcal{G}^a\Sigma=\Delta^a_{cl}\;,\label{6ghost3}
\end{equation}
onde
\begin{equation}
\mathcal{G}^a=\int{d^4x}\left[\frac{\delta}{\delta{c}^a}+
gf^{abc}\left(\bar{c}^b\frac{\delta}{\delta{b}^c}+
\varphi^b_i\frac{\delta}{\delta\omega^c_i}+
\bar{\omega}^b_i\frac{\delta}{\delta\bar{\varphi}^c_i}+
V^b_{i\mu}\frac{\delta}{\delta{N}^c_{i\mu}}+
U^b_{i\mu}\frac{\delta}{\delta{M}^c_{i\mu}}\right)\right]\;,\label{6ghostop1}
\end{equation}
e $\Delta^a_{cl}$ é uma quebra clássica, linear nos campos,
\begin{equation}
\Delta^a_{cl}=gf^{abc}\int{d^4x}\left(\Omega^b_\mu{A}_\mu^c-L^bc^c\right)\;.\label{6clasbr1}
\end{equation}

\item Equações de movimento linearmente quebradas
\begin{eqnarray}
\frac{\delta\Sigma}{\delta\bar{\varphi}^a_i}+\partial_\mu\frac{\delta\Sigma}{\delta{M}^a_{i\mu}}&=&
gf^{abc}A^b_\mu{V}^c_{i\mu}\;,\nonumber\\
\frac{\delta\Sigma}{\delta\omega^a_i}+\partial_\mu\frac{\delta\Sigma}{\delta{N}^a_{i\mu}}-
gf^{abc}\bar{\omega}^b_i\frac{\delta\Sigma}{\delta{b}^c}&=&
gf^{abc}A^b_\mu{U}^c_{i\mu}\;,\nonumber\\
\frac{\delta\Sigma}{\delta\bar{\omega}^a_i}+\partial_\mu\frac{\delta\Sigma}{\delta{U}^a_{i\mu}}-
gf^{abc}V^b_{i\mu}\frac{\delta\Sigma}{\delta\Omega_\mu^c}&=&-
gf^{abc}A^b_\mu{N}^c_{i\mu}\;,\nonumber\\
\frac{\delta\Sigma}{\delta\varphi^a_i}+\partial_\mu\frac{\delta\Sigma}{\delta{V}^a_{i\mu}}-
gf^{abc}\bar{\varphi}^b_i\frac{\delta\Sigma}{\delta{b}^c}-
gf^{abc}\bar{\omega}^b_i\frac{\delta\Sigma}{\delta\bar{c}^c}-
gf^{abc}U^b_{i\mu}\frac{\delta\Sigma}{\delta\Omega_\mu^c}&=&
gf^{abc}A^b_\mu{M}^c_{i\mu}\;.\nonumber\\
\label{6eqs1}
\end{eqnarray}

\item Identidade supersimétrica global
\begin{equation}
\int{d^4x}\left(c^a\frac{\delta\Sigma}{\delta\omega^a_i}+\bar{\omega}^a_i\frac{\delta\Sigma}{\delta\bar{c}^a}+
U^a_{i\mu}\frac{\delta\Sigma}{\delta\Omega^a_\mu}\right)=0\;.\label{6super1}
\end{equation}

\item Identidade supersimétrica global no setor de Gribov
\begin{equation}
\int{d^4x}\left(\varphi^a_i\frac{\delta\Sigma}{\delta\omega^a_j}-
\bar{\omega}^a_j\frac{\delta\Sigma}{\delta\bar{\varphi}^a_i}+
V^a_{i\mu}\frac{\delta\Sigma}{\delta{U}^a_{j\mu}}-
U^a_{j\mu}\frac{\delta\Sigma}{\delta{M}^a_{i\mu}}\right)=0\;.\label{6super2}
\end{equation}
\end{itemize}

\noindent É importante ter em mente que a ação (\ref{6action4}) não
é a ação física da teoria de Yang-Mills com restrição ao primeiro
horizonte, mas sim uma teoria mais geral que atinge a ação física de
Gribov-Zwanziger (\ref{6action2}) no limite apropriado
(\ref{6phys1}). Note ainda que a equação do gap, que fixa o
parâmetro de Gribov (\ref{6gap2}), continua sendo obtida através da
minimização da energia do vácuo. Contudo, na versão local, a equação
do gap é dada por
\begin{equation}
\left<gf^{abc}A^a_\mu\varphi^{bc}_\mu\right>+\left<gf^{abc}A^a_\mu\bar{\varphi}^{bc}_\mu\right>=-8(N^2-1)\gamma^2\;.
\label{6gap3}
\end{equation}
Esta condição deve ser efetuada após o processo de renormalização.
De fato, qualquer cálculo explícito deve ser desenvolvido após a
renormalização, assim como todo e qualquer limite.

Outra importante questão vem da quebra de BRST que ocorre no limite
físico. Esta quebra não é entendida totalmente ainda. Em
\cite{Maggiore:1993wq} esta quebra foi associada a uma quebra
espontânea de BRST devido à presença do horizonte de Gribov. Em
\cite{Zwanziger:2006proc} os valores físicos das fontes são ditos
poderem ser obtidos através do potencial efetivo da ação
(\ref{6action4}), reforçando a idéia da quebra espontânea. Contudo,
esta questão ainda é um tanto obscura e não falaremos neste tópico
até as conclusões finais.

\subsection{Stability}

Finalmente, uma vez que temos uma ação local (\ref{6action4}) e
identidades de Ward (\ref{6st1}-\ref{6ghost3}),
(\ref{6eqs1}-\ref{6super2}) descrevendo as simetrias desta ação,
podemos discutir a estabilidade da mesma. De fato, utilizando os
métodos usuais da renormalização algébrica \cite{book}, Ap.
\ref{ap_ferram}, mostra-se que a ação (\ref{6action4}) é
multiplicativamente renormalizável a todas as ordens em teoria de
perturbações, veja \cite{Zwanziger:1992qr}. Ainda, devido ao grande
número de identidades de Ward, a ação (\ref{6action4}), e,
consequentemente, (\ref{6action1}) e (\ref{6action2}), possuem
apenas duas divergências independentes, como no caso do calibre de
Landau sem considerar o horizonte de Gribov. O contratermo mais
geral pode ser escrito como
\begin{equation}
\Sigma^c=a_0S_{YM}+a_1\int{d^4x}\left(A_\mu^a\frac{\delta\Sigma}{\delta{A}_\mu^a}+
\tilde{\Omega}_\mu^a\partial_\mu{c}^a+\tilde{V}^a_{i\mu}\tilde{M}^a_{i\mu}-\tilde{U}^a_{i\mu}\tilde{N}^a_{i\mu}
\right)\;,\label{6cont1}
\end{equation}
onde as variáveis $\tilde{\Omega}$, $\tilde{V}$, $\tilde{M}$,
$\tilde{U}$ e $\tilde{N}$, devido às identidades (\ref{6eqs1}), são
dadas por
\begin{eqnarray}
\tilde{\Omega}_\mu^a&=&\Omega_\mu^a+\partial_\mu\bar{c}^a-
gf^{abc}\tilde{U}^b_{i\mu}\varphi^c_i-gf^{abc}V^b_{i\mu}\bar{\omega}^c_i\;,\nonumber\\
\tilde{V}^a_{i\mu}&=&V^a_{i\mu}+\partial_\mu\varphi^a_i\;,\nonumber\\
\tilde{M}^a_{i\mu}&=&M^a_{i\mu}+\partial_\mu\bar{\varphi}^a_i\;,\nonumber\\
\tilde{U}^a_{i\mu}&=&U^a_{i\mu}+\partial_\mu\bar{\omega}^a_i\;,\nonumber\\
\tilde{N}^a_{i\mu}&=&N^a_{i\mu}+\partial_\mu\omega^a_i\;.\label{6red1}
\end{eqnarray}
Os fatores de renormalização independentes são dados pela expresão
(\ref{4ren1}). Os campos fantasmas, campos auxiliares e fontes
externas auxiliares se renormalizam através dos mesmos fatores
\begin{equation}
Z_c^{1/2}=Z_{\bar{c}}^{1/2}=Z_\varphi^{1/2}=Z_{\bar{\varphi}}^{1/2}=Z_\omega^{1/2}=Z_{\bar{\omega}}^{1/2}=
Z_V=Z_M=Z_U=Z_N=Z_g^{-1/2}Z_A^{-1/4}\;,\label{6ren2}
\end{equation}
Essa relação entre as fontes auxiliares, a constante de acoplamento
e o glúon vem das identidades de Ward (\ref{6super1}-\ref{6super2}).
Devido a esta propriedade não há necessidade de se introduzir um
parâmetro para a renormalização deste termo, como no caso do
parâmetro LCO.

Note que todas as propriedades do calibre de Landau são preservadas
quando o ho-rizonte é levado em conta, incluindo o teorema de não
renormalização dos campos de Faddeev-Popov, \cite{Blasi:1990xz},
expressado em (\ref{4ren4}) e generalizado em (\ref{6ren2}). Ainda,
o termo de fontes puro não necessita de um parâmetro LCO para
garantir a renormalizabilidade.

Com isso terminamos a análise física sobre as ambigüidades de Gribov
existente no calibre de Landau, salvo os resultados que
apresentaremos daqui para frente no próximos capítulos.

\chapter{Gribov horizon and the operator $A_\mu^aA_\mu^a$}\label{cap7_gribov2}

Neste capítulo estudaremos a compatibilidade do horizonte de Gribov
com o operador de dimensão dois $A_\mu^aA_\mu^a$, os detalhes
técnicos podem ser encontrados em
\cite{Sobreiro:2004us,Sobreiro:2004yj,Dudal:2005na}. Na parte
anterior desta tese, revisamos a condensação deste operador, como
feito originalmente em
\cite{Verschelde:2001ia,Knecht:2001cc,Dudal:2002pq,Browne:2003uv}.
Como resultado este operador condensa, conduzindo a energia do vácuo
para um valor menor que a predição perturbativa. Como consequência
da existência de um valor não trivial para o o condensado
$\left\langle A_\mu^2\right\rangle$, um parâmtero de massa para os
gluons é gerado dinamicamente.

Começaremos apresentando cálculos explícitos utilizando a ação de
Gribov-Zwanziger, (\ref{6action2}), sem considerar os efeitos da
massa dinâmica. Veremos que tal teoria gera uma ação quântica
efetiva que obedece a uma equação homogênea do grupo de
renormalização. Veremos ainda que, na aproximação de um laço, a
energia do vácuo é sempre positiva independentemente do esquema de
renormalização utilizado e da escala escolhida. O que sugere que
cálculos em mais laços sejam necessários. Uma breve discussão sobre
cálculos a dois laços é apresentada.

Em seguida mostraremos que a ação de Gribov-Zwanziger permanece
estável na presença do operador composto $A_\mu^aA_\mu^a$, quando
introduzido via o método LCO, com apenas três divergências
independentes a serem renormalizadas. Tal ação gera uma ação
quântica efetiva que respeita uma equação homogênea do grupo de
renormalização. Desta ação quântica extraímos duas equações de gap
acopladas a serem resolvidas nos parâmetros de Gribov $\gamma$ e do
condensado $\left\langle A_\mu^aA_\mu^a\right\rangle$. Discutiremos
as tentativas de solucionar as equações de gap acopladas e o fato de
não encontrarmos soluções consistentes, tudo no esquema $\MSbar$ de
renormalização.

Como tentativa de se aprimorar os resultados apresentamos uma
expansão otimizada de forma a reduzir a dependência dos resultados
no esquema de renormalização escolhido. Desta forma, a dependência
no esquema de renormalização se traduz através de um único
parâmetro, relacionado à renormalização da constante de acoplamento.
Apesar deste esforço, a energia do vácuo ainda se mostra positiva a
um laço, o que parece ser um resultado geral, presente também quando
o condensado $\left<A_\mu^aA_\mu^a\right>$ é incluído. Este
resultado sugere que cálculos a mais laços devem ser feitos, ou
ainda, que o problema de Gribov, sendo altamente não perturbativo
não pode ser tratado com teoria de perturbações.

Por fim discutiremos as implicações físicas de um possível parâmetro
de Gribov não nulo e faremos comparações com a rede e as informações
obtidas através das equações de Schwinger-Dyson. Os detalhes
técnicos deste capítulo podem ser encontrados em
\cite{Sobreiro:2004us,Sobreiro:2004yj,Dudal:2005na}.

\section{Computations with the Gribov-Zwanziger action}

Antes de discutirmos o operador
$\left\langle{A}_\mu^aA_\mu^a\right\rangle$ na presença do horizonte
de Gribov vamos estudar algumas propriedades quânticas da ação pura
de Gribov-Zwanziger. Para tal utilizaremos a ação física na
representação local, (\ref{6action2}), que nada mais é que a ação de
Gribov-Zwanziger (\ref{6action1}) na representação local. Sendo esta
ação renormalizável, a ação quântica corespondente obedece a uma
equação homogênea do grupo de renormalização, o que nos permite
efetuar cálculos explícitos. Fundamentalmente, vamos estudar a
energia do vácuo a um laço, através da ação quântica efetiva, e
tentar dar uma estimativa quantitativa para o parâmetro de Gribov.

\subsection{One-loop quantum action in $\MSbar$ scheme}

Para se calcular a ação quântica efetiva a um laço,
$\Gamma_\gamma^{(1)}$, associada a ação clássica (\ref{6action2}),
basta-nos considerar a parte quadrática da mesma na definição da
ação quântica. Note ainda que, nesta aproximação, apenas os campos
auxiliares bosônicos e o campo de calibre se misturam, os demais
podem ser ignorados,
\begin{equation}
e^{-\Gamma_\gamma^{(1)}}=\int{DAD\bar{\varphi}D\varphi}\;e^{-S_{quad}}\;,\label{7eff1}
\end{equation}
com $S_{quad}$ dada por
\begin{eqnarray}
S_{quad} &=&\int{d^4x}\left[ \frac{1}{4}\left(
\partial _{\mu
}A_{\nu }^{a}-\partial _{\nu }A_{\mu }^{a}\right) ^{2}+\frac{1}{2\alpha }%
\left( \partial_\mu A_\mu^{a}\right) ^{2}+\overline{\varphi }_{\mu
}^{ab}\partial
^{2}\varphi _{\mu }^{ab}\right.  \nonumber \\
&-&\left. \gamma ^{2}g\left( f^{abc}A_{\mu }^{a}\varphi _{\mu
}^{bc}+f^{abc}A_{\mu }^{a}\overline{\varphi }_{\mu }^{bc}\right)
-d(N^{2}-1)\gamma ^{4}\vphantom{\frac{1}{4}\left( \partial _{\mu
}A_{\nu }^{a}-\partial _{\nu }A_{\mu }^{a}\right) ^{2}}\right] \;,
\label{7quad1}
\end{eqnarray}
lembrando que o limite $\alpha\rightarrow0$ deve ser tomado no final
para que o calibre de Landau seja recuperado. Um elaborado cálculo
nos leva à seguinte expressão não renormalizada, para
$\Gamma_\gamma^{(1)}$,
\begin{equation}
\Gamma^{(1)}=-d(N^{2}-1)\gamma ^{4}+\frac{(N^{2}-1)}{2}\left(
d-1\right) \int \frac{d^{d}p}{\left( 2\pi \right) ^{d}}\ln \left(
p^{4}+2Ng^{2}\gamma ^{4}\right) \;, \label{7eff2}
\end{equation}
onde a regularização dimensional foi utilizada. Da ação efetiva
(\ref{7eff2}) podemos extrair a equação do gap a um laço. Para tal,
derivamos $\Gamma_\gamma^{(1)}$ com relação ao parâmetro de Gribov,
de acordo com a definição da equação do gap (\ref{7gap2}),
\begin{equation}
\frac{\p\Gamma_\gamma^{(1)}}{\p\gamma}=0\;,\label{7gap1}
\end{equation}
de forma que,
\begin{equation}
\frac{N\left( d-1\right) }{d}g^{2}\int \frac{d^{d}p}{\left( 2\pi \right) ^{d}%
}\frac{1}{\left( p^{4}+2Ng^{2}\gamma ^{4}\right) }=1\;.
\label{7gap2}
\end{equation}

Utilizando agora o esquema de renormalização $\MSbar$ na expressão
(\ref{7eff2}) chegamos a ação quântica efetiva renormalizada a um
laço na forma
\begin{equation}
\Gamma_\gamma^{(1)}=-4(N^2-1)\gamma^4-\frac{3(N^2-1)}{32\pi^2}g^2N\gamma^4\left(\ln\frac{2Ng^2\gamma^4}{\overline{\mu}^4}-
\frac{8}{3}\right)\;.\label{7eff3}
\end{equation}
de onde podemos extrair facilmente a equação do gap renormalizada
\begin{equation}
\frac{5}{3}-\ln\frac{2Ng^2\gamma^4}{\overline{\mu}^4}=\frac{128\pi^2}{3g^2N}\;.\label{7gap3}
\end{equation}
Assim, temos a ação quântica a um laço (\ref{7eff3}) e a
correspondente equação do gap (\ref{7gap3}) que determina o valor
físico do parâmetro de Gribov.

\subsection{Renormalization group invariance}

A ação (\ref{6action2}) é, de fato, multiplicativamente
renormalizável, o que faz com que a ação efetiva associada obedeça a
uma equação homogênea do grupo de renormalização. Por segurança, é
conveniente checarmos a invariância da ação efetiva (\ref{7eff3})
sob o grupo de renormalização. Para isso necessitamos saber a
dimensão anômala do parâmetro de Gribov. Da expresão (\ref{6ren2}),
vinculando os fatores de renormalização das fontes externas
auxiliares, é fácil ver que
\begin{equation}
\gamma_{\gamma^2}(g^2)=-\frac{1}{2}\left[\frac{\beta(g^2)}{2g^2}-\gamma_A(g^2)\right]\;,\label{7diman1}
\end{equation}
onde $\gamma_A(g^2)$ representa a dimensão anômala do campo de
gluons $A_\mu^a$ e, obviamente, $\beta(g^2)$ é a função beta
associada a dimensão anômala da constante de acoplamento $g$. Note
que a relação (\ref{7diman1}) é válida a todas as ordens em teorias
de perturbações. Desta forma, é fácil checar que, a um laço,
\begin{equation}
\overline{\mu}\frac{d\Gamma_\gamma^{(1)}}{d\overline{\mu}}=\left[\frac{\beta^{(1)}(g^2)}{2g^2}-\gamma_A^{(1)}(g^2)+
\frac{3Ng^2}{32\pi^2}\right]4(N^2-1)\gamma^4\;.\label{7rg1}
\end{equation}
Ainda, das expressões a um laço da dimensão anômala do glúon e da
função beta, (\ref{5beta1}), é trivial checar que
\begin{equation}
\overline{\mu}\frac{d\Gamma^{(1)}}{d\overline{\mu}}=0\;,\label{7rg2}
\end{equation}
provando assim a invariância sob o grupo de renormalização da ação
quântica a um laço.

\subsection{Vacuum energy analysis}

Antes de discutirmos as possíveis soluções da equação do gap
(\ref{7gap3}), vamos previamente substituir a mesma na expressão da
ação efetiva (\ref{7eff3}). Esta substituição nos fornece a
expressão geral da energia do vácuo a um laço, no esquema de
renormalização $\MSbar$. Encontramos facilmente que,
independentemente da escala escolhida,
\begin{equation}
E_{vac}^{\MSbar}=\frac{3(N^2-1)}{32\pi^2}g^2N\gamma^4>0\;.\label{7vac1}
\end{equation}
Vemos que a energia do vácuo a um laço, no esquema $\MSbar$, é
sempre maior que zero. De fato podemos mostrar que a energia do
vácuo, a um laço, é sempre positiva, independentemente do esquema de
renormalização escolhido e dada pela expressão (\ref{7vac1}). Seja
então a ação efetiva a um laço renormalizada num esquema de
renormalização não massivo geral
\begin{equation}
\Gamma=-4\left(N^2-1\right)\gamma^4-\frac{3Ng^2\gamma^4\left(N^2-1\right)}{32\pi^2}\left(\ln\frac{2Ng^2\gamma^4}{\omu^4}+a\right)\;,
\label{7rengeral1}
\end{equation}
com $a$ uma constante arbitrária. A equação do gap correspondente
seria então
\begin{equation}
-4-\frac{3g^2N}{32\pi^2}\left(\ln\frac{2Ng^2\gamma^4}{\omu^4}+a\right)-\frac{3g^2N}{32\pi^2}=0\;.\label{7rengeral2}
\end{equation}
Assim, substituindo a equação do gap (\ref{7rengeral2}) na ação
quântica (\ref{7rengeral1}) chegamos a expressão da energia do vácuo
a um laço, independente do esquema de renormalização empregado, que
coincide exatamente com a expressão do caso $\MSbar$, (\ref{7vac1}).
Esta expressão é válida para qualquer escala $\omu$ e para qualquer
valor de $a$. Desta forma, concluímos que a energia do vácuo a um
laço é sempre positiva no modelo original de Gribov-Zwanziger.

Ressaltamos a importância do sinal da energia do vácuo, uma vez que
é relacionada ao condensado de gluons
$\left\langle{F_{\mu\nu}^aF_{\mu\nu}^a}\right\rangle$, através da
anomalia do traço. Da expressão da anomalia do traço,
\begin{equation}
\theta_{\mu\mu}=\frac{\beta(g^2)}{2g^2}F_{\mu\nu}^aF_{\mu\nu}^a\;,\label{7traceano1}
\end{equation}
pode-se deduzir a energia do vácuo e conectá-la ao condensado
$\left\langle{F_{\mu\nu}^aF_{\mu\nu}^a}\right\rangle$. Em
particular, para o caso $N=3$, encontra-se
\begin{equation}
\left\langle\frac{g^2}{4\pi^2}F_{\mu\nu}^aF_{\mu\nu}^a\right\rangle=-\frac{32}{11}\Evac\;,\label{7traceano2}
\end{equation}
a dois laços. Desta forma, um valor positivo da energia do vácuo
implica num valor negativo para o condensado
$\left\langle\frac{g^2}{4\pi^2}F_{\mu\nu}^aF_{\mu\nu}^a\right\rangle$,
em contradição com o que é usualmente encontrado,
\cite{Shifman:1978bx,DiGiacomo:1998nu}. Note, ainda, que, até onde
se sabe (quatro laços), a função beta é negativa,
\cite{vanRitbergen:1997va,Czakon:2004bu,Chetyrkin:2004mf}. Portanto,
$\Evac$ e
$\left\langle\frac{g^2}{4\pi^2}F_{\mu\nu}^aF_{\mu\nu}^a\right\rangle$
devem permanecer com sinais opostos a ordens superiores.

Mesmo com o resultado desanimador de uma energia positiva para o
vácuo, podemos ainda tentar encontrar soluções para o parâmetro de
Gribov. Para tal, vamos nos concentrar agora no esquema $\MSbar$, ou
seja, nas equações (\ref{7eff3}) e (\ref{7gap3}). A escolha natural
para a escala de renormalização na equação do gap (\ref{7gap3})
seria $\omu^4=2Ng^2\gamma^4$ de forma a eliminar os logaritimos. O
que nos fornece para o parâmetro de expansão perturbativa o valor
\begin{equation}
\frac{g^2N}{16\pi^2}\bigg|_{\omu^4=2Ng^2\gamma^4}=\frac{8}{5}>1\;.\label{7sol1}
\end{equation}
o que já invalida uma solução confiável.

Podemos, ainda, tentar encontrar uma escala que nos forneça uma
constante de expansão relativamente pequena. Para tal basta impormos
a condição $\frac{g^2N}{16\pi^2}<1$ na equação do gap (\ref{7gap3}),
o que nos leva a
\begin{equation}
\ln\left(\frac{2Ng^2\gamma^4}{\omu^4}\right)<-1\;.\label{7cond1}
\end{equation}
Infelizmente, a equação (\ref{7cond1}) não possui solucão real para
$\omu$. Este resultado implica ainda que, no esquema $\MSbar$, a
técnica de resomar logaritimos dominantes está fora de
aplicabilidade. Ainda, os resultados desta seção (Energia do vácuo
positiva e não termos encontrado uma solução consistente para a
equação do gap) indicam que devemos efetuar cálculos com mais laços
para tentar obter mais informações sobre o parâmetro de Gribov e o
horizonte.

Devemos chamar atenção para o fato de que, recentemente, cálculos
explícitos com métodos numéricos tem sido empregados para extrair
informações sobre as propriedades quânticas da ação de
Gribov-Zwanziger, \cite{Gracey:2005cx,Gracey:2006dr}. Em
\cite{Gracey:2005cx} a equação do gap a dois laços foi calculada e a
singularidade do propagador dos campos fantasmas foi verificada. Em
\cite{Gracey:2006dr} o propagador do campo de gluons foi calculado a
um laço, confirmando sua nulidade no limite infravermelho. Ainda, a
constante de acoplamento invariante pelo grupo de renormalização, a
um laço, se mostrou congelar num valor finito no limite de baixas
energias. Sendo o cálculo feito utilizando o propagador dos campos
fantasmas em união com o propagador do glúon através do teorema de
não renormalizacão dos campos fantasmas, \cite{Blasi:1990xz}.
Notavelmente, este teorema se mostra válido quando o horizonte é
levado em conta, \cite{Dudal:2005na}, fato também evidenciado por
cálculos na rede, \cite{Cucchieri:2004sq}. Contudo, as equações a
dois laços também parecem não fornecer nenhuma solução consistente,
indicando que o problema de Gribov seja alta e exclusivamente não
perturbativo.

\section{Gribov horizon in the presence of the operator $A_\mu^aA_\mu^a$}

Uma vez que discutimos algumas propriedades quânticas da ação de
Gribov-Zwanziger a um laço, podemos iniciar o estudo da mesma na
presença do operador $A_\mu^aA_\mu^a$. Isto significa que na ação
(\ref{6action3}), além dos operadores compostos
$f^{abc}A_\mu^a\varphi_\mu^{bc}$ e
$f^{abc}A_\mu^a\overline{\varphi}_\mu^{bc}$, consideraremos também a
inserção $A_\mu^aA_\mu^a$. Começemos, antes de entrarmos nos
cálculos explícitos, por discutir sua renormalizabilidade.

\subsection{Renormalizability}

Introduziremos o operador $A_\mu^aA_\mu^a$ através do método LCO,
como discutido na segunda parte desta tese e em \cite{Dudal:2005na}.
De acordo com este método o operador $A_\mu^aA_\mu^a$ pode ser
introduzido considerando a ação
\begin{equation}
\Sigma'=\Sigma+S_{LCO}\;,\label{7action1}
\end{equation}
onde $\Sigma$, dada por (\ref{6action4}), é a ação de
Gribov-Zwanziger com o sistema de fontes auxiliares e ação de fontes
externas para definir as transformações BRST não lineares. A ação
LCO, por sua vez, é dada por (\ref{4lco1}). É importante ter em
mente que as fontes auxiliares $\{U,V,M,N\}$, bem como as fontes LCO
$\{\lambda,J\}$ são introduzidas de forma a manter a simetria BRST
da ação de partida, bem como a renormalizabilidade por contagem de
potências. Ainda, o parâmetro LCO, $\zeta$, é necessário para
absorver as divergências da função de correlação
$\left\langle{A}^2(x)A^2(y)\right\rangle$. Ademais, como discutido
na segunda parte desta tese, este parâmetro é fundamental para se
estabelecer a invariância sob o grupo de renormalização, veja
\cite{Verschelde:2001ia}.

Para provar a renormalizabilidade da ação (\ref{7action1})
utilizamos a teoria de renormalização algébrica, \cite{book}. Os
detalhes técnicos da prova podem ser encontrados em
\cite{Dudal:2005na}. O primeiro passo é identificar as identidades
de Ward presentes na ação (\ref{7action1}). De fato, exceto pela
identidade de Slavnov-Taylor, as identidades existentes na ação de
Gribov-Zwanziger, (\ref{6action4}), se preservam identicamente na
presença do termo LCO (\ref{4lco1}). Tais identidades estão listadas
em (\ref{6b1}-\ref{6super2}), onde devemos apenas substituir
$\Sigma$ por $\Sigma'$. O mesmo é válido para a equação de inserção
(\ref{4lambda1}), definindo o operador $A_\mu^aA_\mu^a$,
compativelmente com a simetria $SL(2,\mathbb{R})$, (\ref{4sl1}). A
única identidade que se altera é a identidade de Slavnov-Taylor,
devido a presença do dubleto de fontes LCO, $\{\lambda,J\}$, veja
(\ref{4brs3}). A identidade de Slavnov-Taylor fica na forma
\begin{eqnarray}
\mathcal{S}(\Sigma')&=&\int{d^4}x\left(\frac{\delta\Sigma'}{\delta\Omega_\mu^a}\frac{\delta\Sigma'}{\delta{A}_\mu^a}+
\frac{\delta\Sigma'}{\delta{L}^a}\frac{\delta\Sigma'}{\delta{c}^a}+b^a\frac{\delta\Sigma'}{\delta\bar{c}^a}+
\bar{\varphi}_i^a\frac{\delta\Sigma'}{\delta\bar{\omega}_i^a}+\omega_i^a\frac{\delta\Sigma'}{\delta\varphi_i^a}\right.\nonumber\\
&\;&\;\;\;\;\;\;\;\;\;\;\;\;+\left.M_\mu^{ai}\frac{\delta\Sigma'}{\delta{U}_\mu^{ai}}+N_\mu^{ai}\frac{\delta\Sigma'}{\delta{V}_\mu^{ai}}+
J\frac{\delta\Sigma'}{\delta\lambda}\right)=0\;,\label{7st1}
\end{eqnarray}\\\\ O segundo passo é encontrar o contratermo mais
geral possível de forma a respeitar as identidades de Ward da ação
de partida, $\Sigma'$. De acordo com \cite{Dudal:2005na}, o
contratermo mais geral se escreve como\footnote{Diferentemente de
\cite{Dudal:2005na}, estamos considerando a equação de inserção
(\ref{4lambda1}), que é responsável pelo teorema de não
renormalização do operador $A^2$, definido em (\ref{4ren3}).}
\begin{eqnarray}
\Sigma'^c&=&a_0S_{YM}+a_1\int{d^4}x\left(A_\mu^a\frac{\delta{S}_{YM}}{\delta{A}_\mu^a}+\widetilde{\Omega}_\mu^a\partial_\mu{c}^a+
\widetilde{V}_\mu^{ai}\widetilde{M}_\mu^{ai}-\widetilde{U}_\mu^{ai}\widetilde{N}_\mu^{ai}\right)\nonumber\\
&+&\int{d^4}x\left[\frac{a_1}{2}JA_\mu^aA_\mu^a+\frac{a_8}{2}\zeta{J}^2\right]\;.\label{7contra1}
\end{eqnarray}

O terceiro e último passo é checar a renormalizabilidade
multiplicativa da teoria, mostrando que o contratermo
(\ref{7contra1}) pode ser reabsorvido pela ação clássica
(\ref{7action1}). É importante ressaltar que, devido a validade das
identidades (\ref{6b1}-\ref{6super2}) e (\ref{4lambda1}), no
presente caso, todas as propriedades de renormalização obtidas no
caso de Gribov-Zwanziger puro (Veja capítulo anterior) se preservam
quando incluído o operador $A_\mu^aA_\mu^a$ bem como as propriedades
do setor LCO. Esta preservação é facilmente entendida observando que
o setor de Gribov-Zwanziger não se mistura com o setor LCO devido
aos números quânticos dos campos e fontes envolvidos. Desta forma,
os fatores de renormalização dos campos, fontes e parâmetro são
dados ainda por (\ref{4ren1}), (\ref{6ren2}) e (\ref{4ren3}).

De fato, os teoremas de não renormalização mais importantes aqui
presentes são os que se seguem: O teorema de não renormalização dos
campos de Faddeev-Popov e dos campos auxiliares de Zwanziger,
(\ref{6ren2}), a não renormalização do operador $A_\mu^aA_\mu^a$,
(\ref{4ren3}), de onde extraímos a não renormalização do parâmetro
de Gribov, $\gamma$,
\begin{equation}
Z_{\gamma^2}=Z_g^{-1/2}Z_A^{-1/4}\;,\label{7ren3}
\end{equation}
relação que gera a dimensão anômala do parâmetro de Gribov na forma
(\ref{7diman1}). Relembramos que os fatores de renormalização da
constante de acoplamento e do glúon estão descritas em
(\ref{4ren1}).

Em resumo, a ação de Gribov-Zwanziger se mostra multiplicativamente
renormalizável ainda na presença do operador $A_\mu^aA_\mu^a$,
possuindo apenas três divergências independentes. Esta propriedade
diz que, de fato, a ação quântica efetiva obedece a uma equação
homogênea do grupo de renormalização, o que nos permite tentar obter
estimativas dos valores do parâmetro de Gribov e do condensado
$\left<A_\mu^aA_\mu^a\right>$.

\subsection{One-loop quantum action in $\MSbar$ scheme}

Uma vez provada a renormalizadilidade da ação de Gribov-Zwanziger
podemos partir para o cálculo da ação quântica efetiva a um laço.
Para tal, de acordo com o método LCO \cite{Verschelde:2001ia},
devemos colocar todas as fontes externas em seus respectivos valores
físicos. As fontes de Zwanziger assumem os valores dados em
(\ref{6phys1}) enquanto as demais fontes se anulam. Contudo, devemos
notar que nem todos os termos de fontes podem ser colocados a zero
diretamente. No caso das fontes de Zwanziger, o limite físico pode
ser tomado sem problema algum, pois os termos de fontes puras não se
renormalizam, veja (\ref{6ren2}) e (\ref{7ren3}). No caso da fonte
LCO $J$, o termo quadrático, proporcional a $\zeta$ possui
contratermo correspondente a renormalização das funções de
correlação $\left<A^2(x)A^2(y)\right>$, carregando, assim,
divergências de vácuo. Desta forma, a ação a ser considerada deve
ser
\begin{equation}
S_J=S_{phys}+\int{d^4x}\left(\frac{J}{2}A_\mu^aA_\mu^a-\frac{\zeta}{2}J^2\right)\;,\label{7J1}
\end{equation}
onde $S_{phys}$ é dada pela ação física de Gribov-Zwanziger na forma
local, (\ref{6action2}). Para dar conta do termo quadrático em $J$,
introduzimos os campos de Hubbard-Stratonovich, $\sigma$, através da
introdução da unidade na forma (\ref{4hub1}). Tal transformacão
elimina o termo quadrático em $J$, permitindo assim o procedimento
padrão do cálculo de uma ação efetiva, sem fontes para atrapalhar. A
ação resultante é, portanto,
\begin{equation}
S'_J=S_{phys}+S_\sigma-\int{d^4x}\frac{J}{g}\sigma\;,\label{7J2}
\end{equation}
onde a ação $S_\sigma$ é dada por (\ref{4sigma1}). Como discutido em
\cite{Verschelde:2001ia,Knecht:2001cc,Browne:2003uv,Dudal:2003by,
Dudal:2005na}, o campo auxiliar de Hubbard-Stratonovich aparece
acoplado linearmente a fonte $J$, de forma que o operador
$A_\mu^aA_\mu^a$ está relacionado ao campo $\sigma$ através da
seguinte relação
\begin{equation}
\left\langle{A}_\mu^aA_\mu^a\right\rangle=-\frac{1}{g}\left\langle\sigma\right\rangle\;.\label{7cond2}
\end{equation}
A relação (\ref{7cond2}) nos diz que um valor esperado não nulo de
$\sigma$ resulta num valor não trivial para o condensado
$\left\langle A_\mu^a A_\mu^ a\right\rangle$. Uma vez resolvido o
problema do termo quadrático em $J$, podemos voltar ao trabalho de
calcular a ação efetiva. Assim, para tal, fazemos $J=0$ e
consideramos apenas os termos quadráticos em (\ref{7J2}), de forma
que a ação quântica a um laço se define como
\begin{equation}
e^{-\Gamma^{(1)}}=\int{DAD\bar{\varphi}D\varphi}e^{-S'_{quad}}\;,\label{7eff4}
\end{equation}
com a ação quadrática explícita dada por
\begin{eqnarray}
S'_{quad}&=&\int{d^4x}\left[ \frac{1}{4}\left(
\partial _{\mu
}A_{\nu }^{a}-\partial _{\nu }A_{\mu }^{a}\right) ^{2}+\frac{1}{2\alpha }%
\left( \partial_\mu A_\mu^{a}\right) ^{2}+\overline{\varphi }_{\mu
}^{ab}\partial
^{2}\varphi _{\mu }^{ab}\right.  \nonumber \\
&-&\left. \gamma ^{2}g\left( f^{abc}A_{\mu }^{a}\varphi _{\mu
}^{bc}+f^{abc}A_{\mu }^{a}\overline{\varphi }_{\mu }^{bc}\right)
-d(N^{2}-1)\gamma ^{4}\vphantom{\frac{1}{4}\left( \partial _{\mu
}A_{\nu }^{a}-\partial _{\nu }A_{\mu }^{a}\right)
^{2}}+\frac{\sigma^2}{2g^2\zeta}+\frac{1}{2}\frac{\sigma}{g\zeta}A_\mu^aA_\mu^a\right]\;,\nonumber\\
& &\label{7quad2}
\end{eqnarray}
onde o limite $\alpha\rightarrow0$ deve ser tomado no final para
recuperarmos o calibre de Landau. Lembramos ainda que, não há
integração em $\sigma$ pois estamos calculando o potencial efetivo
nos valores estáveis do mesmo.

Cabem agora algumas importantes observações. Primeiramente, vemos
que tomando o valor trivial do campo de Hubbard-Stratonovich,
$\sigma=0$, recaímos no caso da ação de Gribov-Zwanziger sem o
operador de glúon discutido na seção anterior. Por outro lado, o
limite $\gamma=0$, nos fornece o caso de Yang-Mills sem a restrição
ao horizonte de Gribov, recaindo assim no caso onde temos apenas o
efeito do condensado $\left<A_\mu^aA_\mu^a\right>$, discutido no
início da segunda parte desta tese. Lembramos que estes limites são
válidos apenas na ausência das equações de gap que fixam tais
quantidades a valores físicos da teoria, no caso a condição de
horizonte
\begin{equation}
\frac{\p\Gamma^{(1)}}{\p\gamma^2}=0\;.\label{7gap4}
\end{equation}
e a minimização do potencial de $\sigma$
\begin{equation}
\frac{\p\Gamma^{(1)}}{\p\sigma}=0\;.\label{7gap5}
\end{equation}
Estas condições não excluem as soluções triviais, porém, no caso da
condição de horizonte, tal solução implica na não restrição de
Gribov, o que não é aceitável na presente argumentação. O caso em
que o condensado de gluons é trivial foi estudado na seção anterior,
contudo, este caso se mostrou um tanto obscuro a um laço, uma vez
que a energia do vácuo se mostra positiva e o parâmtero de expansão
muito grande. Aqui vamos desconsiderar desde já a possibilidade de
soluções triviais, procuraremos, portanto, apenas por soluções não
triviais.

Outra importante observação diz respeito ao parâmetro LCO $\zeta$.
As equações de gap (\ref{7gap4}-\ref{7gap5}) não fixam este
parâmetro, deixando-o livre na teoria. Contudo, de acordo com o
discutido na segunda parte desta tese e em
\cite{Verschelde:2001ia,Knecht:2001cc,Browne:2003uv,Dudal:2003by,Dudal:2004rx},
$\zeta$ é fixado de forma que a ação quântica $\Gamma^{(1)}$ deva
obedecer a uma equação homogênea do grupo de renormalização,
\begin{equation}
\left[\omu\frac{\p}{\p\omu}+\beta(g^2)\frac{\p}{\p
g^2}+\gamma_{\gamma^2}(g^2)\gamma^2\frac{\p}{\p\gamma^2}+\gamma_\sigma(g^2)\sigma\frac{\p}{\p\sigma}\right]\Gamma=0\;,\label{7rg3}
\end{equation}
com
\begin{eqnarray}
\omu\frac{\p g^2}{\p\omu}&=&\beta(g^2)\;,\nonumber\\
\omu\frac{\p\gamma^2}{\p \omu} &=& \gamma_{\gamma^2}(g^2)\gamma^2 \;,\nonumber\\
\omu\frac{\p\sigma}{\p\omu}&=&\gamma_{\sigma}(g^2)\sigma\;.\label{7rg4}
\end{eqnarray}
Esta exigência é satisfeita se escolhemos $\zeta$ como uma função da
constante de acoplamento $g^2$, na forma
\begin{equation}
\zeta(g^2)=\frac{\zeta_0}{g^2}+\zeta_1+\zeta_2
g^2+\cdots\;.\label{7rge5}
\end{equation}

Retornando à ação efetiva a um laço (\ref{7eff4}), é fácil calcular
a expresão não renormalizada de $\Gamma^{(1)}$, em $d$ dimensões na
regularização dimensional,
\begin{equation}
\Gamma^{(1)}=-d\left( N^{2}-1\right) \gamma ^{4}+\frac{\sigma ^{2}}{2g^{2}\zeta }+%
\frac{N^{2}-1}{2}(d-1)\int \frac{d^{d}p}{\left( 2\pi \right)
^{d}}\ln \left[ p^{4}+2Ng^{2}\gamma ^{4}+\frac{g\sigma }{g^{2}\zeta
}p^{2}\right]\;.   \label{7eff5}
\end{equation}
de onde extraímos as expressões não renormalizadas das equações de
gap, a partir de (\ref{7gap4}-\ref{7gap5}),
\begin{eqnarray}
\frac{\sigma}{\zeta_0}\left(1-\frac{\zeta_1}{\zeta_0}g^2\right)+\frac{\left(N^2-1\right)}{2}\frac{g(d-1)}{\zeta_0}\int
\frac{d^dp}{\left(2\pi\right)^d}\frac{p^2}{p^4+\frac{g\sigma}{\zeta_0}p^2+2Ng^2\gamma^4}&=&0\;,\nonumber\\
\frac{d-1}{d}g^2N\int\frac{d^dp}{\left(2\pi\right)^d}\frac{1}{p^4+\frac{g\sigma}{\zeta_0}p^2+2Ng^2\gamma^4}&=&1\;.\label{7gap6}
\end{eqnarray}
Vemos que a segunda das (\ref{7gap6}) coincide com a equação do gap
obtida previamente em \cite{Sobreiro:2004us}, enquanto que a
primeira descreve a condensação de $A_\mu^aA_\mu^a$ quando a
restrição à região de Gribov é levada em consideração. Notamos que o
valor do condensado agora é influênciado pelo valor do parâmtero de
Gribov.

De acordo com \cite{Dudal:2005na}, depois de um elaborado cálculo no
esquema de renormalização $\MSbar$ em (\ref{7eff5}), chegamos,
finalmente, a\footnote{Observamos que a expresão (\ref{7eff6})
também é válida, ou seja, real, quando $m^4<\lambda^4$, se
$\ell_+(m,\lambda)$ e $\ell_-(m,\lambda)$, definidos como
\begin{eqnarray}
    \ell_+(m,\lambda)&=&\left(m^2+\sqrt{m^4-\lambda^4}\right)^2
\left(\ln\frac{m^2+\sqrt{m^4-\lambda^4}}{2\omu^2}-\frac{5}{6}\right)\;,\nonumber\\
\ell_-(m,\lambda)&=&\left(m^2-\sqrt{m^4-\lambda^4}\right)^2
\left(\ln\frac{m^2-\sqrt{m^4-\lambda^4}}{2\omu^2}-\frac{5}{6}\right)\;,\label{7l1}
\end{eqnarray}
são complexos conjugados. Onde utilizamos a relação
$\ln(z)=\ln|z|+i\arg(z)$ com $-\pi<\arg(z)\leq\pi$.}
\begin{eqnarray}
\Gamma^{(1)}&=&- \frac{\left( N^{2}-1\right)\lambda
^{4}}{2g^2N}+\frac{\zeta_0m^4}{2g^2}\left( 1-\frac{\zeta _{1}}{\zeta
_{0}}g^{2}\right) \nonumber\\&+&\frac{3\left( N^{2}-1\right)
}{256\pi^2}\left[\left(m^2+\sqrt{m^4-\lambda^4}\right)^2
\left(\ln\frac{m^2+\sqrt{m^4-\lambda^4}}{2\omu^2}-\frac{5}{6}\right)\right.\nonumber\\
&+&\left.\left(m^2-\sqrt{m^4-\lambda^4}\right)^2
\left(\ln\frac{m^2-\sqrt{m^4-\lambda^4}}{2\omu^2}-\frac{5}{6}\right)+\lambda^4\right]\;,\label{7eff6}
\end{eqnarray}
onde a notação foi simplicada para\footnote{Nenhuma confusão é
esperada com relação a fonte LCO $\lambda$ que já foi eliminada para
se efetuar o cálculo da ação efetiva.}
\begin{eqnarray}
\lambda^4&=&8g^2N\gamma^4\;,\\
m^2&=&\frac{g\sigma}{\zeta_0}\;.\label{7def1}
\end{eqnarray}
Nesta notação, as equaçoes de gap (\ref{7gap4}-\ref{7gap5}) passam a
ser expressas, respectivamente, como
\begin{eqnarray}
\frac{\p\Gamma^{(1)}}{\p\lambda}=0\;,\nonumber\\
\frac{\p\Gamma^{(1)}}{\p m^2 }=0\;.\label{7gap7}
\end{eqnarray}
Da expressão da ação efetiva a um laço renormalizada (\ref{7eff6})
podemos então, com o auxílo das (\ref{7gap7}), extrair as equações
do gap renormalizadas, a um laço,
\begin{eqnarray}
0&=&\frac{3Ng^2}{256\pi^2}\left[-2\frac{\left(m^2+\sqrt{m^4-\lambda^4}\right)}{\sqrt{m^4-\lambda^4}}\ln\frac{m^2+
\sqrt{m^4-\lambda^4}}{2\omu^2}\right.\nonumber\\
&+&\left.2\frac{\left(m^2-\sqrt{m^4-\lambda^4}\right)}{\sqrt{m^4-\lambda^4}}\ln\frac{m^2-\sqrt{m^4-\lambda^4}}{2\omu^2}+\frac{20}{3}\right]-1\;,
\label{7gap8a}
\end{eqnarray}
e
\begin{eqnarray}
0&=&\frac{\zeta_0m^2}{g^2}\left(1-\frac{\zeta_1}{\zeta_0}g^2\right)\nonumber\\
&+&\frac{3\left(N^2-1\right)}{256\pi^2}
\left[2\left(m^2+\sqrt{m^4-\lambda^4}\right)\left(1+\frac{m^2}{\sqrt{m^4-\lambda^4}}\right)\ln\frac{m^2+\sqrt{m^4-\lambda^4}}{2\omu^2}\right.\nonumber\\
&+&\left.2\left(m^2-\sqrt{m^4-\lambda^4}\right)\left(1-\frac{m^2}{\sqrt{m^4-\lambda^4}}\right)\ln\frac{m^2-\sqrt{m^4-\lambda^4}}{2\omu^2}-
\frac{8}{3}m^2\right]\;.\nonumber\\
\label{7gap8b}
\end{eqnarray}

Antes de tentarmos resolver as equações de gap
(\ref{7gap8a}-\ref{7gap8b}), vamos checar algumas propriedades da
ação quântica (\ref{7eff6}) sob o grupo de renormalização.

\subsection{Renormalization group}

Nesta seção vamos checar a invariância da ação quântica efetiva a um
laço (\ref{7eff6}) sob o grupo de renormalização. Para tal
necessitamos de algumas dimensões anômalas. Primeiramente, definimos
as dimensões anômalas dos novos parâmteros $m^2$ e $\lambda$,
definidos em (\ref{7def1}), de acordo com
\begin{eqnarray}
\omu\frac{\p m^2}{\p\omu}&=&\gamma_{m^2}(g^2)m^2\;,\nonumber\\
\overline{\mu}\frac{\p\lambda}{\p\overline{\mu}}&=&\gamma_\lambda(g^2)\lambda\;,\label{7diman2}
\end{eqnarray}
Utilizando as relações (\ref{7ren2}-\ref{7ren3}) e a definição
(\ref{7def1}), é fácil deduzir que
\begin{eqnarray}
\gamma_{m^2}(g^2)&=&\left[\frac{\beta(g^2)}{2g^2}-\gamma_A(g^2)\right]\;,\nonumber\\
\gamma_\lambda(g^2)&=&\frac{1}{4}\left[\frac{\beta(g^2)}{2g^2}+\gamma_A(g^2)\right]\;.\label{7diman3}
\end{eqnarray}
A primeira das (\ref{7diman3}) é a relação análoga da dimensão
anômala do operador $A_\mu^aA_\mu^a$, dada em (\ref{4diman1}). Vemos
que, notavelmente, as dimensões anômalas do operador
$A_\mu^aA_\mu^a$ e de $\lambda$ são diretamente proporcionais, a
todas as ordens em teorias de perturbações.

A ação quântica (\ref{7eff6}) deve, portanto, obedecer a seguinte
equação homogênea do grupo de renormalização
\begin{equation}
\left[\omu\frac{\p}{\p\omu}+\beta(g^2)\frac{\p}{\p
g^2}+\gamma_{\lambda}(g^2)\lambda\frac{\p}{\p\lambda}+\gamma_{m^2}(g^2)m^2\frac{\p}{\p{m^2}}\right]\Gamma^{(1)}=0\;.\label{7rg5}
\end{equation}
Assim, utilizando as relações de um laço da dimensão anômala do
glúon e da função beta, dadas por (\ref{7diman1}), é trivial checar
que
\begin{equation}
\omu\frac{d}{d\omu}\Gamma^{(1)}=0\;,
\end{equation}
confirmando a invariância sob o grupo de renormalização da ação
(\ref{7eff6}).

\subsection{Gap equations analysis}

Agora temos todas as ferramentas para resolvermos as equações de gap
(\ref{7gap8a}-\ref{7gap8b}) no esquema $\MSbar$. Contudo, é fácil
mostrar que neste esquema não existe solução para $m^2>0$. Para tal
vamos utilizar os valores explícitos de $\zeta_0$ e $\zeta_1$, dados
por (\ref{4zeta1}), que podem ser encontrados em
\cite{Verschelde:2001ia,Browne:2003uv,Dudal:2003by}. Seja agora a
variável
\begin{equation}
t=\frac{\lambda^4}{m^4}\;,\label{7t1}
\end{equation}
de forma que as (\ref{7gap8a}-\ref{7gap8b}) passam a ser escritas
como
\begin{eqnarray}
\frac{16\pi^2}{g^2N}&=&\frac{3}{8}\left(-2\ln\frac{m^2}{2\omu^2}+\frac{5}{3}+\frac{1}{\sqrt{1-t}}\ln\frac{t}{\left(1+\sqrt{1-t}\right)^2}-\ln
t\right)\;,\nonumber\\
-\frac{24}{13}\left(\frac{16\pi^2}{g^2N}\right)+\frac{322}{39}&=&4\ln\frac{m^2}{2\omu^2}-\frac{4}{3}-\frac{2-t}{\sqrt{1-t}}\ln\frac{t}{\left(1+\sqrt{1-t}
\right)^2}+2\ln t\;.\label{7gapt1}
\end{eqnarray}
Assim, combinando as duas (\ref{7gapt1}) para eliminarmos os
logarítimos e, consequentemente, a dependência na escala $\omu$,
chegamos à condição
\begin{equation}
\frac{68}{39}\left(\frac{16\pi^2}{g^2N}\right)+\frac{122}{39}=\frac{1}{2}\frac{t}{\sqrt{1-t}}\ln\frac{t}{\left(1+\sqrt{1-t}\right)^2}\;.\label{7gapt2}
\end{equation}
O lado esquerdo desta condição, assumindo um resultado consistente
($g^2>0$), é sempre positivo. Contudo, o lado direito de
(\ref{7gapt2}) é sempre negativo para $t>0$. Portanto, não temos
solução para $m^2>0$, independentemente da escala $\omu$.

\begin{figure}[h]
\begin{center}
  \scalebox{1}{\includegraphics{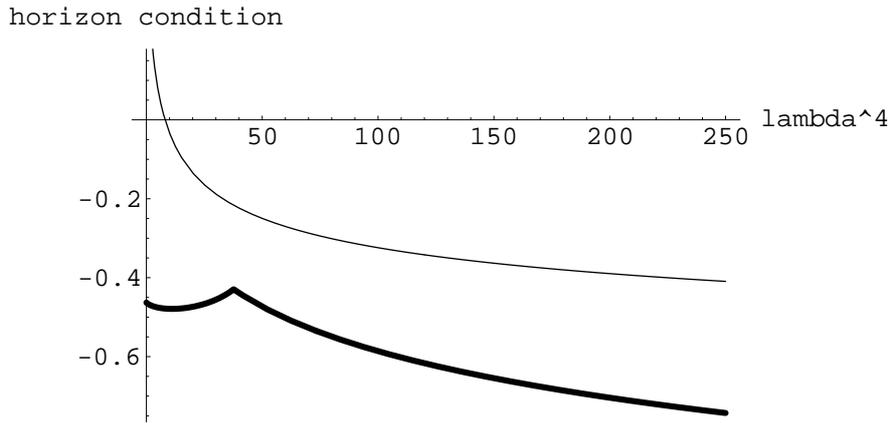}}
\caption{Condição de horizonte como função do parâmetro de Gribov em
unidades de $\lms$.}\label{7fig2}
\end{center}
\end{figure}

Podemos ainda assumir a possibilidade de haver uma solução para
$m^2<0$. Para tal, desenhemos a condição de horizonte (\ref{7gapt2})
como função de $\lambda^4$, veja Fig. \ref{7fig2}. Nesta figura a
curva fina corresponde ao caso  $m^2<0$, enquanto a curva grossa
representa as soluções para  $m^2>0$. Analizando esta figura,
concluímos claramente que não existe solução para $m^2>0$, uma vez
que a condição não atinge o zero. Este resultado geral está de
acordo com o obtido anteriormente, de forma independente de $\omu$.
Contudo, vemos que existe uma única solução para o caso $m^2<0$.

Os valores correspondentes para o parâmetro de expansão, o parâmetro
de Gribov e o condensado, bem como a energia do vácuo são, de fato,
\begin{eqnarray}
\label{msbaropLa}\frac{g^2N}{16\pi^2}&\approx&1.41\;,\nonumber\\
\label{msbaropLb} \lambda^4&\approx&8.063\lms^4\;,\nonumber\\
\label{msbaropLc}m^2&\approx&-0.950\lms^2\;,\nonumber\\
\label{msbaropld}E_\mathrm{vac}&\approx&0.047\lms^4\;.\label{7valores1}
\end{eqnarray}
Estes resultados não são, obviamente, satisfatórios pois o parâmetro
de expansão não é pequeno o suficiente para dar sentido a uma série
perturbativa. Ainda, a energia do vácuo permanece positiva, como no
caso sem condensação discutido na seção anterior. Contudo, é a única
solução possível no caso do esquema de renormalização $\MSbar$. O
próximo passo é tentar otimizar o esquema de renormalização de forma
a eliminar a dependência dos resultados encontrados no esquema de
renormalização escolhido, seja o esquema $\MSbar$ ou qualquer outro
esquema.

\subsection{Optimization of the renormalization scheme}

No caso em que a condensação do operador $A_\mu^aA_\mu^a$ não é
levada em consideração, mostramos que a energia do vácuo a um laço é
sempre positiva, independentemente da escala e do esquema de
renormalização. Contudo, no caso em que a condensação é estudada na
presença do horizonte de Gribov, apenas um estudo no esquema
$\MSbar$ foi efetuado. Vamos, portanto, tentar minimizar a
dependência dos resultados no esquema de renormalização escolhido.
Esse estudo é motivado pelo fato de que cálculos mais precisos (dois
ou mais laços) estão fora da ambição desta tese, dada a complicação
já presente nos cálculos a um laço aqui apresentados. Portanto, a
alteração no esquema de renorma-lização permanecerá na aproximação
de um laço. Assim, de alguma forma, num esquema de renormalização
diferente do $\MSbar$, o parâmetro de expansão poderia apresentar um
valor pequeno o suficiente para dar sentido a uma solução. Ou mesmo
a energia do vácuo poderia vir a assumir um valor negativo. Contudo,
tal otimização do esquema de renormalização consiste num longo
cálculo técnico, o que não é interessante sob o ponto de vista
físico. Desta forma, os detalhes técnicos do método podem ser
encontrados em \cite{Dudal:2005na} ou no Apêndice\footnote{Este
apêndice foi incorporado à tese pois possui definições e notações
necessárias para se entender os resultados que se seguem, bem como,
diferentemente de cálculos a um laço, é um método novo, que não está
presente na literatura usual.} \ref{ap_rensq}. Aqui, vamos nos
concentrar nos principais resultados obtidos após a otimização.

De acordo com o Apêndice \ref{ap_rensq}, o que, essencialmente, foi
feito foi substituir as quantidades $m^2$ e $\lambda^4$, na ação
efetiva (\ref{7eff6}), por seus correspondentes invariantes de
escala e de esquema de renormalização, ordem a ordem, $\wm^2$ e
$\wl^4$, respectivamente. A liberdade na escolha do esquema de
renormalização residual pode ser reduzida a um parâmetro único,
denotado por $b_0$, relacionado à renormalização da constante de
acoplamento. Desta forma, por construção, $\wm^2$ e $\wl^4$ devem
ser inependentes de $b_0$. Do mesmo modo, a energia do vácuo, sendo
uma quantidade física, deve ser independente de $b_0$. Este método
torna possível fixar o parâmtro $b_0$ requerendo uma dependência
mínima neste parâmetro.

Como primeiro exemplo escolhemos $b_0=0$, o que corresponde a
utilizar a constante de acoplamento do esquema $\MSbar$. Desta
forma, com a expansão otimizada, encontramos
\begin{eqnarray}
y&\equiv&\frac{N}{16\pi^2 x}\approx0.340\;,\nonumber\\
\wl^4x^{-2b}&\approx&15.66\lms^4\;,\nonumber\\
\wm^2x^{-a}&\approx&-1.40\lms^2\;,\nonumber\\
E_\mathrm{vac}&\approx&0.11\lms^4\;,\label{7opt1}
\end{eqnarray}
Vemos que a constante de acoplamento $y$ já é relativamente pequena,
menor que $1$, enquanto a energia do vácuo permanece positiva.

Agora, utilizando a quantidade $\Upsilon(b_0)$, como definida em
(\ref{apfunc1}), o valor ótimo para $b_0$ é dado por
$b_0^*\approx0.425$. Utilizando este valor, encontramos\footnote{A
definição de todas as quantidades podem ser encontradas no Apêndice
\ref{ap_rensq}.},
\begin{eqnarray}
y&\equiv&\frac{N}{16\pi^2 x}\approx0.047\;,\nonumber\\
\wl^4x^{-2b}&\approx&2.07\lms^4\;,\nonumber\\
\wm^2x^{-a}&\approx&-0.23\lms^2\;,\nonumber\\
E_\mathrm{vac}&\approx&0.019\lms^4\;,\label{7opt2}
\end{eqnarray}
Vemos que o parâmetro de expansão é satisfatoriamente pequeno neste
caso enquanto que a energia do vácuo permanece positiva. A conclusão
a que chegamos, portanto, é que não somos capazes de encontrar uma
solução satisfatória com energia do vácuo negativa, e
consequentemente, com condensado de glúon
$\left\langle{F}_{\mu\nu}^aF_{\mu\nu}^a\right\rangle$ positivo. Como
os cálculos foram feitos na aproximação de um laço, seria
interessante que fossem efetuados cálculos a dois ou mais laços,
para que uma resposta mais forte pudesse ser dada para o sinal de
$m^2$ e $\Evac$. Da mesma forma, tais cálculos forneceriam valores
numéricos mais confiáveis. Ainda, podemos argumentar sobre a
possibilidade de a presença do horizonte colocar a teoria num nível
altamente não-perturbativo, de forma que um algorítimo analítico
realmente não perturbativo fosse necessário. Infelizmente tal
conclusão não passa de especulação, uma vez que tal técnica
infravermelha não está a nossa disposição.

Lembramos que no esquema de renormalização $\MSbar$, a um laço, a
solução das equações do gap são, necessariamente, tais que
$\left\langle{A_\mu^aA_\mu^a}\right\rangle>0$. Por outro lado, sem a
restrição à região de Gribov, o valor encontrado foi
$\left\langle{A_\mu^aA_\mu^a}\right\rangle<0$, utilizando a técnica
LCO, \cite{Verschelde:2001ia,Browne:2003uv,Dudal:2003by}. Lembramos
também que em
\cite{Boucaud:2001st,Boucaud:2000nd,RuizArriola:2004en,Boucaud:2005rm,Furui:2005bu},
foi encontrado um valor positivo para
$\left\langle{A_\mu^aA_\mu^a}\right\rangle$, utilizando a expansão
em produto de operadores em combinação com o operador
$\left\langle{A_\mu^aA_\mu^a}\right\rangle$. Contudo, quando o
horizonte é levado em consideração, a massa dinâmica deixa de ser
interpretado como uma massa efetiva, uma vez que os pólos do
propagador do glúon passam a depender de uma mistura entre o
parâmetro de Gribov e o condensado
$\left\langle{A_\mu^aA_\mu^a}\right\rangle$. Tal combinação consiste
num número que, possivelmente é complexo, o que torna inválida a
interpretação dos pólos como massa efetiva e do glúon como uma
partícula. Vamos expandir esta questão na próxima seção.

Observamos ainda que, no caso do modelo de Gribov-Zwanziger puro, o
resultado de que a energia do vácuo é sempre positiva continua
valendo, mesmo com a otimização do esquema de renormalização. Veja
\cite{Dudal:2005na}.

\section{Discussion}

Vamos discutir agora as possíveis consequências da existência de
valores não nulos para o parâmetro de Gribov e do condensado
$\left<A_\mu^aA_\mu^a\right>$. Comecemos com os propagadores da
teoria.

\subsection{Propagators}

Se não consideramos o condensado $\left<A_\mu^aA_\mu^a\right>$,
recaímos no caso discutido no capítulo anterior e podemos considerar
apenas a ação de Gribov-Zwanziger (\ref{6action2}). Neste caso, o
propagador do glúon, no nível árvore é dado pela expressão
(\ref{6gluon1}), como pode ser calculado através da aproximação de
Gribov, \cite{Gribov:1977wm}. Em particular, a presença do parâmetro
de Gribov implica na nulidade do propagador do glúon a momento nulo.

Quando incluímos os efeitos da condensação do operador
$A_\mu^aA_\mu^a$, o propagador do glúon se torna, no nível árvore,
\begin{equation}
D_{\mu\nu}^{ab}(q)=\delta^{ab}\frac{\mathcal{D}(q^2)}{q^2}\left(\delta_{\mu\nu}-\frac{q_\mu{q}_\nu}{q^2}\right)=
\delta^{ab}\frac{q^2}{q^4+m^2q^2+\frac{\lambda^4}{4}}\left(\delta_{\mu\nu}-\frac{q_\mu{q}_\nu}{q^2}\right)\;.\label{7prop1}
\end{equation}
Este tipo de propagador, é conhecido como propagador de Stingl,
devido ao autor que utilizou este propagador pela primeira vez como
um anzätz para resolver as equações de Schwinger-Dyson, veja
\cite{Stingl:1985hx} para mais detalhes.

Ainda, ressaltamos que comparações reais de $\mathcal{D}(p^2)$ com
resultados obtidos com simu-lações numéricas na rede
\cite{Bonnet:2001uh,Langfeld:2001cz,Bloch:2003sk}, está ainda fora
de questão. Isso porque deveríamos ir além da aproximação do nível
árvore, por exemplo, incluindo efeitos de polarização a ordens
maiores e/ou tentando um aperfeiçoamento do tratamento via grupo de
renorma-lização. Em geral essas correções também dependem do momento
externo.

Com relação ao propagador dos campos fantasmas, estes podem ser
calculados a um laço, utilizando-se o propagador do glúon na forma
(\ref{6gluon1}) ou (\ref{7prop1}) em conjunto com suas respectivas
equações do gap (\ref{6gap2}) e a segunda das (\ref{7gap6}). Em
ambos os casos, veja
\cite{Gribov:1977wm,Zwanziger:1989mf,Zwanziger:1992qr,Sobreiro:2004us,Sobreiro:2004yj,Dudal:2005na},
encontramos para o propagador dos campos fantasmas um comportamento
infravermelho singular na forma
\begin{equation}
\frac{\delta^{ab}}{N^2-1} \left\langle
c^a\oc^b\right\rangle_{q\approx0}\equiv\left.\frac{1}{q^2}\mathcal{G}(q^2)\right|_{p\approx0}\approx\frac{1}{q^4}\;.\label{7prop2}
\end{equation}
Desta expressão vemos que o propagador fantasma é mais singular que
a predição perturbativa, devido a presença do horizonte. Em
particular, esta propriedade está diretamente relacionada com o
teorema de não renormalização do parâmetro de Gribov, (\ref{7ren3}),
que diz que o termo puro em $\gamma$, proveniente do termo puro nas
fontes, não se renormaliza. Essa propriedade garante este
comportamento do propagador fantasma a mais ordens, veja
\cite{Gracey:2005vu,Gracey:2006dr} para cálculos a dois laços. Esta
propriedade do propagador dos campos fantasmas foram confirmadas
também com cálculos numéricos em simulações na rede, veja
\cite{Bloch:2003sk}.

Observamos ainda que, da mesma forma que o propagador do glúon, uma
discussão mais detalhada, a mais laços, do propagador fantasma deve
ser efetuada de forma a obtermos expressões para se comparar com
outros resultados.

\subsection{Coupling constant}

É usual utilizar como uma definição não perturbativa da constante de
acoplamento forte renormalizada $\alpha_R$,
\cite{vonSmekal:1997is,Bloch:2003sk}, a expressão
\begin{equation}
\alpha_R(q^2)=\alpha_R(\mu)\mathcal{D}(q^2,\mu)\mathcal{G}^2(q^2,\mu)\;,\label{7coup1}
\end{equation}
onde $\mathcal{D}$ e $\mathcal{G}$ são os fatores de forma do
propagador do glúon e dos campos fantasmas, respectivamente. Esta
definção é um tipo de extensão não perturbativa do resultado
perturbativo (\ref{4ren4}). De acordo com os resultados obtidos
através do estudo das equações de Schwinger-Dyson,
\cite{Atkinson:1997tu,Atkinson:1998zc,Alkofer:2000wg,Watson:2001yv,Zwanziger:2001kw,Lerche:2002ep},
tais fatores de forma devem satisfazer a seguinte lei de potências,
no limite infravermelho,
\begin{eqnarray}
\lim_{p\rightarrow0}\mathcal{D}(p^2)&\propto&\left(p^2\right)^\theta\;,\nonumber\\
\lim_{p\rightarrow0}\mathcal{G}(p^2)&\propto&\left(p^2\right)^\omega\;,\label{7form1}
\end{eqnarray}
onde os expoentes infravermelhos $\theta$ e $\omega$ obedecem a
seguinte regra de soma
\begin{equation}
\theta+2\omega=0\;.\label{7sum1}
\end{equation}
Tal regra de soma sugere o aparecimento de um ponto fixo para a
constante de acoplamento renormalizada, (\ref{7coup1}), como já
sugerido por simulações na rede,
\cite{Bloch:2003sk,Furui:2003jr,Furui:2004cx}. No nosso caso, os
resultados para os fatores de forma do glúon e dos campos fantasmas
na aproximação a ordem zero, obedecem a regra de soma (\ref{7sum1}),
com $\theta=2$ e $\omega=-1$. Note que sem o parâmetro de Gribov, a
regra (\ref{7sum1}) desaparece, bem como a evidêncoa do ponto fixo
para a constante renormalizada.

\subsection{Positivity violation}

É comum utilizar o propagador do glúon como indicador do
confinamento das excitações gluônicas através da chamada {\it
violação da positividade}, veja
\cite{Alkofer:2003jj,Cucchieri:2004mf} e referências contidas. O
conceito de {\it positividade} é o que se segue.

O propagador do glúon Euclideano,
$D(q)\equiv\frac{\mathcal{D}(q^2)}{q^2}$ pode, sempre, ser escrito
através da representação espectral
\begin{equation}
D(q)=\frac{\mathcal{D}(q)}{q^2}=\int_0^{+\infty}dM^2\frac{\rho(M^2)}{q^2+M^2}\;.\label{7spectr1}
\end{equation}
A densidade espectral $\rho(M^2)$ deve sempre ser positiva, de forma
a possuir uma representação de K\"{a}llen-Lehmann, permitindo assim,
a interpretação de campos em termos de partículas estáveis. Pode-se
ainda definir o {\it correlator temporal}, \cite{Cucchieri:2004mf},
\begin{equation}
\mathcal{C}(t)=\int_0^{+\infty}dM\rho(M^2)e^{-Mt}\;,\label{7spectr2}
\end{equation}
Desta definição vemos que se $\mathcal{C}(t)>0$, então, sempre,
$\rho(M^2)>0$. Contudo, se para algum $t$, $\mathcal{C}(t)<0$ então
$\rho(M^2)$ não pode ser sempre positivo. Podemos converter a
expressão (\ref{7spectr2}) para possuir argumento dependente do
propagador do glúon (\ref{7spectr1}),
\begin{equation}
\mathcal{C}(t)=\frac{1}{2\pi}\int_{-\infty}^{+\infty}e^{-ipt}D(p)dp\;.\label{7spectr3}
\end{equation}

Utilizando a expressão (\ref{7prop1}) do propagador do glúon,
podemos analizar caso a caso da positividade do glúon:
\begin{itemize}
\item Se $\lambda=0$, então $m^2>0$, e
\begin{equation}
\mathcal{C}(t)=\frac{e^{-mt}}{2m}>0\;.\label{7ct1}
\end{equation}
\item Se $m^2=0$,
\begin{equation}
\mathcal{C}(t)=\frac{e^{-\frac{Lt}{2}}}{2L}\left(\cos\frac{Lt}{2}-\sin\frac{Lt}{2}\right)\;,\label{7ct2}
\end{equation}
que assume valores negativos para certos valores de $t$.
\item Nos demais casos temos
\begin{equation}
\mathcal{C}(t)=\frac{1}{2}\left[\frac{\sqrt{\omega_1}}{\omega_1-\omega_2}e^{-\sqrt{\omega_1}t}
+\frac{\sqrt{\omega_2}}{\omega_2-\omega_1}e^{-\sqrt{\omega_2}t}\right]\;,\label{7ct3}
\end{equation}
onde $\omega_1$ e $\omega_2$ são os pólos do propagador
(\ref{7prop1}), com a convenção de possuir parte real positiva,
\begin{equation}
\omega_i=\frac{m^2+(-1)^{i}\sqrt{m^4-\lambda^4}}{2}\;,\;\;\;i\;\in\;\{1,2\}\;.\label{7polos1}
\end{equation}
Não é difícil deduzir que a expressão (\ref{7ct3}) sempre pode vir a
assumir valores negativos. Veja \cite{Dudal:2005na}.
\end{itemize}
Concluímos que, quando a restrição à região de Gribov é feita, a
função $\mathcal{C}(t)$ sempre exibe violação de positividade,
quando considerado o propagador do glúon ao nível árvore, com ou sem
o condensado $\left\langle{A}_\mu^aA_\mu^a\right\rangle$.

\chapter{Linear covariant gauges}

\section{Motivation and introduction}

Neste capítulo vamos discutir brevemente a generalização do
tratamento das ambigüidades de Gribov para o caso dos calibres
lineares covariantes (CLC). O principal estímulo desta análise é o
fato de este estudo nunca ter sido efetuado anteriormente.
Obviamente, existem estudos feitos nestes calibres, mas não
diretamente tratando a questão das ambigüidades de Gribov. Em
particular, chamamos atenção para trabalhos feitos numericamente na
rede, \cite{Giusti:1996kf,Giusti:1999im,Giusti:2000yc}, onde foi
encontrado evidências de um propagador gluônico (transverso e
longitudinal) suprimido no infravermelho. Também, existem estudos
deste calibre utilizando-se as equações de Schwinger-Dyson,
\cite{Alkofer:2003jr}, onde foi encontrado um propagador do glúon
também suprimido no limite de baixas energias, enquanto que o
propagador fantasma apresenta o típico comportamento singular de
Gribov, \cite{Gribov:1977wm}.

Ainda, os CLC são a generalização natural, mais simples, do calibre
de Landau, o que torna este estudo, no mínimo, interessante.
Ademais, uma visão global das ambigüidades de Gribov,
independentemente do calibre escolhido ainda não está disponível.
Somente recentemente, se iniciou tais estudos, em calibres que não o
Landau ou Coulomb. Em particular, trabalhos foram desenvolvidos no
MAG, \cite{Quandt:1997rg,Bruckmann:2000xd,Capri:2005zj}, e nos CLC,
\cite{Sobreiro:2005vn}. Tal interesse da comunidade física pode ser
visto como o próximo passo a uma compreensão completa da quantização
das teorias de Yang-Mills e teorias de calibre em geral.

O calibre de Landau permite que a restrição à primeira região de
Gribov seja efetuada devido ao fato de o operador de Faddeev-Popov
ser hermitiano, possuindo assim, autova-lores reais. Ao irmos para o
caso mais geral em que consideramos os CLC esta propriedade é
perdida, pois o glúon deixa de ser exclusivamente transverso. Ainda,
ao contrário do calibre de Landau, nos CLC não se conhece um
funcional cuja minimização defina os mesmos. Contudo, ainda podemos
estabelecer propriedades das ambigüidades de Gribov e consequências
físicas para a eliminação das cópias neste calibre,
\cite{Sobreiro:2005vn}, incluíndo ainda os efeitos da geração
dinâmica de massa \cite{Dudal:2003np,Dudal:2003by}. Os efeitos
físicos aparecem nos propagadores. O propagador do glúon se torna
suprimido na região infravermelha, sendo que a componente transversa
é suprimida devido à restrição de domínio do espaço funcional das
configurações e reforçada pela presença da massa dinâmica enquanto a
componente longitudinal é suprimida devido à massa dinâmica apenas.
O propagador dos campos de Faddeev-Popov não apresenta singularidade
do tipo $1/k^4$ pois não está associado ao funcional de restrição.
Em contrapartida, uma função de Green com a singularidade $1/k^4$ é
identificada.

Os detalhes deste capítulo podem ser encontrados em
\cite{Sobreiro:2005vn}.

\section{Identification of a restriction region}

Como no caso de Landau, é possível determinar uma região no espaço
das configurações que pode ser usada para eliminar um grande número
de cópias de Gribov. Primeiramente vamos escrever a ação de
Yang-Mills quantizada nos calibres lineares covariantes,
\begin{equation}
S_{CLC}=S_{YM}+\int
d^{4}x\left( b^{a}\partial A^{a}+\frac{\alpha }{2}b^{a}b^{a}+\overline{c}%
^{a}\mathcal{M}^{ab} c^{b}\right)\;,\label{8eq1}
\end{equation}
cujo limite $\alpha\rightarrow0$ recupera o calibre de Landau.
Lembramos que o operador de Faddeev-Popov não é hermitiano neste
caso. A integral de caminho de Faddeev-Popov, portanto, pode ser
escrita como
\begin{equation}
Z=\int{DA}DbD\bar{c}Dc\exp\left\{-S_{CLC}\right\}\;.\label{8eq2}
\end{equation}
A condição de calibre é dada por
\begin{equation}
\partial_\mu{A}_\mu^a=-\alpha{b}^a\;,\label{8eq3}
\end{equation}
que produz as mesmas equações de cópias que o calibre de Landau,
(\ref{6copias1}-\ref{6copias3}), exceto pelo fato de
$\mathcal{M}^{ab}$ não ser hermitiano.

Vamos agora enunciar algumas propriedades das cópias de Gribov no
caso dos CLC. Tais propriedades nos permitem identificar uma região
relacionada à região de Gribov. Essa região pode ser utilizada para
estabelecer um corte no domínio de integração da integral de
caminho.
\begin{itemize}
\item  Propriedade 1: {\it Se a componente transversa $A_\mu^{aT}$ de uma dada configuração
$A_\mu^a=\left(A_\mu^{aT}+A_\mu^{aL}\right)$ pertence à região de
Gribov $C_0$, então o operador de Faddeev-Popov $\mathcal{M}
^{ab}(A)$ não possui autovalores nulos, \textit{i.e.}, se}
\begin{equation}
A_\mu^{aT}\;\in\;C_0\Rightarrow\mathcal{M}^{ab}\ne0\;.\label{8prop1}
\end{equation}

A prova desta propriedade é feita assumindo-se o contrário, ou seja,
supondo que exista um modo zero do operador de Faddeev-Popov se a
componente transversa do campo de calibre pertencer à região de
Gribov. Feito isso, não é difícil mostrar que, para calibres
próximos ao de Landau\footnote{Tudo o que será feito daqui para
frente nos CLC será na aproximação $\alpha\ll1$.}, $\alpha\ll1$,
essa suposição é impossível. Com esta propriedade podemos definir
uma região, $\Omega$, da seguinte forma
\begin{equation}
\Omega\equiv\left\{A_\mu^a\;\big|\;A_\mu^a=\;A_\mu^{aT}+A_\mu^{aL},\;A_\mu^{aT}\in\,C_0
\right\}\;,\label{8r1}
\end{equation}
ou, equivalentemente
\begin{equation}
\Omega\equiv\left\{A_\mu^a\;\big|\;A_\mu^a=\;A_\mu^{aT}+A_\mu^{aL},\;\mathcal{M}^{abT}>0\right\}\;.\label{8r2}
\end{equation}
onde,
\begin{equation}
\mathcal{M}^{abT}=-\partial_\mu\left(\delta^{ab}\partial_\mu-gf^{abc}A_\mu^{cT}\right)\;.\label{8op1}
\end{equation}
Outra forma de escrever a propriedade 1 é, portanto,
\begin{equation}
A_\mu^{aT}\;\in\;C_0\Rightarrow{A}_\mu^a\;\in\;\Omega\;.
\end{equation}
Desta propriedade segue, trivialmente a seguinte.

\item Propriedade 2: {\it Se a componente transversa $A_\mu^{aT}$ de uma configuração
$A_\mu^a$ pertence à região de Gribov $C_0$, então o campo $A_\mu^a$
não possui cópias de Gribov infinitesimais.}

A prova é imediatamente trivial, uma vez que se entende a
propriedade anterior. Esta propriedade nos diz que a restrição à
região $\Omega$ elimina todas a cópias que podem ser obtidas através
de transformações infinitesimais.

\item Propriedade 3: {\it No limite $\alpha$ que vai a zero, a região $\Omega$ se reduz à
região de Gribov $C_0$.}
\begin{equation}
\lim_{\alpha\rightarrow0}\Omega=C_0\;.\label{8lim1}
\end{equation}

Esta propriedade é também imediata pois, no limite
$\alpha\rightarrow0$, a componente longitudinal de $A_\mu^a$ vai a
zero. Desta forma podemos entender a região $\Omega$ como uma
deformação da região de Gribov.
\end{itemize}

\section{Implementation of the restriction}

As propriedades discutidas na seção anterior sugerem e motivam uma
restrição no domínio de integração considerando apenas o domínio em
$\Omega$, definido em (\ref{8r1}-\ref{8r2}). A integral de caminho
se torna, portanto,
\begin{equation}
Z=\int{DA}DbD\bar{c}Dc\exp\left\{-S_{CLC}\right\}\mathcal{V}(\Omega)\;.\label{8eq4}
\end{equation}
onde, $\mathcal{V}(\Omega)$ é responsável pela restrição à região
$\Omega$.
\begin{figure}[ht]
\centering \epsfig{file=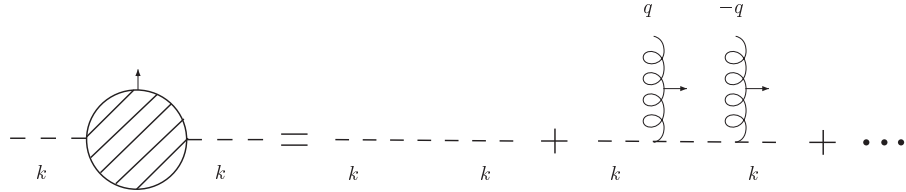,width=12cm}
\caption{Propagador fantasma com contribuição de gluons transversais
externos.}\label{8fig1}
\end{figure}Como no caso de Landau utilizaremos a condição de
ausência de pólos para caracterizar $\mathcal{V}(\Omega)$,
\cite{Gribov:1977wm}. Contudo, como a definição de $\Omega$,
(\ref{8r1}-\ref{8r2}), não depende do operador de Faddeev-Popov,
devemos utilizar outra quantidade. Fortuitamente, a definição
(\ref{8r1}-\ref{8r2}), depende apenas das configurações transversas,
através do operador $\mathcal{M}^{abT}$, que claramente é
hermitiano. Assim, repetimos a análise do propagador do ghost para o
caso do operador $\left(\mathcal{M}^{abT}\right)^{-1}$, representada
na Fig. \ref{8fig1}\footnote{A notação diagramática da decomposição
do glúon em componentes transversal e longitudinal está sendo
utilizada da seguinte forma: linhas de gluons com uma seta
transversal à direção de propagação do glúon representam gluons
transversais enquanto linhas de gluons com setas tangenciais à
direção de propagação representam as componentes longitudinais do
glúon.}, obtendo assim, a condição de restrição
\begin{equation}
\frac{Ng^2}{4(N^2-1)}\int\frac{d^4k}{(2\pi)^4}\frac{A_\mu^{aT}(k)A_\mu^{aT}(-k)}{k^2}<1\;.\label{8nopole1}
\end{equation}
Esta condição implica no seguinte funcional de restrição
\begin{equation}
\mathcal{V}(C_0)=\exp\left\{-Ng^2\gamma^4\int\frac{d^4k}{(2\pi)^4}\frac{A_\mu^{aT}(k)A_\mu^{aT}(-k)}{k^2}+
4(N^2-1)\gamma^4\right\}\;,\label{8rest1}
\end{equation}
onde o parâmetro de Gribov, novamente, é determinado através do
requerimento de que a energia do vácuo dependa minimamente deste
parâmetro. À primeira ordem, a equação do gap coincide com a equação
do gap no calibre de Landau,
\begin{equation}
\frac{3}{4}Ng^2\int\frac{d^4k}{(2\pi)^4}\frac{1}{k^4+2Ng^2\gamma^4}=1\;.\label{8gap1}
\end{equation}
Note que este é um resultado esperado uma vez que a condição de
ausência de pólos leva em consideração apenas a contribuição
transversa do operador de Faddeev-Popov.

\section{Propagators}

Um cálculo simples do propagador do glúon no nível árvore, com a
integral de caminho (\ref{8eq4}), levando em consideração o
funcional de restrição (\ref{8rest1}), nos leva a
\begin{equation}
D_{\mu\nu}^{ab}(k)=\delta^{ab}\left[\frac{k^2}{k^4+2Ng^2\gamma^4}\left(\delta_{\mu\nu}-
\frac{k_{\mu}k_\nu}{k^2}\right)+\frac{\alpha}{k^2}\frac{k_\mu{k}_\nu}{k^2}\right]\;.\label{8gluon1}
\end{equation}
Primeiramente notamos que a componente transversa  do propagador
(\ref{8gluon1}) é suprimida no limite infravermelho enquanto a
componente longitudinal se mantém singular. Ainda, no limite
$\alpha\rightarrow0$, os resultado do calibre de Landau são
recuperados, (\ref{6gluon1}).

\begin{figure}[ht]
\centering \epsfig{file=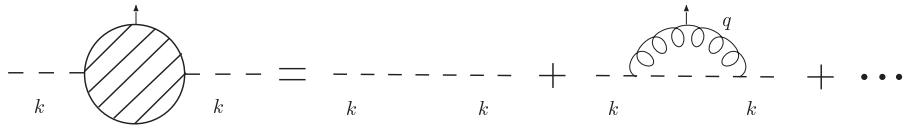,width=12cm}
\caption{Propagador fantasma com contribuição de gluons
transversais.}\label{8fig2}
\end{figure}

\begin{figure}[ht]
\centering \epsfig{file=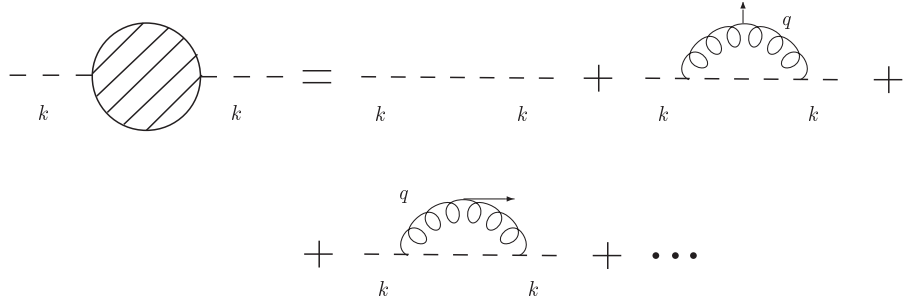,width=12cm}
\caption{Propagador fantasma com contribuição de gluons transversais
e longitudinais.}\label{8fig3}
\end{figure}

Com relação à quantidade $\left(\mathcal{M}^{abT}\right)^{-1}$,
esperamos que seja bastante singular na região infravermelha, devido
ao efeito da restrição. De fato, um cálculo a um laço utilizando a
integral de caminho (\ref{8eq4}), levando em consideração o
funcional de restrição (\ref{8rest1}) e fazendo uso da equação do
gap (\ref{8gap1}), é fácil mostrar que
\begin{equation}
\frac{1}{N^2-1}\left[\left(\mathcal{M}^T(k)\right)^{-1}\right]^{aa}\bigg|_{k\rightarrow0}\propto\frac{1}{k^{4}}\;,
\label{8T1}
\end{equation}
conforme esquematizado na Fig. \ref{8fig2}. Evidenciando a presença
de forças de longo alcance, necessárias para estabelecer o
confinamento.
Para ilustrarmos o fato de que a restrição não se relaciona com o
limite infravermelho do propagador dos campos fantasmas, podemos
calcular o mesmo utilizando o esquema de renormalização
$\overline{MS}$, representado na Fig. \ref{8fig3}
\begin{equation}
\frac{1}{N^2-1}\mathcal{G}^{aa}(k)\bigg|_{k\rightarrow0}\propto
\frac{1}{k^4}\frac{1}{\left(1-\alpha\frac{2\gamma^2}{3\pi{k}^2}\log\frac{k^2}{\bar{\mu}^2}\right)}\;,\label{8gh1}
\end{equation}
onde $\bar{\mu}$ é a massa de renormalização no esquema
$\overline{MS}$.
Note que o termo logarítimico vem de contribuições de configurações
longitudinais, como é evidente pela presença de $\alpha$. Portanto,
não temos uma singularidade do tipo Gribov-Zwanziger no propagador
dos campos fantasmas.

\section{Inclusion of dynamical mass generation effects}

De acordo com \cite{Dudal:2003by}, a geração dinâmica de massa nos
CLC é descrita através da ação
\begin{equation}
S=S_{CLC}+S_\sigma\;,\label{8m1}
\end{equation}
where $S_{CLC}$ é dada por (\ref{8eq1}) e o termo $S_\sigma$ em
(\ref{8m1}) contém o campo auxiliar de Hubbard-Stratonovich,
$\sigma$, e é dado por
\begin{equation}
S_\sigma=\int
d^4x\left[\frac{\sigma^2}{2g^2\zeta}+\frac{1}{2}\frac{\sigma}{g\zeta}A_\mu^aA_\mu^a+
\frac{1}{8\zeta}\left(A_\mu^aA_\mu^a\right)^2\;\right]\;.\label{8m2}
\end{equation}
A introdução do campo $\sigma$ permite o estudo da condensação do
operador $A_\mu^aA_\mu^a$, gerando massa para o glúon na forma
\begin{equation}
m^{2}=\frac{\left\langle\sigma\right\rangle}{g\zeta}\;. \label{8m4}
\end{equation}
É importante ressaltar o fato de que a ação
$\left(S_{YM}+S_\sigma\right)$ é invariante de calibre, de acordo
com as transformações
\begin{eqnarray}
\delta{A}_\mu^a&=&-D_\mu^{ab}\omega^b\;,\nonumber\\
\delta\sigma&=&gA_\mu^a\partial_\mu\omega^a\;.\label{8gauge1}
\end{eqnarray}
Desta forma, a ação (\ref{8m1}), nos conduz a uma integral de
caminho ainda contaminda por cópias de Gribov. Ainda, como as cópias
de Gribov estão relacionadas apenas com a condição de calibre
(\ref{8eq3}), o termo de massa não afeta as equações de cópias
(\ref{6copias1}-\ref{6copias3}), assim como a condição de ausência
de pólos (\ref{8nopole1}). Esta análise nos leva, portanto, ao mesmo
funcional de restrição (\ref{8rest1}) considerado para o caso não
massivo. Portanto, nesta aproximação o problema das ambigüidades de
Gribov comuta com a geração dinâmica de massa. A integral de caminho
resultante é
\begin{equation}
Z=\int{DA}DbD\bar{c}Dc\exp\left\{-S\right\}\mathcal{V}(\Omega)\;.\label{8eq2a}
\end{equation}

Contudo, a presença da massa dinâmica afeta a forma do propagador do
glúon, que, é alterado para
\begin{equation}
D_{\mu\nu}^{ab}(k)=\delta^{ab}\left[\frac{k^2}{k^4+m^2k^2+2Ng^2\gamma^4}\left(\delta_{\mu\nu}-
\frac{k_{\mu}k_\nu}{k^2}\right)+
\frac{\alpha}{k^2+\alpha{m}^2}\frac{k_{\mu}k_\nu}{k^2}\right]\;.\label{8gluon2}
\end{equation}
Vemos que a componente transversa se mantém suprimida na região
infravermelha, onde a massa reforça essa supressão. A componenete
longitudinal, que antes não apresentava supressão infravermelha
(\ref{8gluon1}), agora, devido à massa, se torna suprimida no limite
de baixas energias. Note ainda que o parâmetro de Gribov continua
sendo determinado por uma equação do gap, agora, na forma
\begin{equation}
\frac{3}{4}Ng^2\int\frac{d^{4}q}{\left(2\pi\right)^4}\frac{1}{q^4+m^2q^2+2Ng^2\gamma^4}=1\;.
\label{8gap2}
\end{equation}
Esta equação é diferente da anterior, (\ref{8gap1}), devido à
presença da massa. No entanto, é determinada da mesma forma,
requerendo que a energia do vácuo seja mínima com relação ao
parâmetro de Gribov. Na aproximação atual (um laço) o termo de massa
já mostra sua influência nessa condição. Note ainda que esta
expressão coincide com o caso de Landau com geração dinâmica de
massa, \cite{Sobreiro:2004us,Sobreiro:2004yj,Dudal:2005na}, veja
expressão (\ref{8gap1}).\ Por fim, temos a quantidade
$(\mathcal{M}^T)^{-1}$, que permanece exibindo o típico
comportamento singular de Gribov, (\ref{8T1}). Esse comportamento
não é afetado, portanto, pela presença da massa dinâmica.

\section{Discussion}

Nas teorias de Yang-Mills quantizadas nos calibres lineares
covariantes, fomos capazes de identificar uma região no espaço
funcional das configurações a qual é livre de cópias de Gribov
infinitesimais. Tal região, denotada $\Omega$, foi definida na
expressão (\ref{8r1}-\ref{8r2}). Ainda, fomos capazes de identificar
uma função de Green associada à fronteira de $\Omega$ que pode ser
utilizada para implementar a região na integral de caminho
(\ref{8eq2}). Não obstante, os efeitos de geração dinâmica de massa
puderam ser levados em conta pois a ação efetiva que descreve a
condensação do operador $A_\mu^a{A}_\mu^a$, (\ref{8m1}), é, de fato,
invariante de calibre. Implementando a restrição à região $\Omega$
na integral de caminho (\ref{8eq2a}), que descreve as teorias de
Yang-Mills nos CLC com geração dinâmica de massa, temos como efeito
físico a modificação dos propagadores da teoria. Note que estes
resultados são válidos apenas na aproximação de CLC próximos ao de
Landau, $\alpha\ll1$.

No caso do propagador do glúon (\ref{8gluon2}), temos que ambas
componentes, transversal e longitudinal, são suprimmidas no limite
infravermelho. A componente transversal exibe um comportamento tipo
Stingl \cite{Stingl:1985hx,Stingl:1985hx2}, onde a supressão ocorre
devido à presença do parâmetro de Gribov e é reforçada pela presença
da massa. A componente longitudinal, por sua vez, também exibe
supressão infravermelha, contudo, sua causa ocorre exclusivamente
devido à massa. Este resultado está em acordo qualitativo com
resultados obtidos através de simulações numéricas na rede
\cite{Giusti:1996kf,Giusti:1999im,Giusti:2000yc}, bem como
resultados provenientes de soluções das equações de Schwinger-Dyson
\cite{Alkofer:2003jr}.

No caso do propgador dos campos fantasmas, o comportamento singular
infravermelho $1/k^4$, presente no calibre de Landau, é perdido nos
CLC. Esse efeito é compreensível uma vez que se entende que o
operador de Faddeev-Popov não mais se relaciona com a região de
Gribov neste caso. Ao invés do propagador dos campos de
Faddeev-Popov, a função de Green singular, que evidencia a presença
de forças de longo alcance foi definida em (\ref{8op1}) e consiste
no propagador dos campos fantasmas com contribuição a um laço apenas
da componente transversa do glúon. Note que, até nosso conhecimento,
não existem dados na rede sobre o propagador fantasma ou da função
$\mathcal{M}^T$. A última, também está indisponível no caso das
equações de Schwinger-Dyson. Contudo, no caso das equações de
Schwinger-Dyson o propagador fantasma foi calculado utilizando-se um
método de corte de vértices nus, \cite{Alkofer:2003jr}. Como
resultado, foi encontrado um propagador singular tipo Gribov,
$1/k^4$. Este resultado está em desacordo com o resultado aqui
apresentado, (\ref{8gh1}). Mas, o resultado da Schwinger-Dyson,
\cite{Alkofer:2003jr}, consiste num {\it Ansätz} ousado pois o
método de corte de vértices nus se mostra válido apenas no calibre
de Landau,
\cite{vonSmekal:1997is,Atkinson:1997tu,Atkinson:1998zc,Watson:2001yv,Zwanziger:2001kw,Lerche:2002ep},
pois utiliza-se o teorema de não-renormalização do vértice
glúon-fantasma-antifantasma, \cite{Blasi:1990xz,Cucchieri:2004sq},
que só é estabelecido no calibre de Landau.

Assim, dois aspectos não-perturbativos das teoria de Yang-Mills
foram estudados no caso dos CLC, a massa dinâmica devido a
condensação do operador $A_\mu^aA_\mu^a$ e as ambigüidades de
Gribov, relacionadas com a quantização das teorias de Yang-Mills.
Contudo, muitos aspectos ainda merecem atenção. Primeiramente,
lembramos que nossos resultados baseiam-se na aproximação
$\alpha\ll1$. Ainda, o método utilizado para se implementar a
restrição consiste basicamente na aproximação de Gribov
\cite{Gribov:1977wm}, e um aperfeiçoamento, como o feito por
Zwanziger \cite{Zwanziger:1989mf,Zwanziger:1992qr} é necessário para
que cálculos explícitos possam ser efetuados.

\part{\ FINAL}

\chapter{Conclusions}

Antes de adentramos nas conclusões desta tese, lembramos que a lista
de resultados obtidos está apresentada de maneira sistemática no
Capítulo 2.

\section{Mass condensates conclusions}

Estudamos simultaneaneamente a condensação dos operadores locais
compostos de dimensão dois $A_\mu^aA_\mu^a$ e $f^{abc}\oc^ac^b$ no
calibre de Landau \cite{Capri:2005vw}. Esta análise extende a já
existente para o caso do condensado de gluons
$\left\langle{A}_\mu^aA_\mu^a\right\rangle$
\cite{Verschelde:2001ia,Dudal:2002pq,Dudal:2003by} e do condensado
Overhauser $\left\langle{f}^{abc}\oc^bc^c\right\rangle$
\cite{Lemes:2002rc,Dudal:2003dp}. Exceto pela renormalizabilidade do
método LCO, todos os resultados obtidos foram no caso do grupo
$SU(2)$. Neste caso, o condensado Overhauser foi considerado na
direção Abeliana do espaço de cor, $\left\langle
\varepsilon^{3bc}\oc^a c^b\right\rangle$.

Utilizando a técnica LCO para construir a ação quântica efetiva a um
laço, mostramos que ambos condensados são favorecidos dinamicamente
conforme a diminuição da energia do vácuo. A renormalizabilidade da
teoria resultante foi provada a todas as ordens em teoria de
perturbações através da técnica de renormalização algébrica
\cite{book}. Além disso, tal ação efetiva obedece a uma equação
homogênea do grupo de renormalização.

Efeitos devido a presença de condensados não triviais foram
apresentadas. Em parti-cular, analizando a condição de calibre
(\ref{5ident2}) ao nível quântico, foi possível demonstrar que o
propagador do glúon permanece transverso, a todas as ordens em
teoria de perturbações. Mostramos também, através de uma análise via
a identidade de Slavnov-Taylor (\ref{5b11}), que, a um laço, o
condensado Overhauser quebra a transversalidade da polarização do
vácuo.

Ainda, o condensado $\left\langle{A}_\mu^aA_\mu^a\right\rangle$ gera
uma massa efetiva para o glúon, no nível árvore, como é evidente da
expressão (\ref{5c1}). Da mesma forma, o propagador fantasma é
afetado no nível árvore devido a presença do condensado
$\left\langle\varepsilon^{3bc}\oc^ac^b\right\rangle$, como
caracterizado em (\ref{5c2}). Ademais, determinando a correção de um
laço à massa efetiva do glúon, encontramos que o condensado
Overhauser induz à quebra da degenerescência entre as massas
Abeliana e não Abeliana. A massa não Abeliana se mostra maior que a
massa Abeliana. Este resultado pode ser interpretado como uma
evidência analítica da dominância Abeliana no calibre de Landau,
veja (\ref{5f1}-\ref{5f1bis}).

Vale ressaltar que evidências da dominância Abeliana no calibre de
Landau foram encontradas através de simulações numéricas na rede
\cite{Suzuki:2004dw,Suzuki:2004uz}, onde foi identificado um efeito
Meissner dual Abeliano. Contudo, este último resultado foi
invalidado no trabalho \cite{Chernodub:2005gz}, de mesmos autores e
outros. Por outro lado, em \cite{Cucchieri:2005yr}, foram
encontradas evidências do condensado Overhauser, no caso $SU(2)$, na
rede, através de um estudo do operador de Faddeev-Popov. Os dados
foram ajustados de acordo com nossa predição (\ref{5c2}) e assumindo
um pequeno valor para o condensado. O resultado encontrado para o
propagador dos campos fantasmas foi na forma $\sim p^{-4}$, de
acordo com nosso resultado analítico (\ref{5c2}).

\section{Gribov ambiguities conclusions}

\subsection{Condensates and the Gribov horizon}

Com relação às ambigüidades de Gribov fizemos um estudo no caso
geral do grupo $SU(N)$ da ação de Gribov-Zwanziger
\cite{Zwanziger:1989mf,Zwanziger:1992qr}. Uma análise detalhada da
condensação do operador $A_\mu^aA_\mu^a$ em conjunto com a restrição
à região de Gribov foi exposta
\cite{Sobreiro:2004us,Sobreiro:2004yj,Dudal:2005na}.

Antes de discutirmos a inclusão do operador $A_\mu^aA_\mu^a$ na ação
de Gribov-Zwanziger, fizemos uma pequena discussão sobre a condição
de horizonte a um e dois laços no caso puro de Gribov-Zwanziger.
Como resultado, mostramos que a energia do vácuo, a um laço, é
sempre positiva, independentemente da escala ou do esquema de
renormalização. Tentativas de resultados explícitos diretamente no
esquema de renormalização $\MSbar$ fracassaram a um e dois laços.
Ressaltamos que um valor positivo para a energia do vácuo implica
num valor negativo para o condensado de gluons
$\left\langle{F}_{\mu\nu}^a{F}_{\mu\nu}^a\right\rangle$, através da
anomalia do traço.

Incluindo o operador $A_\mu^aA_\mu^a$ na ação de Gribov-Zwanziger
através do formalismo LCO, fomos capazes de mostrar, a todas as
ordens em teoria de perturbações, que o modelo permanece
renormalizável. A ação efetiva correspondente, construída com a
técnica LCO \cite{Verschelde:2001ia}, a um laço no esquema $\MSbar$,
obedece a uma equação homogênea do grupo de renormalização.
Mostramos formalmente que as equações acopladas, definindo o
horizonte e o condensado $\langle{A}_\mu^aA_\mu^a\rangle$, não
possuem solução para $\langle{A}_\mu^aA_\mu^a\rangle<0$. Não
obstante, para $\langle{A}_\mu^aA_\mu^a\rangle>0$, apenas uma
solução é possível. Contudo, um valor explícito consistente não foi
encontrado.

Numa tentativa de otimizar os resultados, fizemos uma redução na
dependência no esquema de renormalização. Desta forma, fomos capazes
de definir um esquema de renormalização aprimorado, dependendo
apenas da escala de renormalização $\omu$ e do parâmetro $b_0$,
associado à renormalização da constante de acoplamento $g$. Ainda,
definimos as quantidades $\wm$ (associada ao condensado
$\langle{A}_\mu^aA_\mu^a\rangle$) e $\wl$ (associado ao parâmetro de
Gribov), as quais são independentes de escala e do esquema de
renormalização ordem a ordem. Resolvendo as equações de gap
numericamente, não fomos capazes de encontrar uma solução com
energia de vácuo negativa. Contudo, soluções consistentes puderam
ser extraídas no caso limite em que recaímos no esquema $\MSbar$ bem
como no caso em que a dependência no esquema é mínima.

Este resultado indica que cálculos a ordens maiores devem ser
efetuados para uma conclusão sobre a energia do vácuo. Contudo, isto
está fora da ambição do trabalho aqui apresentado. Ademais, o
presente resultado pode ser uma indicação de que a restrição à
região de Gribov lança a teoria na região infravermelha profunda, e
a teoria perturbativa ou o grupo de renormalização nada podem fazer
neste caso. Necessitaríamos de um esquema de cálculo realmente não
perturbativo.

Os parâmetros de massa $\wm$ e $\wl$ são de natureza não
perturbativa e aparecem nos propagadores do glúon e dos campos de
Faddeev-Popov. Mesmo sem estimativas numéricas explícitas, fomos
capazes de prover uma análise qualitativa sobre as consequências
físicas da presença destes parâmetros. Em primeiro lugar, chamamos a
atenção para o caso do condensado $\langle{A}_\mu^aA_\mu^a\rangle$
sem a presença do horizonte. Neste caso, a massa dinâmica encontrada
é positiva, $m^2>0$. Quando incluímos o horizonte, a massa se mostra
negativa $m^2<0$. Aparentemente isto seria um problema relacionado a
presença de táquions na teoria. Contudo, devemos lembrar que os
pólos do propagador são afetados também pelo parâmetro de Gribov,
como exposto em (\ref{7prop1}). Desta forma, os pólos se mostram
complexos, indicando, não a presença de táquions, mas sim que o
glúon não faz parte do espectro físico da teoria, como deve ser para
objetos com carga de cor. Esta hipótese se confirma quando
entendemos que o propagador do glúon (\ref{7prop1}) viola o
princípio de positividade.

Com relação ao propagador dos campos fantasmas, a um laço, levando
em conta a condição de horizonte, este se mostra singular no limite
infravermelho na forma $\sim k^{-4}$. Esta propriedade pode ser
interpretada como o surgimento de forças de longo alcance dentro da
teoria, necessárias para o confinamento. Note ainda que, estes
resultados estão de acordo com os resultados encontrados na rede
\cite{Cucchieri:2005yr,Cucchieri:2006xi}, assim como os resultados
provenientes das equações de Schwinger-Dyson \cite{Fischer:2006ub}.

Devemos chamar a atenção para a importância que os condensados
fantasmas têm nesta tese, sendo fundamentais para a dominância
Abeliana no calibre de Landau, bem como para estabilizar o vácuo da
QCD. Desta forma um estudo mais completo considerando o condensado
de gluons e condensados fantasmas com a ação de Gribov-Zwanziger
pode vir a fornecer informações mais precisas sobre o vácuo da QCD e
os pólos dos propagadores. Vale ressaltar que, uma análise
qualitativa desta proposta levaria a quebra de degenerescência das
massas Abeliana e não Abeliana, contudo, uma vez que os pólos se
tornam complexos, não sabemos como a dominância Abeliana se
apresentaria neste contexto. Ao menos, longe do horizonte,
$\gamma=0$, esta análise foi feita em detalhes nesta tese.

\subsection{Gribov ambiguities in other gauges}

As ambigüidades de Gribov possuem um bom entendimento no calibre de
Landau e consequentemente, devido a semelhança técnica, no calibre
de Coulomb. Em outros calibres pouco foi desenvolvido. Desta forma,
apresentamos o primeiro trabalho analítico sobre as ambigüidades de
Gribov nos calibres lineares covariantes para pequenos valores de
$\alpha$. Ressaltamos que além destes calibres, trabalhos foram
feitos apenas no calibre máximo Abeliano
\cite{Quandt:1997rg,Bruckmann:2000xd,Capri:2005tj,Capri:2006vv,Dudal:2006ib,Capri:2006cz}.
Contudo, nesta tese nos atemos ao calibre de Landau e sua
generalização mais simples, os calibres lineares covariantes.

Como no calibre de Landau, a componente transversa do propagador do
glúon apresenta supressão infravermelha. Além disso, a componente
longitudinal se mostra imutável na aproximação de Gribov. Contudo,
na presença do condensado
$\left\langle{A}_\mu^aA_\mu^a\right\rangle$, este tem o efeito de
suprimir a componente lontidunal, de acordo com a expressão
(\ref{8gluon2}). Vemos que no caso do setor transverso existe a
violação do princípio da positividade, o que não ocorre no setor
longitudinal. Contudo, perturbativamente o setor longitudinal se
cancela, e esperamos que o mesmo ocorra não perturbativamente. Estes
resultados estão em acordo qualitativo com o simulações na rede
\cite{Giusti:1996kf,Giusti:1999im,Giusti:2000yc} e com análises
feitas através das equações de Schwinger-Dyson
\cite{Alkofer:2003jr}.

Com relação ao propagador fantasma, encontramos que, ao invés do
propagador fantasma, a função de Green que apresenta comportamento
singular infravermelho é dada por
$\left[\left(\mathcal{M}^T(k)\right)^{-1}\right]^{aa}\propto{k}^{-4}$,
definido em (\ref{8T1}). Lembramos que esta quantidade só
cor-responde ao propagador dos campos fantasmas no caso limite
$\alpha=0$.

Apesar de tudo, muitos aspectos dos calibres lineares covariantes
ainda merecem ser investigados. Uma lista parcial seria: 1) A
existência de um funcional minimizante ca-racterizando a condição
dos calibres lineares covariantes. Tal funcional poderia ser muito
útil para investigar as cópias de Gribov não infinitesimais; 2) Como
os propagadores se comportam para o caso de $\alpha$ geral; 3) Seria
possível encontrar uma ação tipo Gribov-Zwanziger, local e
renormalizável? Tal ação possibilitaria uma investigação sobre as
correções quânticas nestes calibres.

\section{Final Remarks}

Nesta tese encontramos evidências da geração dinâmica de massa
devido a condensação de operadores de massa, evidências da
dominância Abeliana no calibre de Landau, evidências do confinamento
e um entendimento das ambigüidades de Gribov, generalizando-as a
outros calibres. Além de evidências podemos dizer que confirmamos a
complexidade do setor infravermelho da QCD, assim como a necessidade
do tratamento completo das ambigüidades de Gribov. Em especial, um
método de cálculo totalmente não perturbativo parece ser necessário
para um compreendimento completo desta faceta da Natureza. Isso
mostra que esta investigação está longe de ter chegado ao fim e
muito trabalho ainda precisa ser feito.

Esperamos, portanto, que esta tese inspire mais trabalhos sobre o
comportamento infravermelho das teorias de Yang-Mills, em particular
sobre a QCD, assim como nós estaremos nos dedicando à continuidade
desta pesquisa.

\part{APPENDICES}

\appendix

\chapter{Conventions}\label{ap_nota}

Este apêndice é dedicado a estabelecer as notações e convenções
utilizadas nesta tese. Vamos seguir um modelo evolutivo ao invés de
simplesmente enumerar as notações e convenções utilizadas, tentando
assim, deixar este apêndice mais amigável para leitura.

A simetria de calibre das teorias de Yang-Mills é descrita por um
grupo de Lie compacto semi-simples cujos elementos são matrizes
unitárias de determinante positivo, o chamado grupo $SU(N)$,
\begin{equation}
SU(N)=\left\{U(x)\;\big|\;U(x)^\dagger{U}(x)=1,\;\det{U(x)}=+1\right\}\;.\label{apsun}
\end{equation}
Sejam $\lambda^a$ os geradores do grupo. Tais geradores, obviamente,
obedecem à álgebra de um grupo de Lie
\begin{equation}
\left[\lambda^a,\lambda^b\right]=f^{abc}\lambda^c\;,\label{apalg}
\end{equation}
onde $f^{abc}$, um objeto totalmente antissimétrico, representa as
constantes de estrutura do grupo. Os geradores são escolhidos
antihermiteanos, $\lambda^a=-\lambda^{a\dagger}$, com condição de
normalização dada por $tr(\lambda^a\lambda^b)=\delta^{ab}$. Note que
$a\;\in\;\{1,2,\ldots,N^2-1\}$.

A ação de Yang-Mills em um espaço-tempo Euclideano quadridimensional
é
\begin{equation}
S_{YM}=\frac{1}{4}\int{d^4x}F^a_{\mu\nu}F^a_{\mu\nu}\;,\label{apym}
\end{equation}
onde o tensor intensidade de campo é dado por
\begin{equation}
F^a_{\mu\nu}=\partial_\mu{A}_\nu^a-\partial_\nu{A}_\mu^a+gf^{abc}A^b_\mu{A}^c_\nu\;,\label{apF1}
\end{equation}
e $g$ é a constante de acoplamento. A conecção $A^a_\mu$ representa
o campo de gluons. A ação (\ref{apym}) é invariante sob
transformações de calibre locais com relação ao grupo $SU(N)$ na
forma
\begin{equation}
A^U_\mu=A_\mu+U^\dagger{D}_\mu{U}\;,\label{aptransf1}
\end{equation}
onde a derivada covariante é definida por
\begin{equation}
D_\mu{U}=\partial_\mu{U}+g[A_\mu,U]\;,\label{apderiv1}
\end{equation}
e a conecção expandida na álgebra é dada por
$A_\mu=A_\mu^a\lambda^a$. Para transformações de calibre
infinitesimais podemos expandir os elementos do grupo de acordo com
\begin{equation}
U(x)=\exp[\omega(x)]=\exp[\omega^a(x)\lambda^a]\approx1+\omega+O(\omega^2)\;.\label{apexp}
\end{equation}
Assim, as transformações infinitesimais são descritas por
(\ref{aptransf1})
\begin{equation}
\delta{A}_\mu^a=D_\mu^{ab}\omega^b\;,\label{aptransf2}
\end{equation}
de forma que
\begin{equation}
\delta{S_{YM}}=0\;.
\end{equation}
A derivada covariante na representação adjunta em (\ref{aptransf2})
se define como
\begin{equation}
D_\mu^{ab}=\delta^{ab}\partial_\mu-gf^{abc}A^c_\mu\;.\label{apderiv2}
\end{equation}
Estaremos sempre utilizando a representação adjunta ao longo desta
tese.

\chapter{Tools}\label{ap_ferram}

Este apêndice é dedicado às principais técnicas utilizadas neste
trabalho. As técnicas serão apresentadas brevemente. Como mencionado
antes, não é o objetivo deste apêndice ou desta tese rever os
princípios da Teoria Quântica de Campos e da Teoria de
Renormalização, para tal existe uma vasta literatura disponível, por
exemplo, \cite{Itzykson} e \cite{Collins}, respectivamente.

Começaremos com o princípio geral da ação quântica
\cite{Schwinger:1951xk,Lowenstein:1971vf,Lowenstein:1971jk,Lam:1972mb,Clark:1976ym},
fundamental para uma teoria quântica de campos. Após isso,
revisaremos a quantização pelo método de BRST
\cite{Becchi:1975nq,Tyutin:1975qk} seguida de um resumo sucinto da
teoria de renormalização algébrica \cite{book}. Então estaremos
aptos a discutir os principais pontos do grupo de renormalização
\cite{Wilson:1974sk}. Por fim, abordaremos a técnica de operadores
compostos locais (técnica LCO) \cite{Verschelde:2001ia}.

\section{Quantum action principle}\label{appaq}

Antes de desenvolvermos o método de quantização BRST da ação
(\ref{apym}) e estudarmos as simetrias da teoria no formalismo
funcional (Identidades de Ward) vamos discutir o chamado princípio
de ação quântica
\cite{Schwinger:1951xk,Lowenstein:1971vf,Lowenstein:1971jk,Lam:1972mb,Clark:1976ym}.

A simetria de calibre gozada pela ação de Yang-Mills (\ref{apym})
produz uma ambigüidade. Qualquer valor esperado a ser calculado com
essa ação irá contar cada configuração do campo de calibre infinitas
vezes devido ao fato de que cada campo está relacionado a infinitos
outros através das transformações de calibre (\ref{aptransf1}). Para
resolver este problema devemos fixar o calibre, ou seja, introduzir
um vínculo para filtrar os graus de liberdade espúrios existentes
devido à simetria de calibre. Faremos isso em detalhe na próxima
seção, por hora vamos assumir que saibamos a ação com o calibre
fixado, digamos, $\Sigma$, onde esta ação seria composta pela ação
de Yang-Mills mais um termo que descreva a fixação de calibre e as
demais propriedades necessárias,
\begin{equation}
\Sigma=S_{YM}+S_{fc}\;.
\end{equation}
A integral de caminho, ou gerador funcional das funções de Green
seria
\begin{equation}
Z[j,J]=\int{D\phi}\exp{\left(-\Sigma-\int{d^4x}j^a_\mu{A}_\mu^a-\int{d^4x}J^A\mathcal{O}^{A}\right)}\;.\label{appath0}
\end{equation}
Nesta expressão $D\phi$ é a medida quântica da integral de caminho e
$j^a_\mu$ e $J^A$ são campos clássicos, ou fontes externas, chamadas
fontes de Schwinger. O índice $A$ é um multi-índice geral. O objeto
$\mathcal{O}^A$ é um operador composto local. O gerador funcional
conexo se define por
\begin{equation}
Z[j,J]=e^{-W[j,J]}\;.\label{apcon}
\end{equation}
E a ação quântica, ou funcional de vértice é a transformada de
Legendre de $W[j,J]$,
\begin{equation}
\Gamma[A,J]=W[j,J]-\int{d^4x}j^a_\mu{A}_\mu^a\;.\label{apvert}
\end{equation}
Note que a condição de controrno assumida é tal que o valor esperado
do vácuo é das variáveis de campo são nulas
\begin{equation}
\frac{\delta\Gamma}{\delta{A}^a_\mu}\bigg|_{A^a_\mu=0}=0\;.\label{apcond}
\end{equation}
É sabido que o funcional gerador $Z$ obedece à equação
renormalizada, veja por exemplo \cite{Itzykson,Ryder,Peskin},
\begin{equation}
j^a_\mu{Z}+\Delta^a_\mu\cdot{Z}=0\;,\label{apeq1}
\end{equation}
onde a inserção\footnote{A notação de inserção é na forma
\begin{equation}
\Delta^a_\mu\cdot{Z}=\int{D\phi}\Delta^a_\mu\exp{\left(-\Sigma-\int{d^4x}j^a_\mu{A}_\mu^a\right)}\;.
\end{equation}} $\Delta^a_\mu$ é um operador composto local cuja
contribuição clássica coincide com as equações de movimento
clássicas. A equação (\ref{apeq1}) para a ação quântica fica como
\begin{equation}
\frac{\delta\Gamma}{\delta{A}^a_\mu}-\Delta^a_\mu\cdot\Gamma=0\;,\label{apeq2}
\end{equation}

A derivação da expressão (\ref{apeq2}) pode ser repetida para o
funcional gerador das funções de vértice com a inserção
$\mathcal{O}^A$,
\begin{equation}
\frac{\delta\Gamma}{\delta{J}^A}=\mathcal{O}^A\cdot\Gamma\;.\label{apins}
\end{equation}
Assim, seguindo os passos anteriores encontramos
\begin{equation}
\frac{\delta\Gamma}{\delta{J}^A}\frac{\delta\Gamma}{\delta{A}^a_\mu}-\Delta^{aA}_\mu\cdot\Gamma=0\;,\label{apeq3}
\end{equation}
onde $\Delta^{aA}_\mu$ é um operador local composto. A expressão
(\ref{apeq3}) nada mais é que o princípio de ação quântica (PAQ) na
representação do funcional de vértice.

Não é muito difícil entender o conteúdo físico do PAQ (\ref{apeq3}).
De fato, para nossos fins, o PAQ diz que uma variação não linear do
campo $A_\mu^a$ dá origem a uma inserção de um operador local
composto. E como este operador não é um campo fundamental, este deve
ser acoplado a uma fonte externa $J^A$. Assim, a expressão
(\ref{apins}) define o valor esperado de $\mathcal{O}^A$, contudo,
não temos um contorno para $\mathcal{O}^A$ como temos para $A^a_\mu$
descrito pela expressão (\ref{apcond}). Por tanto, operadores locais
compostos devem estar acoplados a uma fonte externa pois não sabemos
como estes se comportam sob correções quânticas. Essas fontes são
introduzidas para controlar estes operadores. Para operadores
fundamentais, como são os campos fundamentais da teoria, não há
necessidade de se inserir uma fonte pois temos a condição
(\ref{apcond}) para controlá-los.

\section{BRST quantization}

\subsection{gauge fixing}

Como mencionado na seção anterior, devemos nos livrar dos graus de
liberdade não físicos da ação de Yang-Mills (\ref{apym}). Para tal,
devemos introduzir um vínculo na teoria. Um método eficiente e
elegante de se introduzir um vínculo à uma ação invariante de
calibre é o chamado método de quantização BRST
\cite{Becchi:1975nq,Tyutin:1975qk}. Vamos rever este método para o
caso das teorias de Yang-Mills.

Para cada gerador do grupo de simetria a ser quebrada, $\lambda^a$,
introduz-se um campo fantasma de Grassmann $c^a$ obedecendo a uma
equação isomórfica à equação de estrutura de Maurer-Cartan
\cite{Reinhardt:2004mm},
\begin{equation}
sc^a=\frac{g}{2}f^{abc}c^cc^c\;,\label{apbrs1}
\end{equation}
sendo esta a transformação BRST para o campo fantasma. O campo
fantasma nada mais é que os conhecidos campos fermiônicos de
Faddeev-Popov \cite{Faddeev:1967fc}. Para o glúon, a transformação
de BRST ocorre como
\begin{equation}
sA_\mu^{a}=-D_\mu^{ab}c^b\;.\label{apbrs2}
\end{equation}
Note que esta transformação é isomórfica à uma transformação de
calibre infinitesimal (\ref{aptransf2}). O vínculo de calibre
covariante, digamos,
\begin{equation}
\partial_\mu{A}^a_\mu=f^a\;,\label{apgaugefix0}
\end{equation}
é introduzido juntamente com um dubleto BRST
\begin{eqnarray}
s\bar{c}^a&=&b^a\;,\nonumber\\
sb^a&=&0\;,\label{apbrs3}
\end{eqnarray}
o qual é inofensivo ao conteúdo físico da teoria \cite{book}. Na
expressão (\ref{apgaugefix0}), $f^a$ pode depender dos campos $c$,
$\bar{c}$ and $b$. Ainda, de forma a ser possível descrever a
simetria BRST como uma equação funcional, de acordo com o PAQ,
introduzimos também fontes invariantes de BRST, $\Omega^a_\mu$ e
$L^a$, para controlar as transformações BRST não lineares,
(\ref{apbrs1}) e (\ref{apbrs2}). Uma simples analise dimensional
sobre os campos e o operador $s$ nos permite definir\footnote{Note
que existe uma liberdade na escolha das dimensões ultravioletas dos
campos de Faddeev-Popov e do operador BRST. Esta liberdade vem do
fato de que as transformações BRST fornecem duas relações
independentes entre as dimensões destes campos e do operador $s$,
portanto, temos um sistema indefinido de equações. Esta questão será
importante para a definição dos condensados fantasmas discutidos na
segunda parte desta tese. Voltaremos a discutir esta questão no
estudos dos condensados fantasmas.} a dimensão ultravioleta dos
campos e fontes de acordo com a tabela\footnote{Lembramos que, em
$d$ dimensões, a constante de acoplamento possui dimensão
ultravioleta dada por $[g]=2-\frac{d}{2}$, implicando na condição de
renormalizabilidade $2\le{d}\le4$.} \ref{aptable1}.
\begin{table}[t]
\centering
\begin{tabular}{|c|c|c|c|c|c|c|}
\hline
campos / fontes & $A$ & $c$ & $\bar{c}$ & $b$ & $\Omega$ & $L$ \\
\hline
dimensão & $1$ & 0 & $2$ & $2$ & $3$ & $4$ \\
número fantasma & 0 & 1 & $-1$ & 0 & $-1$ & $-2$ \\
\hline
\end{tabular}
\caption{Dimensão e número fantasma dos campos e fontes.}
\label{aptable1}
\end{table}
Assim, a ação covariante mais geral possível apropriada para uma
teoria quântica consistente é dada por
\begin{equation}
\Sigma=S_{YM}+S_{gf}+S_{ext}\;,\label{apaction2}
\end{equation}
onde,
\begin{eqnarray}
S_{gf}&=&s\int{d^4x}\;\bar{c}^a\left(\partial_\mu{A}^a_\mu+\frac{\alpha}{2}b^a+\frac{\beta}{2}gf^{abc}\bar{c}^bc^c\right)
\nonumber\\
 &=&\int{d^4x}\left[b^a\left(\partial_\mu{A}^a_\mu+\frac{\alpha}{2}b^a+\beta{g}{f}^{abc}\bar{c}^bc^c\right)+
 \bar{c}^a\partial_\mu{D}_\mu^{ab}c^b+\frac{\beta}{4}g^2f^{abc}f^{cde}\bar{c}^a\bar{c}^bc^dc^e\right]\;,\nonumber\\
 \label{apgaugefix1}
\end{eqnarray}
e
\begin{eqnarray}
S_{ext}&=&s\int{d^4x}\left(-\Omega^a_{\mu}A^a_\mu+L^ac^a\right)\nonumber\\
 &=&\int{d^4x}\left(-\Omega^a_{\mu}D_\mu^{ab}c^b+\frac{g}{2}f^{abc}L^ac^bc^c\right)\;.\label{apextx}
\end{eqnarray}
Note que a ação (\ref{apaction2}) possui todos os termos possíveis
de dimensão ultravioleta quatro que respeitem a simetria global de
cor.

Duas observações devem ser feitas. Primeiro, é trivial ver que $s$ é
um operador nilpotente, $s^2=0$, portanto, a ação (\ref{apaction2})
é BRST invariante, $s\Sigma=0$. Segundo, a introdução dos parâmetros
de calibre $\alpha$ e $\beta$ depende da escolha do calibre a ser
fixado (\ref{apgaugefix0}) e, em alguns casos estes termos são
necessários para fins de renormalização.

Geometricamente, este método pode ser entendido da seguinte forma: A
ação de Yang-Mills (\ref{apym}) descreve a dinâmica da conecção,
$A^a_\mu$, definida no fibrado principal de grupo de simetria
$SU(N)$ imerso numa variedade Euclideana de quatro dimensões.
Contudo, a geometria existente neste espaço topológico é maior.
Podemos definir neste espaço as uma-formas de Maurer-Cartan $c^a$, a
derivada exterior na órbita de calibre $s$, e dubletos para
definirmos uma seção no fibrado principal. Assim, partindo desta
geometria completa podemos derivar a ação de Yang-Mills como foi
feito acima, utilizando apenas argumentos geométricos. Fisicamente,
os campos fantasmas aparecem de forma a garantir a unitariedade da
matriz $S$, eliminando os graus de liberdade espúrios da teoria.

Obviamente, como já foi mencionado e será amplamente explorado, a
quantização BRST não é completamente eficiente devido à existência
das cópias de Gribov \cite{Gribov:1977wm}. Contudo, nada mais
falaremos sobre este assunto neste apêndice, uma vez que a terceira
parte desta tese é devotada a este tema.

\subsection{Ward identities}

Um ponto importante para se estabelecer a renormalizabilidade de uma
teoria é saber as simetrias nela existentes. Assim, vamos escrever a
simetria BRST e as demais simetrias na forma de uma equação
funcional compativelmente com o PAQ. No caso da simetria BRST, a
equação funcional é a chamada identidade de Slavnov-Taylor
\begin{equation}
\mathcal{S}(\Sigma)=\int{d^4x}\left(\frac{\delta\Sigma}{\delta\Omega^a_\mu}\frac{\delta\Sigma}{\delta{A}^a_\mu}+
\frac{\delta\Sigma}{\delta{L}^a}\frac{\delta\Sigma}{\delta{c}^a}+b^a\frac{\delta\Sigma}{\delta\bar{c}^a}\right)=0\;.
\label{apst1}
\end{equation}
Além disso, dependendo do calibre escolhido, outras simetrias podem
surgir. Tais simetrias podem envolver os campos fantasmas e demais
objetos que não o campo de calibre. Essas simetrias darão lugar a
outras identidades, as chamadas identidades de Ward,
\begin{equation}
\mathcal{W}^\mathcal{A}(\Sigma)=\Delta^\mathcal{A}_{cl}\;,\label{apward1}
\end{equation}
onde $\mathcal{A}$ é um multi-índice geral e
$\Delta^\mathcal{A}_{cl}$ uma possível quebra clássica, linear nos
campos. É um conhecido resultado que esta quebra não evolui ao nível
quântico \cite{book} devido à condição (\ref{apcond}). Por este
motivo, podemos utilizar identidades linearmente quebradas como
simetrias válidas. Ainda, a simetria discreta de Faddeev-Popov surge
naturalmente na quantização BRST,
\begin{equation}
\mathcal{Q}\;\Sigma=0\;,\label{apfpcharg}
\end{equation}
onde $\mathcal{Q}$ é um operador discreto cujas transformações não
nulas são
\begin{eqnarray}
c^a&\rightarrow&c^a\;,\nonumber\\
\bar{c}^a&\rightarrow&-\bar{c}^a\;,\nonumber\\
\Omega_\mu^a&\rightarrow&-\Omega_\mu^a\;,\nonumber\\
L^a&\rightarrow&-2L^a\;.
\end{eqnarray}
Esta simetria define um novo número quântico, o chamado número
fantasma. Estes números estão tabelados na tabela \ref{aptable1}
juntamente com as dimensões ultravioletas dos campos e fontes.

\section{Algebraic renormalization}

Com a ação completa (\ref{apaction2}) e as identidades de Ward
(\ref{apst1}-\ref{apfpcharg}) podemos dar início ao estudo da
estabilidade da ação (\ref{apaction2}) sob correções radiativas.
Para tal, a renormalização algébrica \cite{book} é a teoria
apropriada para se trabalhar. O primeiro passo é a procura por
anomalias, ou seja, checar a validade das identidades de Ward no
nível quântico. Em seguida, utilizando as identidades de Ward,
determina-se o contratermo mais geral possível. E finalmente,
testa-se a estabilidade da ação sob correções quânticas. Uma vez
estabelecida a renormalizabilidade de uma teoria podemos definir o
grupo de renormalização, último assunto antes de adentramos na seção
seguinte, a técnica LCO.

\subsection{Anomalies}

Provar a validade das identidades (\ref{apst1}-\ref{apward1}) ao
nível quântico equivale a mostrar que não existem anomalias na
teoria, {\it i.e.} se uma simetria é perdida conforme aparecem as
correções quânticas então dizemos que existe uma anomalia associada
a esta simetria. O método geral de verificar se uma simetria da ação
clássica sobrevive ao processo de quantização é escrever uma
possível quebra, de ordem $n$ em $\epsilon$, para a identidade de
Ward associada a esta simetria. Assume-se que a identidade é válida
até ordem $n-1$ em $\epsilon$. Desta forma,
\begin{equation}
\mathcal{W}^{\mathcal{A}}(\Gamma)=\Delta^\mathcal{A}_{cl}+\epsilon^n\Delta^\mathcal{A}+O(\epsilon^{n+1})\;,\label{apbr1}
\end{equation}
onde, (\ref{apbr1}) engloba também a identidade de Slavnov-Taylor
(\ref{apst1}). A quantidade $\Gamma$ representa a ação quântica,
\begin{equation}
\Gamma=\Sigma+\epsilon\Sigma^{(1)}+\epsilon^2\Sigma^{(2)}+\epsilon^3\Sigma^{(3)}+\ldots=\Sigma+\Sigma^c
\label{apquantumaction}
\end{equation}
Ainda, $\epsilon$ é o parâmetro de expansão, considerado pequeno.
Substituindo a expressão (\ref{apquantumaction}) em (\ref{apbr1})
chegamos a
\begin{equation}
\mathcal{W}^{\mathcal{A}}_\Gamma\Sigma^c=\epsilon^n\Delta^\mathcal{A}\;,\label{apbr11}
\end{equation}
onde fizemos uso da identidade clássica (\ref{apward1}), portanto,
mostrando que a quebra vem das correções quânticas. O novo operador
$\mathcal{W}^{\mathcal{A}}_\Gamma$ é um operador linear associaao a
$\mathcal{W}^{\mathcal{A}}$, chamado operador linearizado da
simetria\footnote{Veremos um exemplo explícito na próxima seção para
o caso do operador de Slavnov-Taylor.} $\mathcal{W}^{\mathcal{A}}$.
No caso de $\mathcal{W}^{\mathcal{A}}$ ser linear, o operador
linearizado coincide com $\mathcal{W}^{\mathcal{A}}$.

Voltemos a expressão (\ref{apbr1}). Suponhamos então que a quebra
$\Delta^{\mathcal{A}}$ possa ser escrita como
\begin{equation}
\Delta^\mathcal{A}=\mathcal{W}_\Sigma^{\mathcal{A}}\tilde{\Delta}\;,\label{apbr2}
\end{equation}
Neste caso, redefinindo a ação quântica como
\begin{equation}
\Gamma\rightarrow\Gamma-\epsilon^n\tilde{\Delta}\;,\label{apnew}
\end{equation}
e aplicando o operador $\mathcal{W}^{\mathcal{A}}$ na nova ação
(\ref{apnew}), temos
\begin{equation}
\mathcal{W}^{\mathcal{A}}(\Gamma-\epsilon^n\tilde{\Delta})=\mathcal{W}^{\mathcal{A}}(\Gamma)
-\epsilon^n\Delta^\mathcal{A}+O(\epsilon^{n+1})=\Delta^\mathcal{A}_{cl}+O(\epsilon^{n+1})\;,\label{apbr111}
\end{equation}
onde (\ref{apbr1}) foi utilizada. Assim, se $\tilde{\Delta}$ pode
ser absorvida em $\Gamma$ como um contratermo, ordem a ordem em
teoria de perturbações, então não existe anomalia da simetria em
questão. A demostração a ordem $n$ implica na validade a todas as
ordens,por recursão. O método de demonstração em si é totalmente
análogo ao que abordaremos no caso da estabilidade da ação discutida
a seguir. De agora em diante assumiremos que não existem anomalias.

\subsection{Counterterm and quantum stability}

Nos resta ainda provar a estabilidade da ação (\ref{apaction2}) com
relação a correções quânticas. Assim, sem anomalias, consideramos a
ação quântica corrigida a todas as ordens (\ref{apquantumaction}).
Então, impomos as identidades (\ref{apst1}-\ref{apfpcharg}) à ação
quântica (\ref{apquantumaction}),
\begin{eqnarray}
\mathcal{S}(\Gamma)&=&0\;,\nonumber\\
\mathcal{W}^{\mathcal{A}}(\Gamma)&=&\Delta_{cl}^{\mathcal{A}}\;,\nonumber\\
\mathcal{Q}\;\Gamma&=&0\;.\label{apward2}
\end{eqnarray}
Fazendo uso das identidades (\ref{apst1}-\ref{apfpcharg}) em
(\ref{apward2}) chegamos, ordem a ordem, a
\begin{eqnarray}
\mathcal{B}_{\Gamma^{(n-1)}}\Sigma^{(n)}&=&0\;,\nonumber\\
\mathcal{W}_{\Gamma^{(n-1)}}^{\mathcal{A}}\Sigma^{(n)}&=&0\;,\nonumber\\
\mathcal{Q}\Sigma^{(n)}&=&0\;,\label{apcons1}
\end{eqnarray}
onde, $\mathcal{B}_{\Gamma^{(n-1)}}$ é o operador linearizado de
Slavnov-Taylor de ordem $(n-1)$,
\begin{equation}
\mathcal{B}_{\Gamma^{(n-1)}}=\int{d^4x}\left(\frac{\delta\Gamma^{(n-1)}}{\delta\Omega^a_\mu}
\frac{\delta}{\delta{A}^a_\mu}+
\frac{\delta\Gamma^{(n-1)}}{\delta{A}^a_\mu}\frac{\delta}{\delta\Omega^a_\mu}+
\frac{\delta\Gamma^{(n-1)}}{\delta{L}^a}\frac{\delta}{\delta{c}^a}+
\frac{\delta\Gamma^{(n-1)}}{\delta{c}^a}\frac{\delta}{\delta{L}^a}+b^a\frac{\delta}{\delta\bar{c}^a}\right)\;.
\label{aplinst1}
\end{equation}
Note que as expressões (\ref{apcons1}) valem a partir da primeira
ordem, pois a ordem zero temos as já estabelecidas identidades
clássicas (\ref{apst1}-\ref{apfpcharg}). O operador
$\mathcal{W}_{\Gamma^{(n-1)}}^{\mathcal{A}}$ é a versão linearizada
de $\mathcal{W}^{\mathcal{A}}(\cdot)$, da mesma forma que
(\ref{aplinst1}). E, $\Gamma^{(n-1)}$ é a ação quântica
(\ref{apquantumaction}) truncada na ordem $\epsilon^{(n)}$
\begin{equation}
\Gamma^{(n)}=\Sigma+\epsilon\Sigma^{(1)}+\epsilon^2\Sigma^{(2)}+\ldots+\epsilon^{(n-1)}\Sigma^{(n-1)}+O(\epsilon^n)\;.
\label{apquantumaction1}
\end{equation}
É fácil ver que a imposição das identidades (\ref{apward2}), valendo
ordem a ordem, pode ser feita de maneira recursiva, {\it i.e.}
provando a primeira ordem, a prova valerá a todas as ordens por
indução. Portanto, basta impormos
\begin{eqnarray}
\mathcal{B}_{\Sigma}\Sigma^{(1)}&=&0\;,\nonumber\\
\mathcal{W}_{\Sigma}^{\mathcal{A}}\Sigma^{(1)}&=&0\;,\nonumber\\
\mathcal{Q}\Sigma^{(1)}&=&0\;.\label{apcons2}
\end{eqnarray}
Resolver a primeira das (\ref{apcons2}) é um problema de cohomologia
\cite{book}. E a solução é dada por uma parte trivial e uma parte
não trivial,
\begin{equation}
\Sigma^{(1)}=\Delta_0+\mathcal{B}_\Sigma\Delta^{-1}\;.\label{apcount1}
\end{equation}
onde $\Delta_0$ é a parte não trivial da cohomologia, com dimensão
ultravioleta quatro e número fantasma nulo,
\begin{eqnarray}
\mathcal{B}_\Sigma\Delta_0&=&0\;,\nonumber\\
\Delta_0&\ne&\mathcal{B}_\Sigma(\cdot)\;.\label{apcons3}
\end{eqnarray}
É sabido que dubletos BRST não pertencem a parte não trivial da
cohomologia \cite{book}. Desta forma,
$\Delta_0=\Delta_0(A,c,\Omega,L)$. Contudo, é fácil mostrar que a
cohomologia não trivial não carrega dependência nas fontes externas,
implicando em $\Delta_0=\Delta_0(A,c)$. Ainda, devido a simetria
discreta de Faddeev-Popov, não há espaço para termos dependentes nos
campos de Faddeev-Popov. Logo,
\begin{equation}
\Delta_0=\Delta_0(A)\;.
\end{equation}
A simples aplicação da condição (\ref{apcons3}) nos leva a
\begin{equation}
\Delta_0=a_0S_{YM}\;.
\end{equation}
Falta ainda encontrar $\Delta^{-1}$, o objeto mais geral, polinomial
nos campos e fontes, com dimensão ultravioleta igual a quatro e
número fantasma unitário negativo. Assim,
\begin{equation}
\Delta^{-1}=\int{d^4x}\left(a_1\partial_\mu\bar{c}^a{A}_\mu^a+a_2L^ac^a+a_3\frac{\alpha}{2}\bar{c}^ab^a+
a_4\frac{\beta}{2}f^{abc}\bar{c}^a\bar{c}^bc^c+a_5\Omega^{a\mu}A_\mu^a\right)\;,\label{aptriv1}
\end{equation}
onde $a_i$ são, assim como $a_0$, parâmetros independentes. Ademais,
a segunda das condições (\ref{apcons2}) devem ser utilizadas,
promovendo possíveis cancelamentos e vínculos entre estes parâmetros
independentes.

Após determinarmos o contratermo mais geral, ainda devemos checar a
estabilidade da ação (\ref{apaction2}). Isso é feito redefinindo
multiplicativamente os campos, parâmetros e fontes,
\begin{eqnarray}
\Phi_0&=&Z^{1/2}_\Phi\Phi\;,\nonumber\\
\mathcal{J}_0&=&Z_\mathcal{J}\mathcal{J}\;,\nonumber\\
\xi_0&=&Z_\xi\xi\;,\label{apren1}
\end{eqnarray}
onde,
\begin{eqnarray}
\Phi&\in&\left\{A,b,c,\bar{c}\right\}\;,\nonumber\\
\mathcal{J}&\in&\left\{\Omega,L\right\}\;,\nonumber\\
\xi&\in&\left\{g,\alpha,\beta\right\}\;,\label{apren2}
\end{eqnarray}
e então reabsorver o contratermo na ação clássica na forma
\begin{equation}
\Gamma^{(1)}=\Sigma(\Phi_0,\mathcal{J}_0,\xi_0)=\Sigma(\Phi,\mathcal{J},\xi)+\epsilon\Sigma^{(1)}(\Phi,\mathcal{J},\xi)\;.\label{apren3}
\end{equation}
A validade de (\ref{apren3}) a primeira ordem implica na validade da
mesma a todas as ordens. Se isso for possível, então a ação
quantizada no calibre escolhido é dita multiplicamente
renormalizável pelo menos a todas as ordens na teoria de
perturbações.

\subsection{Renormalization group}

Vamos relembrar os principais pontos referentes ao grupo de
renormalização (GR). Como este é um assunto bem conhecido, vamos
diretamente ao grupo de renormalização e sua interpretação física,
evitando o tratamento via a equação de Callan-Symanzik. Para uma
discussão mais detalhada sugerimos a literatura existente sobre o
assunto \cite{book,Collins,Zinn}.

Uma teoria renormalizável consiste numa teoria onde seja possível
filtrar consistentemente as divergências ultravioletas nela
existentes. Os contratermos são ajustados de forma a eliminar tais
divergências. Contudo, qualquer parte finita que estes contratermos
possuam é aceitável. Cada forma de escolher o contratermo consiste
num esquema de renormalização diferente. Uma mudança no esquema de
renormalização é compensada por uma mudança na parte finita do
contratermo. Desta forma a teroria permanece inalterada enquanto a
parametrização utilizada depende do esquema de renormalização
empregado. Sucintamente, o conteúdo físico de uma teoria
renormalizável é invariante sob as transformações entre
parametrizações. Tal invariância caracteriza o chamado grupo de
renormalização. Uma forma de caracterizar tal liberdade é através da
introdução da massa de renormalização $\mu$. O que significa que
podemos descrever a invariância sob o GR através de uma equação
diferencial, a equação do grupo de renormalização (EGR).

Para estudarmos a invariância de uma quantidade física sob a mudança
da parametrização da teoria basta fazermos uso do gerador de tal
variação,
\begin{equation}
d(\mu)=\mu\frac{d}{d\mu}\;.\label{scale1}
\end{equation}
Pensemos numa quantidade de uma teoria, digamos, $Q$. Podemos
estudar a resposta de $Q$ a uma mudança na parametrização da teoria
de acordo com
\begin{equation}
d(\mu)Q=\Delta\cdot Q\;,\label{scale2}
\end{equation}
onde $\Delta$, de acordo com o PAQ, representa uma inserção em $Q$.
Ainda, obviamente, os campos e demais parâmetros (constantes de
acoplamentos, campos, parâmetros de calibre, massas...) de uma
teoria podem vir a depender (dependendo do esquema de renormalização
utilizado) da massa de renormalização $\mu$. Desta forma, a variação
em $\mu$ se expande e obtemos a chamada equação do grupo de
renormalização
\begin{equation}
\left(\mu\frac{\partial}{\partial\mu}+
\beta\frac{\partial}{\partial{g^2}}-\gamma_mm^2\frac{\partial}{\partial{m^2}}-\gamma_\Phi\frac{\partial}{\partial{\Phi}}
-\gamma_\alpha\frac{\partial}{\partial{\alpha}}\right)Q=\Delta\cdot
Q\;,\label{CS1}
\end{equation}
onde $\Phi$ representa os campos da teoria, $\alpha$ os parâmetros
de calibre, $g$ as constantes de acoplamento e $m$ as massas. As
quantidades $\gamma_\Phi$, $\beta$ e $\gamma_\alpha$ e $\gamma_m$
representam suas respectivas renormalizações,
\begin{eqnarray}
\beta&=&\mu\frac{\partial{g^2}}{\partial\mu}\;,\nonumber\\
\gamma_\Phi&=&-\mu\frac{\partial\Phi}{\partial\mu}\;,\nonumber\\
\gamma_m&=&-\frac{\mu}{m^2}\frac{\partial{m^2}}{\partial\mu}\;,\nonumber\\
\gamma_\alpha&=&-\mu\frac{\partial\alpha}{\partial\mu}\;.\label{gamma1}
\end{eqnarray}
A função beta nos diz como a constante de acoplamento varia com a
mudança de escala enquanto as dimensões anômalas $\gamma_\Phi$,
$\gamma_\alpha$ e $\gamma_m$ nos informam como as amplitudes de
campo, parâmetros de calibre e massas, respectivamente, respondem à
mudança de escala. Por sua vez, a EGR (\ref{CS1}) descreve como uma
quantidade responde a uma mudança de escala de renormalização.

No caso em que $Q$ define uma quantidade física, digamos $Q=\Gamma$,
teremos a chamada invariância sob as EGR, ou seja, o conteúdo físico
de uma teoria deve permanecer inalterado se alteramos o esquema de
renormalização. A compensação vem de mudanças simultâneas dos
parâmetros da teoria. É claro que quantidades físicas não devem
depender dos campos, portanto,
\begin{equation}
\left(\mu\frac{\partial}{\partial\mu}+
\beta\frac{\partial}{\partial{g^2}}-\gamma_mm^2\frac{\partial}{\partial{m^2}}-\gamma_\alpha\frac{\partial}{\partial{\alpha}}\right)\Gamma=0\;.\label{CS2}
\end{equation}

\section{Local composite operator technique}

Nesta seção veremos como introduzir operadores locais compostos na
ação, de maneira consistente com o PAQ e a simetria BRST e ainda de
forma a respeitar o GR. Na seção onde estudamos o princípio de ação
quântica, seção \ref{appaq}, vimos como um operador local composto
deve ser introduzido de maneira a respeitar o PAQ, contudo, nada foi
dito sobre a simetria BRST. Por simplicidade e objetividade, vamos
nos restringir ao calibre de Landau.

O objetivo da técnica de operadores compostos locais (Técnica
LCO\footnote{Do inglês: {\it Local composite operator technique}}) é
calcular o potencial efetivo associado a um operador\footnote{A
menos que o contrário seja dito, nos referiremos sempre a um
operador local composto simplesmente por operador.} e, juntamente
com o GR, procurar por mínimos não triviais desses potenciais.
Infelizmente, apenas operadores de dimensão ultravioleta dois
permitem cálculos reais, pois apenas estes operadores podem ser
associados a campos auxiliares através da transformação de
Hubbard-Stratonovich. Por outro lado, tais operadores podem vir a
gerar uma massa dinâmica para outros campos, em particular o campo
de calibre. Comecemos então analisando o método geral de incluir
operadores de dimensão dois em uma ação BRST invariante.

\subsection{Dimension two operators}

Seja $\mathcal{O}^{A}$ um operador de dimensão ultravioleta dois e
número fantasma $q$. O índice $A$ é um multi-índice geral. Seja
também $\Sigma$ uma ação invariante sob transformações de BRST
gerais, onde $s$ é o operador de BRST. De acordo com o PAQ devemos
introduzir uma fonte $J^A$ acoplada ao operador. Para que esta fonte
pertença a cohomologia trivial de BRST, introduzimo-la como um
dubleto BRST,
\begin{eqnarray}
s\lambda^A&=&J^A\nonumber\\
sJ^A&=&0\;.\label{apbrs4}
\end{eqnarray}
Os números quânticos do operador e das fontes estão disponíveis na
tabela \ref{aptable2}.
\begin{table}[t]
\centering
\begin{tabular}{|c|c|c|c|}
\hline
operador / fontes & $\mathcal{O}$ & $\lambda$ & $J$ \\
\hline
dimensão & $2$ & 2 & $2$ \\
número fantasma & $q$ & $-q-1$ & $-q$ \\
\hline
\end{tabular}
\caption{Dimensão e número fantasma do operador e fontes LCO.}
\label{aptable2}
\end{table}
A ação, levando em conta o operador $\mathcal{O}^A$, é definida por
\begin{eqnarray}
\Sigma_{LCO}&=&\Sigma+s\int{d^4x}\left(\lambda^A\mathcal{O}^{A}-f(q)\frac{\zeta}{2}\lambda^AJ^A\right)\;,\nonumber\\
 &=&\Sigma+\int{d^4x}\left[J^A\mathcal{O}^{A}+(-1)^{q+1}\lambda^As\mathcal{O}^{A}-f(q)\frac{\zeta}{2}J^AJ^A\right]\;.
 \label{aplco0}
\end{eqnarray}
O termo quadrático na fonte $J^A$ é permitido por contagem de
potência. Como este termo só aparece para o caso em que $q=0$,
introduzimos a função $f(q)$ definida por
\begin{equation}
f(q)=\lim_{\epsilon\rightarrow0}\int_{-\epsilon}^{\epsilon}{dz}\delta{(z-q)}\;.
\end{equation}
Desta forma, $\zeta$, chamado parâmetro LCO, é introduzido para
absorver as divergências associadas às funções de Green
$\left<\mathcal{O}^{A}(x)\mathcal{O}^{B}(y)\right>$. Note ainda que,
no caso em que conside-ramos mais de um operador, introduzimos mais
dubletos. Estas fontes podem se misturar e gerar mais termos
quadráticos nas fontes. Aqui, por simplicidade, consideraremos
apenas um dubleto.

A ação (\ref{aplco0}) deve ser renormalizável, portanto, o método de
renormalização algébrica deve ser aplicado. Vamos assumir, portanto,
que a ação LCO (\ref{aplco0}) seja renormalizável. Uma importante
relação que resulta deste método diz respeito a renormalização do
operador composto $\mathcal{O}^A$,
\begin{equation}
Z_J=Z_{\mathcal{O}}\;,\label{apz1}
\end{equation}
que será de extrema importância na análise do potencial efetivo via
grupo de renorma-lização, mais adiante.

\subsection{Hubbard-Stratonovich fields}

O problema que surge agora é o termo quadrático na fonte LCO. Como o
objetivo é calcular o potencial efetivo associado ao operador
$\mathcal{O}^A$ devemos eliminar o termo quadrático nas fontes de
forma a podermos aplicar os métodos usuais da TQC. Este termo só
ocorre para o caso em que $q=0$, o que acarreta em $f(q)=1$. Como os
outros casos não possuem este problema, vamos considerar o caso
$q=0$. Consideremos então a integral de caminho, com $\lambda^A=0$,
\begin{equation}
Z[J]=\int{D\phi}\exp\left\{-\Sigma-
\int{d^4x}\left[J^A\mathcal{O}^{A}-\frac{\zeta}{2}J^AJ^A\right]\right\}\;.\label{appath1}
\end{equation}
As quantidades em $Z$ são as quantidades já renormalizadas, de
acordo com (\ref{apren1}), mas, por simplicidade, vamos manter a
notação não renormalizada. A eliminação do termo quadrático em $J^A$
é feita através dos campos de Hubbard-Stratonovich. Estes campos são
introduzidos através da unidade escrita na forma
\begin{equation}
1=\mathcal{N}\int{D\sigma}\exp\left\{-\frac{1}{2\zeta}\int{d^4x}
\left(\frac{\sigma^A}{g}+\mathcal{O}^A-\zeta{J}^A\right)\left(\frac{\sigma^A}{g}+\mathcal{O}^A-
\zeta{J}^A\right)\right\} \;,\label{aphs0}
\end{equation}
Onde $\sigma^A$ são os campos de Hubbard-Stratonovich e
$\mathcal{N}$ o fator de normalização apropriado. Inserindo a
unidade (\ref{aphs0}) na integral de caminho (\ref{appath1})
chegamos a, após normalização,
\begin{equation}
Z[J]=\int D\sigma\exp\left\{-\Sigma-
\int{d^4x}\left[\frac{\sigma^A\sigma^A}{2g^2\zeta}+
\frac{\sigma^A\mathcal{O}^A}{g\zeta}+\frac{\mathcal{O}^A\mathcal{O}^{A}}{2\zeta}
-\frac{J^A\sigma^A}{g}\right]\right\}\;.\label{appath1a}
\end{equation}
Não temos mais o termo quadrático na fonte LCO. Note que a
identificação entre o operador $\mathcal{O}^A$ e o campo $\sigma^A$
é imediata
\begin{equation}
\left<\mathcal{O}^A\right>=-\frac{1}{g}\left<\sigma^A\right>\;.
\end{equation}

Uma vez que o termo quadrático foi eliminado podemos tomar $J^A=0$ e
calcular o potencial efetivo para o campo $\sigma^A$. Contudo,
algumas considerações ainda devem ser feitas a cerca do grupo de
renormalização.

\subsection{Renormalization group}

Para entender como o termo LCO se comporta perante o grupo de
renormalização pensemos numa quantidade física, por praticidade,
consideremos a ação quântica $\Gamma$. De acordo com
\cite{Verschelde:2001ia}, não é difícil deduzir então que
\begin{equation}
\left[\mu\frac{\p}{\p\mu}+\beta(g^2)\frac{\p}{\p{g^2}}-\gamma_{\mathcal{O}}(g^2)\int{d^4x}\;J^A\frac{\delta}{\delta{J^A}}+
\gamma_\zeta(g^2,\zeta)\frac{\p}{\p\zeta}\right]\Gamma=0\;,\label{ap4rge2}
\end{equation}
onde a função beta foi definida em (\ref{gamma1}) e as demais
dimensões anômalas são definidas como
\begin{eqnarray}
\gamma_{\mathcal{O}}&=&\mu\frac{\p}{\p\mu}\log{Z}_J\;,\nonumber\\
\gamma_\zeta&=&\mu\frac{\p}{\p\mu}\zeta\;.\label{gamma2}
\end{eqnarray}
Em particular, sem perda de generalidade, podemos definir $\zeta$
como função apenas de $g$ de forma que uma solução particular de
$\zeta$ seja na forma de uma série de Laurent
\begin{equation}
\zeta=\frac{\zeta_0}{g^2}+\zeta_1+g^2\zeta_2+g^4\zeta_3+\ldots\;.\label{ap4zeta1}
\end{equation}
Esta escolhe implica que $\zeta$ obedeça a uma EGR independente.
Portanto, $\Gamma$ obedece a uma equação mais simplificada,
\begin{equation}
\left[\mu\frac{\p}{\p\mu}+\beta(g^2)\frac{\p}{\p{g^2}}-\gamma_{\mathcal{O}}(g^2)\int{d^4x}\;J^A\frac{\delta}{\delta{J^A}}\right]\Gamma=0\;,
\label{ap4rge2a}
\end{equation}
Note ainda que, para conhecermos a contribuição de ordem $n$ para
$\zeta$ necessitamos conhecer os demais parâmetros a ordem $n+1$.

\subsection{Effective action and condensates}

Por fim, uma vez que temos como calcular os parâmetros da teoria
através do grupo de renormalização, podemos calcular a ação efetiva.
Tal cálculo é efetuado a partir da definição
\begin{equation}
e^{-\Gamma}=Z[\sigma_*]\;,
\end{equation}
onde a dependência em $\sigma$ aparece devido ao fato de o cálculo
ser feito a $\sigma$ constante, ou seja, em configurações estáveis
dos campos de Hubbard-Stratonovich. Tal definição é possível
considerando as flutuações em torno de configurações estáveis,
$\sigma=\sigma_*+\tilde{\sigma}$. Desta forma, procuramos soluções
estáveis da ação efetiva de acordo com
\begin{equation}
\frac{\partial\Gamma}{\p\sigma_*}=0\;.
\end{equation}
Uma solução não nula para $\sigma_*$ indica um valor de espera não
trivial para $\mathcal{O}$, o que indica a condensação deste
operador. Obviamente, tal solução deve obedecer a algumas
exigências, como, por exemplo, possuir um parâmetro de expansão
pequeno o suficiente para dar sentido a uma série perturbativa.

O resultado físico que se colhe da condensação é que o termo
$\frac{\sigma^A\mathcal{O}^A}{g\zeta}$ em (\ref{appath1a}) pode vir
a alterar os propagadores da teoria, por exemplo, se
$\mathcal{O}=A^2$, teremos um termo de massa para o glúon.

\chapter{Optimization of the renormalization scheme}\label{ap_rensq}

Neste apêndice discutiremos o processo de otimização do esquema de
renormalização e dependência na escala aplicado ao caso da ação
quântica a um laço da teoria de Gribov-Zwanziger na presença do
operador $A_\mu^aA_\mu^a$, apresentado no capítulo
\ref{cap7_gribov2}. Os detalhes podem ser encontrados em
\cite{Dudal:2005na}.

\section{Preliminaries}

Vimos que, na ação de Gribov-Zwanziger na presença do operador
$A_\mu^aA_\mu^a$, existem apenas três fatores de renormalização a
serem fixados, estes são $Z_g$, $Z_A$ e $Z_\zeta$. Aparentemente,
necessitamos de três condições de renormalização para fixar um
esquema de renormalização particular. Contudo, de acordo com
\cite{Dudal:2005na}, analizando a ação {\it bare} associada a
expressão (\ref{7J1}), não é difícil ver que vale a relação
\begin{equation}
Z_\zeta\zeta=\omu^{\varepsilon}\zeta_oZ_J^2\;.\label{aprel1}
\end{equation}
Como $\zeta_o$ é uma quantidade independente da escala e do esquema
de renormalização e como $\zeta$ sempre aparece, na ação, na
combinação $Z_\zeta\zeta$, segue que apenas $Z_g$ e $Z_A$ são
relevantes para a ação efetiva, pois $Z_J$ sempre pode ser expresso
em termos destes dois fatores. Desta forma, necessitamos apenas de
duas condições de renormalização para fixar o esquema.

Mudaremos do esquema de renormalização $\MSbar$ para outro esquema
não massivo através das seguintes
transformações\footnote{Quantidades com uma barra se referem ao
esquema $\MSbar$.}
\begin{eqnarray}
\og^2&=&g^2\left(1+b_0g^2+b_1g^4+\cdots\right)\;,\nonumber\\
\ol&=&\lambda\left(1+c_0g^2+c_1g^4+\cdots\right)\;,\nonumber\\
\om^2&=&m^2\left(1+d_0g^2+d_1g^4+\cdots\right)\;,\label{aptrans1}
\end{eqnarray}
onde $b_i$, $c_i$ e $d_i$ são parâmetros que caracterizam o novo
esquema. Note que, considerando as relações (\ref{7diman3}), entre
as dimensões anômalas do parâmetro de Gribov e do condensado podemos
facilmente deduzir que
\begin{equation}
\gamma_\lambda(g^2)=\frac{1}{4}\left[\frac{\beta(g^2)}{g^2}-\gamma_{m^2}(g^2)\right]\;,\label{apdiman1}
\end{equation}
implicando numa relação, válida a todas as ordens em teoria de
perturbações, entre os coeficientes $b_i$, $c_i$ e $d_i$. De acordo
com \cite{Dudal:2005na}, esta relação é dada por
\begin{equation}
c_0=\frac{1}{4}(b_0-d_0)\;,\label{aprel2}
\end{equation}
onde foi utilizada a parametrização
\begin{eqnarray}
\beta(g^2)&=&-2\left(\beta_0g^4+\beta_1g^6+\cdots\right)\;,\nonumber\\
\gamma_{m^2}(g^2)&=&\gamma_0g^2+\gamma_1g^4+\cdots\;,\nonumber\\
\gamma_\lambda(g^2)&=&\lambda_0g^2+\lambda_1g^4+\cdots\;,\label{aprel3}
\end{eqnarray}

Agora, efetuando as transformações (\ref{aptrans1}) na ação efetiva
$\MSbar$, (\ref{7eff6}), encontramos
\begin{eqnarray}
\Gamma^{(1)}&=&- \frac{\left( N^{2}-1\right)\lambda
^{4}}{2g^2N}\left(1+4c_0g^2-b_0g^2\right)+\frac{\zeta_0m^4}{2g^2}\left(
1-\frac{\zeta _{1}}{\zeta _{0}}g^{2}+2d_0g^2-b_0g^2\right)
\nonumber\\&+&\frac{3\left( N^{2}-1\right)
}{256\pi^2}\left[\left(m^2+\sqrt{m^4-\lambda^4}\right)^2
\left(\ln\frac{m^2+\sqrt{m^4-\lambda^4}}{2\omu^2}-\frac{5}{6}\right)\right.\nonumber\\
&+&\left.\left(m^2-\sqrt{m^4-\lambda^4}\right)^2
\left(\ln\frac{m^2-\sqrt{m^4-\lambda^4}}{2\omu^2}-\frac{5}{6}\right)+\lambda^4\right]\;,\label{apeff1}
\end{eqnarray}
enquanto as equações de gap se tornam
\begin{eqnarray}
\frac{\p\Gamma}{\p\lambda}&=&-\frac{2\left(N^2-1\right)}{g^2N}\lambda^3\left(1+4c_0g^2-b_0g^2\right)+
\frac{3\left(N^2-1\right)\lambda^3}{256\pi^2}
\left[\frac{20}{3}\vphantom{-4\frac{\left(m^2+\sqrt{m^4-\lambda^4}\right)}{\sqrt{m^4-\lambda^4}}
\ln\frac{m^2+\sqrt{m^4-\lambda^4}}{2\omu^2}}\right.\nonumber\\
&-&4\frac{\left(m^2+\sqrt{m^4-\lambda^4}\right)}
{\sqrt{m^4-\lambda^4}}\ln\frac{m^2+\sqrt{m^4-\lambda^4}}{2\omu^2}\nonumber\\
&+&\left.4\frac{\left(m^2-\sqrt{m^4-\lambda^4}\right)}
{\sqrt{m^4-\lambda^4}}\ln\frac{m^2-\sqrt{m^4-\lambda^4}}{2\omu^2}\right]=0\;,\label{apgap1}
\end{eqnarray}
e
\begin{eqnarray}
\frac{\p\Gamma}{\p m^2}&=&
\frac{\zeta_0m^2}{g^2}\left(1-\frac{\zeta_1}{\zeta_0}g^2+2d_0g^2-b_0g^2\right)\nonumber\\&+&\frac{3\left(N^2-1\right)}{256\pi^2}
\left[2\left(m^2+\sqrt{m^4-\lambda^4}\right)\left(1+\frac{m^2}{\sqrt{m^4-\lambda^4}}\right)\ln\frac{m^2+\sqrt{m^4-\lambda^4}}{2\omu^2}\right.\nonumber\\
&+&\left.2\left(m^2-\sqrt{m^4-\lambda^4}\right)\left(1-\frac{m^2}{\sqrt{m^4-\lambda^4}}\right)\ln\frac{m^2-\sqrt{m^4-\lambda^4}}{2\omu^2}-
\frac{8}{3}m^2\right]=0\;.\nonumber\\
&&\label{apgap2}
\end{eqnarray}

Podemos utilizar as equações de gap (\ref{apgap1}-\ref{apgap2})
juntamente com a relação (\ref{aprel2}), simi-larmente como foi
feito para se encontrar a relação (\ref{7gapt2}), encontrando
\begin{equation}
\frac{68}{39}\left(\frac{16\pi^2}{g^2N}\right)+\frac{122}{39}+\frac{16\pi^2}{N}\left(\frac{12}{13}b_0-\frac{176}{39}d_0\right)=
\frac{1}{2}\frac{t}{\sqrt{1-t}}\ln\frac{t}{\left(1+\sqrt{1-t}\right)^2}\;.\label{apcond1}
\end{equation}
Desta expressão fica evidente que uma solução com $m^2>0$ pode
existir, dependendo dos valores dos parâmetros $d_0$ (renormalização
da massa) e $b_0$ (renormalização da constante de acoplamento).
Desta forma, para minimizar a liberdade existente na escolha destes
parâmetros vamos realizar uma otimização na dependência do esquema
de renormalização, como ilustrado em
\cite{VanAcoleyen:2001gf,Dudal:2002zn}.

\section{Optimization of the renormalization scheme}

Consideramos uma quantidade $\varrho$ que corre de acordo com
\begin{equation}
\omu\frac{d\varrho}{d\omu}=\gamma_\varrho(g^2)\varrho\;,\label{apro1}
\end{equation}
onde
\begin{equation}
\gamma_\varrho(g^2)=\gamma_{\varrho,0}g^2+\gamma_{\varrho,1}g^4+\cdots\;.\label{apro2}
\end{equation}
Podemos associar a $\varrho$ uma quantidade $\widehat{\varrho}$ que
não depende da escolha do esquema de renormalização e é independente
da escala. Tal quantidade é definida como
\begin{equation}
\wrho=\mf_\varrho(g^2)\varrho\;,\label{apro3}
\end{equation}
com
\begin{equation}
\omu\frac{d}{d\omu}\mf_\varrho(g^2)=-\gamma_\varrho(g^2)\mf_\varrho(g^2)\;.\label{apro4}
\end{equation}
Desta forma $\wrho$ não dependerá da escala $\omu$. É fácil checar
ainda que $\widehat{\varrho}$ é invariante sob transformações do
tipo (\ref{aptrans1}), \cite{VanAcoleyen:2001gf,Dudal:2002zn}. A
equação (\ref{apro4}) pode ser resolvida com a ajuda da relação
\begin{equation}
\omu\frac{d}{d\omu}\mf_\varrho(g^2)\equiv\beta(g^2)\frac{d}{dg^2}\mf_\varrho(g^2)\;.\label{apro5}
\end{equation}
De forma que a equação (\ref{apro4}) pode ser resolvida numa série
em $g^2$,
\begin{equation}
\mf_\varrho(g^2)=(g^2)^{\frac{\gamma_{\varrho,0}}{2\beta_0}}\left[1+\frac{1}{2}
\left(\frac{\gamma_{\varrho,1}}{\beta_0}-\frac{\beta_1\gamma_{\varrho,0}}{\beta_0^2}\right)g^2+\cdots\right]\;.\label{apro6}
\end{equation}

Vamos agora substituir as variáveis $\MSbar$, $\om^2$ e $\ol$, por
seus correspondentes independentes de escala e do esquema de
renormalização, $\wm^2$ e $\wl$, na ação efetiva (\ref{7eff6}).
Invertendo a relação (\ref{apro6}), temos
\begin{eqnarray}
\om^2 &=& (\og^2)^{-\frac{\gamma_{0}}{2\beta_0}}\left[1-\frac{1}{2}
\left(\frac{\overline{\gamma}_{1}}{\beta_0}-\frac{\beta_1\gamma_{0}}{\beta_0^2}\right)\og^2+\cdots\right]\wm^2\;, \\
\label{opt6bis} \ol&=&
(\og^2)^{-\frac{\lambda_{0}}{2\beta_0}}\left[1-\frac{1}{2}
\left(\frac{\overline{\lambda}_{1}}{\beta_0}-\frac{\beta_1\lambda_{0}}{\beta_0^2}\right)\og^2+\cdots\right]\wl\;.\label{apinv1}
\end{eqnarray}
Substituindo (\ref{apinv1}) e (\ref{aptrans1}) na ação efetiva
(\ref{7eff6}) chegamos a\footnote{Onde utilizamos a notação mais
simplificada
\begin{eqnarray}\label{opt6tris}
  a &=& -\frac{\gamma_{0}}{2\beta_0}\;,\;\;\;\;\;\;\;\;\;\;\;\;\;\;\;\;\;\;\;\;\;\;\;\;b=-\frac{\lambda_{0}}{\beta_0}\;, \\
  A &=& -\left(\frac{\overline{\gamma}_{1}}{\beta_0}-\frac{\beta_1\gamma_{0}}{\beta_0^2}\right)\;,\;\;\;\;\;\;\;
  B=-2\left(\frac{\overline{\lambda}_{1}}{\beta_0}-\frac{\beta_1\lambda_{0}}{\beta_0^2}\right)\;,
\end{eqnarray}}
\begin{eqnarray}
\Gamma^{(1)}&=&-\frac{\left( N^{2}-1\right)}{2N}(g^2)^{2b}\wl
^{4}\left(\frac{1}{g^2}+B-b_0+2bb_0\right)\nonumber\\&+&\frac{\zeta_0}{2}\wm^4(g^2)^{2a}\left(
\frac{1}{g^2}+A-b_0+2ab_0-\frac{\zeta _{1}}{\zeta
_{0}}\right)+\frac{3\left( N^{2}-1\right) }{256\pi^2}\times
\nonumber\\&&\left[\left(\wm^2(g^2)^a+\sqrt{\wm^4(g^2)^{2a}-\wl^4(g^2)^{2b}}\right)^2
\left(\ln\frac{\wm^2(g^2)^{a}+\sqrt{\wm^4(g^2)^{2a}-\wl^4(g^2)^{2b}}}{2\omu^2}-\frac{5}{6}\right)\right.\nonumber\\
&+&\left.\left(\wm^2(g^2)^{a}-\sqrt{\wm^4(g^2)^{2a}-\wl^4(g^2)^{2b}}\right)^2
\left(\ln\frac{\wm^2(g^2)^{a}-\sqrt{\wm^4(g^2)^{2a}-\wl^4(g^2)^{2b}}}{2\omu^2}-\frac{5}{6}\right)\right.\nonumber\\
&+&\left.\wl^4(g^2)^{2b}\right]\;. \label{apeff2}
\end{eqnarray}
Portanto, construímos uma ação dependente apenas das quantidades
independentes da escala e do esquema de renormlaização $\wl$ e
$\wm^2$ e da constante de acoplamento $g^2(\omu)$. A liberdade no
esquema de renormalização reside agora no parâmetro $b_0$, e na
escala $\omu$. O último passo é eliminar a dependência na escala.
Para tal, ao invés de utilizarmos uma expansão em $g^2$, fazemos uma
expansão em potências inversas da quantidade
\begin{equation}
x\equiv\beta_0\ln\frac{\omu^2}{\Lambda^2}\;,\label{apx1}
\end{equation}
veja \cite{VanAcoleyen:2001gf,Dudal:2002zn}. A constante de
acoplamento $g^2$ pode ser substituida por $x$ uma vez que $g^2$ é
determinado explicitamente por\footnote{A relação entre a escala
$\Lambda$ e $\Lambda_{\MSbar}$ é dada, de acordo com
\cite{Celmaster:1979km}, por
\begin{equation}
\Lambda= e^{-\frac{b_0}{2\beta_0}}\lms\;.\label{aplambda1}
\end{equation}}
\begin{equation}
g^2=\frac{1}{x}\left(1-\frac{\beta_1}{\beta_0}\frac{\ln\frac{x}{\beta_0}}{x}+\cdots\right)\;.\label{apg1}
\end{equation}
Com isso, chegamos, finalmente a,
\begin{eqnarray}
\Gamma^{(1)}&=&-\frac{\left( N^{2}-1\right)}{2N}x^{-2b}\wl
^{4}\left(x+B+(1-2b)\left(\frac{\beta_1}{\beta_0}\ln\frac{x}{\beta_0}-b_0\right)\right)\nonumber\\
&+&\frac{\zeta_0}{2}\wm^4x^{-2a}\left(x+A-\frac{\zeta _{1}}{\zeta
_{0}}+(1-2a)\left(\frac{\beta_1}{\beta_0}\ln\frac{x}{\beta_0}-b_0\right)\right)+\frac{3\left(
N^{2}-1\right) }{256\pi^2}\times
\nonumber\\&&\left[\left(\wm^2x^{-a}+\sqrt{\wm^4x^{-2a}-\wl^4x^{-2b}}\right)^2
\left(\ln\frac{\wm^2x^{-a}+\sqrt{\wm^4x^{-2a}-\wl^4x^{-2b}}}{2\omu^2}-\frac{5}{6}\right)\right.\nonumber\\
&+&\left.\left(\wm^2x^{-a}-\sqrt{\wm^4x^{-2a}-\wl^4x^{-2b}}\right)^2
\left(\ln\frac{\wm^2x^{-a}-\sqrt{\wm^4x^{-2a}-\wl^4x^{-2b}}}{2\omu^2}-\frac{5}{6}\right)+\wl^4x^{-2b}\right]
\;,\nonumber\\
\label{apeff3}
\end{eqnarray}
Observamos que esta expressão é válda até a ordem
$\left(\frac{1}{x}\right)^0$. As equações de gap resultantes são
agora dadas por
\begin{eqnarray}
\label{apgap3} \frac{1}{\wl^3}\frac{\p\Gamma}{\p\wl}&=&0\;,\\
\label{apgap4}\frac{1}{\wm^2}\frac{\p\Gamma}{\p\wm^2}&=&0\;.
\end{eqnarray}

Em princípio resolveríamos as equações de gap
(\ref{apgap3}-\ref{apgap4}) para duas quantidades $\wm_*$ e $\wl_*$,
como funções de $\omu$ e $b_0$. Contudo, por construção, $\wm$ e
$\wl$ são independentes da escala e do esquema de renormalização,
ordem a ordem. Desta forma temos um método interessante de fixar
estes parâmetros, escolhendo as soluções $\wm_*(\omu,b_0)$ e
$\wl_*(\omu,b_0)$ de forma a dependerem minimamente de $b_0$ e
$\omu$. Para facilitar os cálculos escolhemos a seguinte
escala\footnote{Graças a independencia na escala das quantidades
$\widehat{\;\;}$. Ainda, observamos que o outro logaritimo
$\ln\frac{\wm^2x^{-a}-\sqrt{\wm^4x^{-2a}-\wl^4x^{-2b}}}{2\omu^2}$, é
inofensível pois sempre aparece na combinação $u\ln u$. Assim, se o
argumento do logarítimo for pequeno, este nunca será grande, devido
ao limite $\left.u\ln u\right|_{u\approx0}\approx0$.}
\begin{equation}
\omu^2=\left|\frac{\wm^2x^{-a}+\sqrt{\wm^4x^{-2a}-\wl^4x^{-2b}}}{2}\right|\;,\label{apmu1}
\end{equation}
de modo que agora, temos apenas a liberdade no esquema de
renormalização através do parâmetro $b_0$, que ainda resta a ser
fixado. De fato, fixamos $b_0$ de forma que $\wl_*$, $\wm_*$ e
$E_\mathrm{vac}$ dependam, ao mesmo tempo, minimamente de $b_0$.
Esta condição é atingida requerendo que a função
\begin{equation}
\Upsilon(b_0)\equiv\left|\frac{\p\wl_*^4}{\p
b_0}\right|+\left|\frac{\p\wm_*^4}{\p b_0}\right|+\left|\frac{\p
E_\mathrm{vac}}{\p b_0}\right|\;,\label{apfunc1}
\end{equation}
dependa minimamente de $b_0$. Mesmo que não consigamos a condição
ideal, $\Upsilon=0$, podemos nos satisfazer com a condição mais
fraca de que $\Upsilon$ seja o menor possível, analogamente ao
\emph{princípio de mínima percepção}\footnote{Do inglês:
\emph{principle of minimal sensitivity}.} \`{a} la Stevenson
\cite{Stevenson:1981vj}.

\begin{figure}[h]
\begin{tabular}{cc}
  \scalebox{0.95}{\includegraphics{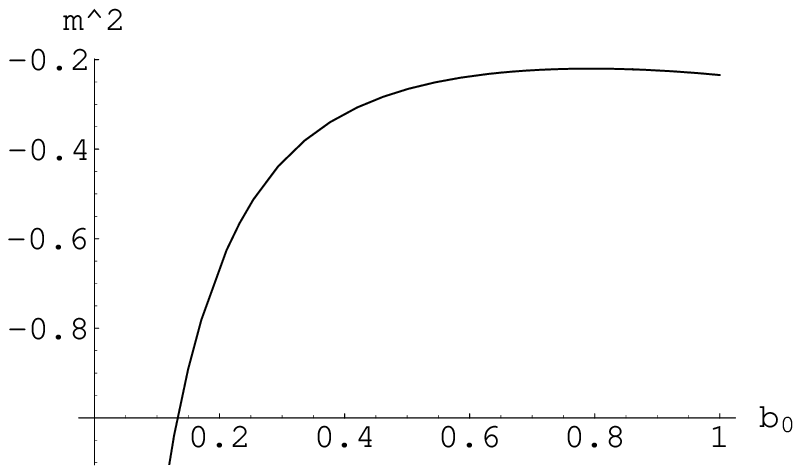}} & \scalebox{0.95}{\includegraphics{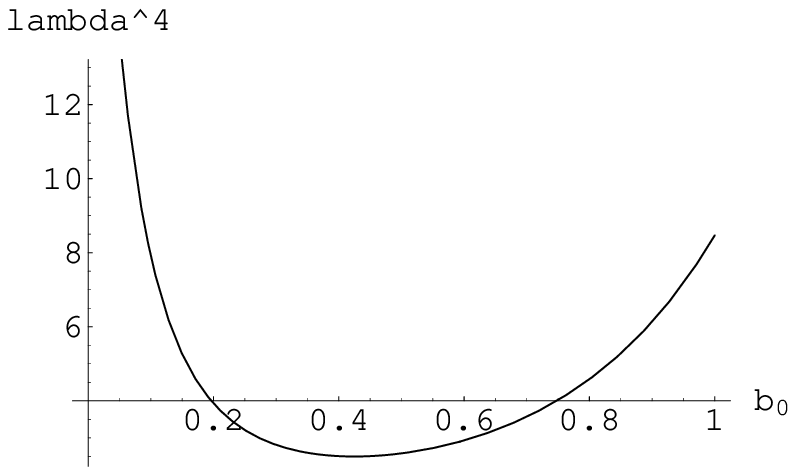}}
\end{tabular}
\caption{Quantidades $\wm^2$ e $\wl^4$ como funções de $b_0$, em
unidades de $\lms$.}\label{apfig1}
\end{figure}

\begin{figure}[h]
\begin{tabular}{cc}
  \scalebox{0.95}{\includegraphics{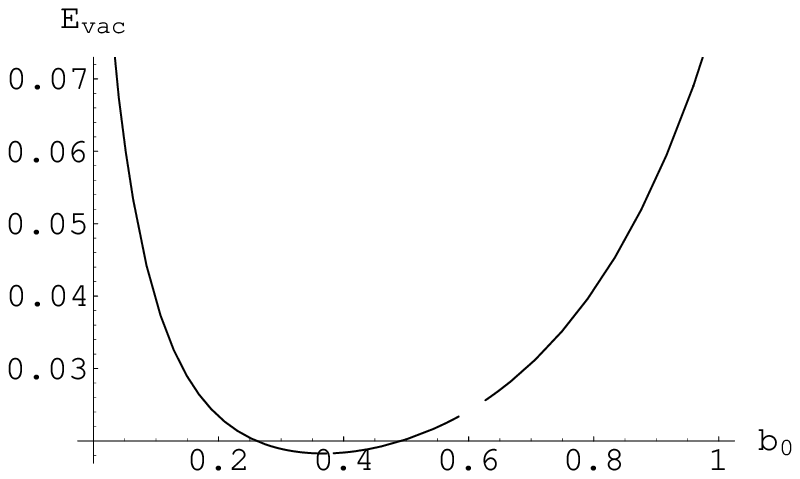}} & \scalebox{0.95}{\includegraphics{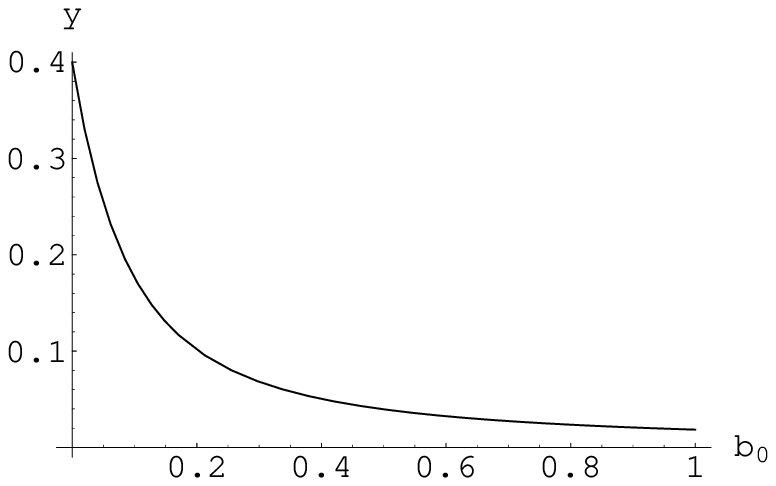}}
\end{tabular}
\caption{Energia do vácuo $\Evac$ e o parâmetro de expansão $y$ como
funções de $b_0$, em unidades de $\lms$.}\label{apfig2}
\end{figure}

Colocamos os gráficos de $\wm^2$ e $\wl^4$ como função de $b_0$ na
Fig. \ref{apfig1} enquanto que $\Evac$ e o parâmetro de expansão
$y\equiv\frac{N}{16\pi^2 x}$ na Fig. \ref{apfig2}.

\section{Numerical results}

Para começarmos com os cálculos explícitos necessitamos de alguns
fatores numéricos. De \cite{Gracey:2002yt}, temos que
\begin{eqnarray}
\beta_1&=&\frac{34}{3}\left(\frac{N}{16\pi^2}\right)^2
\;,\;\;\;\;\;\;\;\;\;\;\gamma_0 =
-\frac{3}{2}\frac{N}{16\pi^2}\;,\;\;\;\;\;\;\;\;\;\;\gamma_1=-\frac{95}{24}\left(\frac{N}{16\pi^2}\right)^2\;,\label{apnum1}
\end{eqnarray}
e, da relação (\ref{apdiman1}),
\begin{eqnarray}
\lambda_0=-\frac{35}{24}\frac{N}{16\pi^2}\;,\;\;\;\;\;\;\;\;\;\;\lambda_1=-\frac{449}{96}\left(\frac{N}{16\pi^2}\right)^2\;.\label{apnum2}
\end{eqnarray}

Pouparemos o leitor das técnicas e truques algébricos utilizados
para resolver as equações de gap (\ref{apgap3}-\ref{apgap4}), os
detalhes podem ser encontrados em \cite{Dudal:2005na}. Lembramos
ainda que as equações de gap foram resolvidas numericamente, com a
ajuda dos \emph{softwares} Maple V e Mathematica. Apenas observamos
que o caso em que a cosntante de acoplamento coincide com aquela
utilizada no esquema $\MSbar$, dada pela escolha $b_0=0$, fornece os
resultados listados em (\ref{7opt1}). Quando resolvemos a condição
de mínimo para a função $\Upsilon$, em (\ref{apfunc1}), encontramos
para o valor ótimo $b_0^*\approx0.425$. Neste caso as soluções estão
em (\ref{7opt2}).

\bibliographystyle{unsrt}
%\bibliography{bib}

\end{document}